\documentclass[a4paper, twoside]{article}

\usepackage{amsmath}
\usepackage{amssymb}
\usepackage{bm}
\usepackage{caption}
\usepackage{enumitem}
\usepackage{epigraph}
\usepackage{fancyhdr}
\usepackage{float}
\usepackage{graphicx} 
\usepackage{hyperref}
\usepackage{lettrine}
\usepackage{pdfpages}
\usepackage{physics}
\usepackage{titlesec}
\usepackage{xcolor}
\usepackage[top=2cm, bottom=1.5cm, left=2.5cm, right=2cm]{geometry}

\newcommand{\red}[1]{\textcolor{red}{#1}}
\newcommand{\blue}[1]{\textcolor{blue}{#1}}

\definecolor{pubcol}{RGB}{67, 83, 52}  
\newcommand{\publicationcolor}[1]{\textcolor{pubcol}{#1}}

\definecolor{colcol}{RGB}{237, 123, 123}  
\newcommand{\collaborationcolor}[1]{\textcolor{colcol}{#1}}

\definecolor{confcol}{RGB}{131, 96, 150}  
\newcommand{\conferencescolor}[1]{\textcolor{confcol}{#1}}

\definecolor{oraltalk}{RGB}{191,147,13}  
\newcommand{\oraltalkcolor}[1]{\textcolor{oraltalk}{#1}}

\definecolor{poster}{RGB}{138, 149, 151}  
\newcommand{\postercolor}[1]{\textcolor{poster}{#1}}

\title{Hybrid Quantum-Classical Algorithms}

\titleformat{\section}[display]
  {\normalfont\Huge\bfseries\raggedleft}
  {\textcolor{gray}{\scalebox{3}{\thesection}}}
  {1em}
  {}
\titlespacing*{\section}{0pt}{2ex}{6ex} 

\titleformat{\part}[display]
  {\normalfont\Huge\bfseries\centering}
  {}
  {0pt}
  {}

\begin{document}
\pagenumbering{roman}

\includepdf[pages=-]{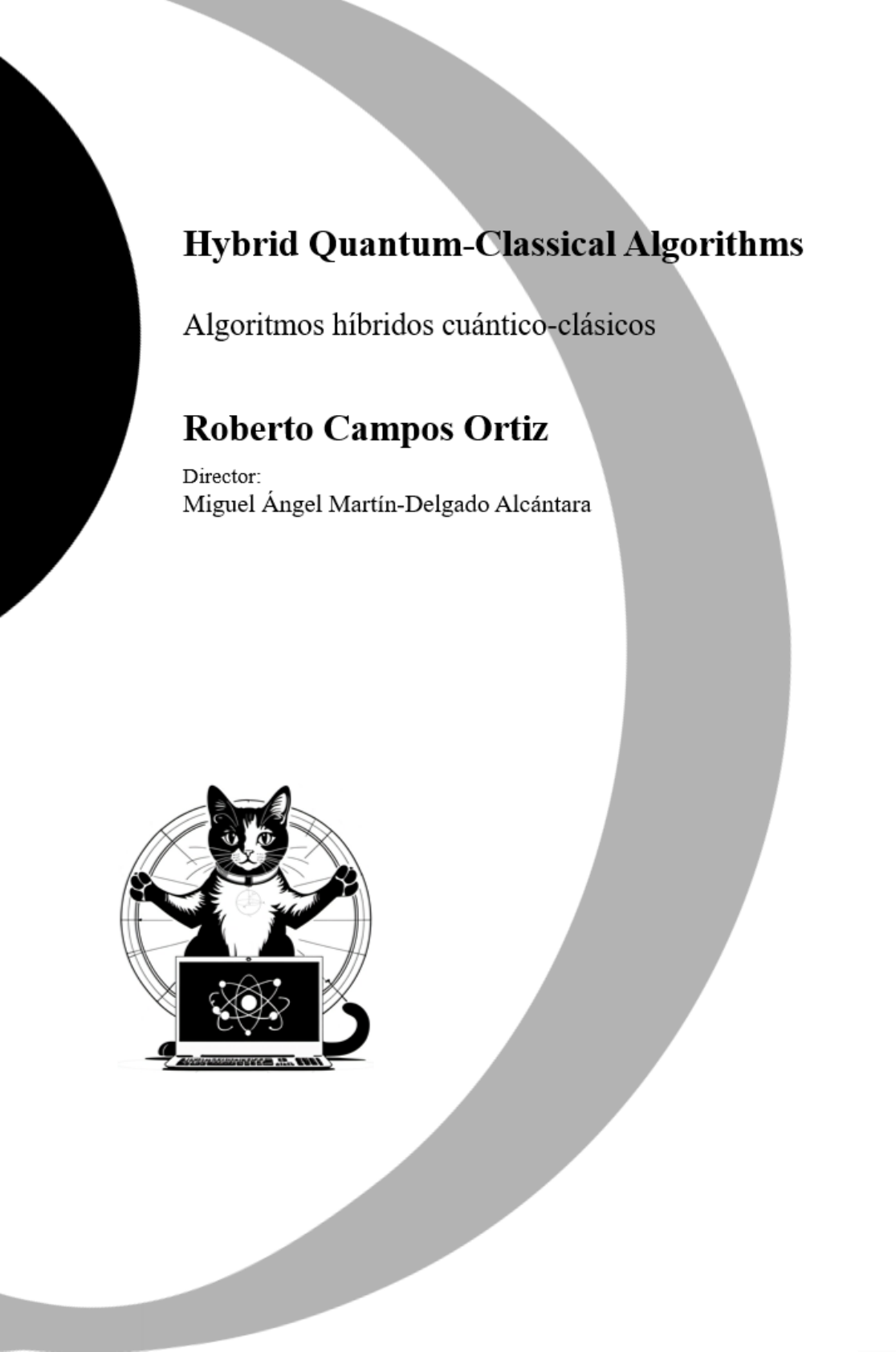}





\newpage\null\thispagestyle{empty}\newpage
\newpage\null\thispagestyle{empty}\newpage

\newpage
\section*{Acknowledgements}
\addcontentsline{toc}{section}{Acknowledgements}
\thispagestyle{empty}

\vspace{1cm}

Una tesis doctoral queda lejos de la mayoría de mi familia y amigos. Así que estos agradecimientos no se los dedico por haber contribuido al proceso concreto, académico y sesudo, sino que me gustaría agradecerles haberme ayudado, en estos 31 años, a llegar a ser la persona que ha podido terminar esta tesis. 

Me gustaría agradecerles a mis padres, Cory y Julio, haberme criado con amor, cariño, dedicación, inteligencia, sabiduría y una larga lista más de adjetivos positivos. Ellos me enseñaron a entender y amar la vida y a saber que no soy inferior a nadie. También fomentaron en mí la curiosidad y el placer del conocimiento, dos requisitos imprescindibles para hacer un doctorado. En resumen, lograron enseñarme la felicidad de ser uno mismo.

Mi agradecimiento más profundo es para Ángeles, quiero agradecerle todo, sin ninguna palabra más, porque cualquier palabra se quedaría pequeña para expresar nuestra relación. Estoy orgulloso de la persona que soy y hay mucho de ella en mí. Ignoro el futuro, pero sé que con ella, todo es mejor.

A Miguel Ángel Martín-Delgado, mi director de tesis, quiero agradecerle, primero, haber sabido explicarme con paciencia y entusiasmo los fundamentos de esta ciencia. Y, segundo, haberme enseñado cosas tan valiosas como liderazgo, perspectiva y, espero, cierta clarividencia para entender un mundo tan cambiante como el de la computación cuántica. Este agradecimiento es extensible a Carmen Page, nuestro apoyo en la sombra científica, siempre pendiente de nosotros para lo que haga falta.

Otra persona que no puede faltar en estos agradecimientos es Aitor Ibarra, mi otro director de tesis, más moral que académico. Mi primer contacto con Quasar y un apoyo constante durante la tesis. Me quedo con su incansable energía para sacar proyectos adelante contra todo y todos, aunque fuesen de alto riesgo. Ha sido un referente de cómo la tenacidad puede convertir una quimera en realidad. 

También quiero agradecer a dos empresas su apoyo. Primero, a Quasar y, a Ignacio de la Calle, el tutor de la empresa por ser la otra mitad de este doctorado industrial, por entender que la investigación también puede ser rentable. También, me gustaría agradecer el soporte del programa de Doctorados Industriales de la Comunidad de Madrid 2019 con referencia IND2019/TIC17146. Segundo a Zapata Computing, especialmente al que fue todo el grupo de QML y a Alejandro Perdomo como cabeza visible. El tiempo que estuve con ellos me sentí uno más de la familia, llevándome a partes iguales conocimiento científico y aprendizaje vital.

Por último, pero no por ello menos importante, quiero agradecer a toda la gente del departamento. Con ellos he compartido el día a día de este doctorado, esas personas con las que la rutina del trabajo siempre es agradable. No puedo poner la lista con todos los nombres que debería, pero sí mencionar algunos en especial. A Pablo Moreno, mi compañero durante una parte importante del doctorado, por enseñarme tanto con tanta paciencia. A Gabriel Escrig, por ser compañero y amigo, por el trabajo y las horas juntos, por la física y por la filosofía compartida. A Sara Giordano, por ser amiga y guía, por las muchas horas de charla y risas, tanta investigación como amistad. Y a Pablo Rabán, por ser un compañero de despacho que se convirtió en amigo, siempre atento, con una sonrisa, una canasta o un chiste, por ser tan grande.

Quiero incluir, en estos agradedicimientos, una última reflexión. Habrá gente que no aparezca explícitamente, pero, si en estos 31 años, has pasado tiempo conmigo y me has visto sonreir, de una forma u otra, el tiempo juntos me ha ayudado a ser como soy, por eso, gracias a ti también. Gracias a todos.

\newpage\null\thispagestyle{empty}\newpage

\newpage
\section*{Abstract}
\thispagestyle{empty}
\addcontentsline{toc}{section}{Abstract}

\lettrine[lines=1, findent=2pt]{\resizebox{!}{1.2\baselineskip}{H}}{}ybrid algorithms combine classical and quantum computing to create a whole that improves the performance of classical algorithms. There are many ways to understand this combination, in this thesis two are studied. First, it is an approach that combines classical and quantum modules communicating with each other to create a search \& sample optimization algorithm. The second one is a classical algorithm that studies the cost and performance of quantum algorithms applied to chemistry.

Hybrid algorithms are interesting due to the limitations of both classical and quantum computing. Some hard problems solved by classical computing are limited in performance and scalability because of the exponentially growing number of states to be evaluated. On the other hand, currently, quantum computing is limited because quantum hardware is not yet developed enough to execute algorithms that process large amounts of data. By combining both technologies, it is possible to obtain an algorithm capable of processing large volumes of data with greater speed.

The first proposed algorithm in this work, quantum Metropolis Solver, QMS, is an adaptation of a quantum walk to a quantum Metropolis-Hastings algorithm, applied to problems of industrial interest such as artificial intelligence, space exploration, energy sector or quantum chemistry. In these use cases, QMS has proven to have an advantage over its classical counterpart, being tested on both simulator and quantum hardware.

The second proposed hybrid algorithm, TFermion, follows a different paradigm. It is a classical algorithm that analyzes the cost of T-type gate of quantum algorithms applied to quantum chemistry. Knowing this cost is crucial to compare the ability of algorithms to be executed on real quantum hardware. The main algorithm of TFermion is to generalize this calculation to other methods and apply it to any molecule. Then, it has been applied to a real problem, the design of more efficient electric batteries.

\newpage\null\thispagestyle{empty}\newpage

\newpage
\section*{\publicationcolor{Publications}, \collaborationcolor{Industrial Collaborations} and \conferencescolor{Conferences}}
\thispagestyle{empty}
\addcontentsline{toc}{section}{Publications, Industrial Collaborations and Conferences}
\markboth{Publications, Industrial Collaborations and Conferences}{Publications, Industrial Collaborations and Conferences}

\begin{itemize}
    \item[\publicationcolor{\textbf{P1*}}] \publicationcolor{R. Campos, P.A.M. Casares, M.A. Martin-Delgado, \textit{Quantum Metropolis Solver: A Quantum Walks Approach to Optimization Problems}, Quantum Machine Intelligence \textbf{5} 00119 (2023)} \footnote{IMPACT FACTOR (IF) 2024: Quantum Machine Intelligence. \textbf{IF: 5.64}, Class. Quantum Grav. \textbf{IF: 3.52}, Quantum Sci. Technol. \textbf{IF: 6.7}, Quantum. \textbf{IF: 6.4}, Phys. Rev. A. \textbf{IF: 2.408}}
    
    \item[\publicationcolor{\textbf{P2*}}] \publicationcolor{P.A.M. Casares, R. Campos, M.A. Martin-Delgado, \textit{QFold: quantum walks and deep learning to solve protein folding}, Quantum Sci. Technol. \textbf{7} 025013}
    
    \item[\publicationcolor{\textbf{P3*}}] \publicationcolor{P.A.M. Casares, R. Campos, M.A. Martin-Delgado, \textit{TFermion: A non-Clifford gate cost assessment library of quantum phase estimation algorithms for quantum chemistry}, Quantum \textbf{6} 768}
    
    \item[\publicationcolor{\textbf{P4}}] \publicationcolor{G. Escrig, R. Campos, P.A.M. Casares, M.A. Martin-Delgado, \textit{Parameter estimation of gravitational waves with a quantum metropolis algorithm}, Class. Quantum Grav. \textbf{4} 045001 (2023)}

    \item[\publicationcolor{\textbf{P5}}] \publicationcolor{G. Escrig, R. Campos, P.A.M. Casares, H. Qi, M.A. Martin-Delgado, \textit{Quantum Bayesian Inference with Renormalization for Gravitational Waves}, \url{https://arxiv.org/abs/2403.00846}}.

    \item[\publicationcolor{\textbf{P6}}] \publicationcolor{A. Delgado, P.A.M. Casares, R. Dos Reis, M. Shokrian Zini, R. Campos, N. Cruz-Hernandez, A. Voigt, A. Lowe, S. Jahangiri, M. A. Martin-Delgado, J. E. Mueller, J. M. Arrazola, \textit{Simulating key properties of lithium-ion batteries with a fault-tolerant quantum computer}, Phys. Rev. A \textbf{3} 032428}
    
    \vspace{1cm}

    \textbf{*} Publications that the original paper appear in the thesis.

    \vspace{1cm}

   \underline{\textbf{Awards}}

    \begin{itemize}
        \item [AW1] \textbf{3MT}: Accesit award in the contest 3 Minutes Thesis contest organized by Universidad Complutense de Madrid. 3MT is an outreach competition in which Ph.D. students had to explain their research in three minutes to a non-expert audience with only one slide, from a static position.
    \end{itemize}   
    
\end{itemize}

\newpage
\begin{itemize}
    \item[\collaborationcolor{IC1}] \collaborationcolor{Company QuasarS.R.:} Quasar is the industrial collaborator associated with this thesis through the framework of an Industrial PhD program supervised by Aitor Ibarra and Ignacio de la Calle. Quasar is a company operating within the space sector, with a specialized focus on providing artificial intelligence solutions for the European Space Agency.
    
    \item[\collaborationcolor{IC2}] \collaborationcolor{Company Zapata Computing:} An internship took place at this company from June to November 2022. This engagement was with the quantum artificial intelligence team, led by Alejandro Perdomo-Ortiz, and focused on an in-depth exploration of quantum generative models. Zapata Computing is a specialized company in the development of industrial-level quantum software.
    
    \item[\collaborationcolor{IC3}] \collaborationcolor{Company Xanadu:} This collaboration arose from the publication \ref{pub:p3} in conjunction with the quantum algorithms and chemistry group at Xanadu, led by Juan Miguel Arrazola. The outcome of this collaborative effort, which also involved the automobile company Volkswagen, led to the publication of \cite{delgado2022}. Xanadu is a comprehensive quantum computing company engaged in the development of hardware (HW) and software (SW) solutions.
    
    \item[\collaborationcolor{IC4}] \collaborationcolor{LIGO Project:} The partnership with the LIGO collaboration and Queen Mary University of London began due to their interest in the publication of \cite{escrig2023}. The main goals were to improve the quality of the data and employ data representation methods similar to those used by LIGO researchers. LIGO's focus involves building advanced interferometers for measuring and analyzing gravitational waves. This collaboration ends with the publication of \cite{escrig2024}.
    
    \item[\collaborationcolor{IC5}] \collaborationcolor{Company Repsol:} Collaboration with this corporation was initiated after their initial interest in the \ref{pub:p1} publication. The core focus of this collaboration is the formulation of a quantum algorithm for the optimization of green hydrogen production. Repsol is a prominent Spanish conglomerate specializing in the manufacturing and distribution of fuels.

    \item[\collaborationcolor{IC6}] \collaborationcolor{Company Amazon-AWS:} Collaboration with the quantum computing team at AWS, QFold library, publication \ref{pub:p2} was integrated as a case study for drug design. AWS provides an industrial platform for the execution of quantum algorithms designed by various entities. Their choice to spotlight QFold from its perceived relevance to companies and hospitals engaged in cancer treatment research.
    
    \item[\collaborationcolor{IC7}] \collaborationcolor{Nebrija University:} This affiliation emerged as a result of the industrial component of this doctoral endeavor, Quasar. Nebrija University expressed interest in establishing a Spanish startup network encompassing research domains linked to quantum computing. This connection culminated in diverse collaborative activities, including didactic assistance. Nebrija University stands as one of the pioneering Spanish institutions to introduce a dedicated master's program exclusively focused on quantum computing.

    \item[\collaborationcolor{IC8}] \collaborationcolor{Wikimedia Foundation:} Wikimedia Foundation created a program to develop Wikipedia content. The objective was to identify ten Wikipedia entries on a specific topic that had not already been created and have five students develop those entries under the supervision of an expert. Ten entries on quantum computing were chosen and supervised within this thesis.

\end{itemize}

\newpage
\begin{itemize}
    \item [\conferencescolor{\textbf{C1-}}] ICE-6 (\postercolor{\textbf{poster}}), May, Online, 2020
    \item [\conferencescolor{\textbf{C2-}}] Quantum Techniques in Machine Learning, November, Online, 2020
    \item [\conferencescolor{\textbf{C3-}}] Quantum Chemistry with Qiskit, November, Online, 2020
    \item [\conferencescolor{\textbf{C4-}}] IEEE Quantum Week (\oraltalkcolor{\textbf{seminar}}), October, Online, 2021
    \item [\conferencescolor{\textbf{C5-}}] Quantum Matter (\oraltalkcolor{\textbf{talk}}), November, Bilbao, Spain, 2021
    \item [\conferencescolor{\textbf{C6-}}] QHACK, February, Online, 2022
    \item [\conferencescolor{\textbf{C7-}}] APS March Meeting (\oraltalkcolor{\textbf{talk}}), March, Chicago, USA, 2022
    \item [\conferencescolor{\textbf{C8-}}] EQAI Summer School (\postercolor{\textbf{poster}}), September, Udine, Italy, 2022
    \item [\conferencescolor{\textbf{C9-}}] QHACK, February, Online, 2023
    \item [\conferencescolor{\textbf{C10-}}] APS March Meeting (\oraltalkcolor{\textbf{talk}}), March, Las Vegas, USA, 2023
    \item [\conferencescolor{\textbf{C11-}}] Quantum Matter (\postercolor{\textbf{poster}}), May, Madrid, Spain, 2023
    \item [\conferencescolor{\textbf{C12-}}] HDCRS Summer School, May, Reikiavik, Iceland, 2023
    \item [\conferencescolor{\textbf{C13-}}] Tech Talk European Space Agency, (\oraltalkcolor{\textbf{talk}}), June, Online, 2023
    \item [\conferencescolor{\textbf{C14-}}] QTYR, July, Madrid, Spain, 2023
    \item [\conferencescolor{\textbf{C15-}}] QTS12 (\oraltalkcolor{\textbf{talk}}), July, Praga, Czech Republic, 2023
    \item [\conferencescolor{\textbf{C16-}}] Granada Seminar (\oraltalkcolor{\textbf{talk}}), September, Granada, Spain, 2023
    \item [\conferencescolor{\textbf{C17-}}] MaDQuantum, (\postercolor{\textbf{poster and stand}}), September, Madrid, Spain, 2023
    \item [\conferencescolor{\textbf{C18-}}] Quantum Techniques in Machine Learning, (\postercolor{\textbf{poster}}), November, Geneva, Switzerland, 2023

\end{itemize}

\newpage\null\thispagestyle{empty}\newpage

\newpage
\section*{Detailed summary in English}
\addcontentsline{toc}{section}{Detailed summary in English}
\thispagestyle{empty}
\markboth{Detailed summary in English}{Detailed summary in English}

\large \textbf{Introduction}
\\

The current situation of quantum computing is not much different from that of a giant with feet of clay. Without being able to demonstrate a practical quantum advantage, there is strong investment at both the industrial and academic levels to develop quantum algorithms. This giant, consisting of purely theoretical algorithms, promises of exponential advantages and infinite applications, sits on shaky ground with only small quantum devices incapable of sustaining the giant and muddied by the high processing power of classical computing. This thesis tries to combine mud and giant to strengthen the base of the structure, while sacrificing some height of the colossus.

Beyond metaphor, the point of view of this work is to try to design algorithms following a hybrid paradigm in which classical and quantum computing cooperate to solve the problem. Another strength of the algorithms presented is that the quantum modules have an approach closer to that of information processing than to that of problem definition by means of formulas and evolution. Due to this combination, algorithms that can work on industrially relevant problems following the paradigm of hybrid classical-quantum computing have been created.

This paradigm follows the philosophy of taking advantage of the benefits offered by both technologies and combining them. It is accepted that this combined architecture may not be the final solution for quantum computing, but it is an intermediate step that in the short and medium term can greatly benefit and accelerate the development of quantum computing, even allowing to have practical applications earlier than expected.
\\

\large \textbf{Part I: Quantum Sample and Search (QS\&S)}
\\

The first part explains the concept of search \& sample algorithms as algorithms that can traverse a state space to find a distribution of the data or the state with minimum or maximum value. This type of algorithm is useful when it is difficult to analyze the state space with another method, and the only solution is to run an evaluation function to know the value of many or all of the states.

These algorithms are classically used to solve optimization problems and are the basis for other more complex algorithms, such as artificial intelligence algorithms. However, the scalability of these algorithms is poor because they require visiting a high number of states and this number grows exponentially in the problems to be solved. Therefore, classically, there is a limitation that must be solved in a quantum way.

Quantumly, there are different ways to create a search \& sample algorithm. In fact, there are several proposals in the literature. However, the ones that stand out above the rest and that are going to be used in this work are the quantum walks algorithms. These quantum walks have the ability to perform a search over a state space polynomially faster than their classical counterparts. On these, a quantum Metropolis-Hastings algorithm can be built that maintains the advantage, but with greater capabilities.

Using this quantum Metropolis-Hastings algorithm, an algorithm called quantum Metropolis Solver, QMS, has been built. QMS, the algorithm around which the whole part \ref{pt:qss} is structured, is a hybrid algorithm that combines a quantum Metropolis-Hastings circuit with classical preprocessing of the problem to achieve quantum advantage in problems of industrial interest and reduced size. QMS is run on quantum simulators running on classical hardware to test its performance, but it has also been tested on quantum hardware to confirm the results obtained in the simulator.

QMS has the ability to solve any optimization problem formulated using a problem description and a state evaluation function. Therefore, it has been applied to different use cases such as hypothesis search in artificial intelligence, gravitational wave parameter estimation for space exploration or protein folding in quantum chemistry.
\\

\large \textbf{Part II: Quantum Chemistry}
\\

The second part explores hybrid algorithms applied to quantum chemistry, specifically, to the prediction of the fundamental state of a quantum system and the calculation of its energy. In this case, the hybrid paradigm is applied from another perspective. Instead of having a quantum and a classical module connected to each other, this alternative has a classical algorithm that analyzes different quantum algorithms to estimate their cost in number of gates. In this way, it is possible to analyze what resources are necessary in a quantum device to be able to execute these algorithms, what difference there is between the available resources and the real needs, which approaches are more efficient, etc. 

To do this, first, it is necessary to explain classical quantum chemistry algorithms and their limitations. These limitations can be overcome with quantum algorithms. However, these quantum algorithms are expensive, especially for current quantum hardware. Therefore, it is necessary to analyze the gate cost of each of the processes that perform the algorithms, the Hamiltonian simulation, the encoding, the mapping, etc. 

This analysis has been generalized for a set of quantum algorithms and any possible molecule in a software tool called TFermion. This tool analyzes in detail the T-gate cost of algorithms that are state of the art in quantum chemistry.

To show an use case of TFermion in the industrial environment, the problem of designing lithium-ion batteries has been chosen. The problem is focused on the cathode design. For this purpose, quantum algorithms can be used to estimate the fundamental state more accurately, and it is interesting to study the cost of quantum gates of these algorithms.

\newpage
\section*{Resumen detallado en Español}
\addcontentsline{toc}{section}{Resumen detallado en Español}
\thispagestyle{empty}
\markboth{Resumen detallado en Español}{Resumen detallado en Español}

\large \textbf{Introducción}
\\

La situación actual de Computación Cuántica no difiere mucho de la de un gigante con pies de barro. Sin que se haya podido demostrar una ventaja cuántica en un problema real, ya se ha desarrollado una fuerte inversión tanto a nivel industrial como académico para diseñar algoritmos cuánticos. Este gigante, formado por, algoritmos puramente teóricos, promesas de ventajas exponenciales e infinitud de aplicaciones, se asienta sobre un terreno inestable en el que sólo existen dispositivos cuánticos de pequeño tamaño incapaces de sostener al gigante y embarrado por la alta capacidad de procesamiento de la computación clásica. Esta tesis trata de combinar barro y gigante para fortalecer la base de la estructura, amén de sacrificar cierta altura del coloso.

Más allá de la metáfora, el punto de vista de este trabajo es tratar de diseñar algoritmos siguiendo un paradigma híbrido en el que computación clásica y cuántica cooperen para resolver problemas industriales. Otro punto fuerte de los algoritmos presentados es que los módulos cuánticos tienen un enfoque más cercano al del procesamiento de la información que al de la definición del problema mediante fórmulas y evoluciones. Gracias a esta combinación, se llega a algoritmos que pueden trabajar sobre problemas relevantes a nivel industrial siguiendo el paradigma de la computación híbrida clásico-cuántica.

Este paradigma sigue la filosofía de aprovechar las ventajas que ofrecen ambas tecnologías y combinarlas. Se acepta que esta arquitectura combinada puede no ser la solución final para la computación cuántica, pero sí un paso intermedio que a corto y medio plazo puede ser beneficioso y acelerar el desarrollo de la computación cuántica, permitiendo, incluso, tener aplicaciones prácticas antes de lo esperado.
\\

\large \textbf{Parte 1: Búsqueda y Muestreo Cuántico}
\\

En la primera parte de la tesis, se explica el concepto de los algoritmos de búsqueda y muestreo como algoritmos que son capaces de recorrer un espacio de estados para encontrar una distribución de los datos o el estado con mínimo o máximo valor. Este tipo de algoritmos son útiles cuando es dificil analizar con otro método el espacio de estados y la única solución es ejecutar una función de evaluación para conocer el valor de muchos o todos los estados.

Estos algoritmos son muy utilizados clásicamente para resolver problemas de optimización y son la base de otros algoritmos más complejos como los de inteligencia artificial. Sin embargo, la escalabilidad de estos algoritmos es mala porque requieren visitar un alto número de estados y, este número, crece de más rápido que la capacidad de cómputo de los ordenadores clásicos. Por ello, clásicamente existe una limitación que se intenta resolver de manera cuántica.

Cuánticamente, existen diferentes formas de crear un algoritmo de búsqueda y muestreo. De hecho, en la literatura, hay diversas propuestas. Sin embargo, la que sobresale por encima del resto y que va a ser utilizada en este trabajo son los algoritmos de caminos cuánticos. Estos caminos cuánticos tienen la capacidad de hacer una búsqueda sobre un espacio de estados polinómicamente más rápida que sus homólogos clásicos. Sobre estos, se puede construir un algoritmo de Metropolis-Hastings cuántico que mantiene la ventaja pero con mayores capacidades.

Utilizando como base este algoritmo de Metropolis-Hastings cuántico, se ha construido un algoritmo llamado Quantum Metropolis Solver, QMS. En torno a este algoritmo, se estructuran un conjunto de capítulos de esta primera parte. Se trata de un algoritmo híbrido que combina un circuito cuántico de Metropolis-Hastings con preprocesamiento clásico del problema para lograr ventaja cuántica en problemas de interés industrial y tamaño reducido. QMS se ejecuta en simuladores cuánticos que corren en hardware clásico para probar su rendimiento pero, también ha sido probado en hardware cuántico para confirmar los resultados obtenidos en el simulador.

QMS tiene la capacidad de resolver cualquier problema de optimización formulado mediante una descripción del problema y una función de evaluación de los estados. Por ello, ha sido aplicado a diferentes casos de uso como búsqueda de hipótesis en inteligencia artificial, estimación de parámetros de las ondas gravitatorias para la exploración espacial o el problema del plegamiento de proteínas.
\\


\large \textbf{Parte 2: Química cuántica}
\\

En esta segunda parte se exploran los algoritmos híbridos aplicados a la química cuántica, concretamente a la predicción del estado fundamental de un sistema cuántico y el cálculo de su energía. En este caso, el paradigma híbrido se aplica desde otra perspectiva. En vez de tener un módulo cuántico y otro clásico conectados entre ellos, se tiene un algoritmo clásico que analiza diferentes algoritmos cuánticos para estimar su coste en número de puertas. De esta manera, se puede analizar qué recursos son necesarios en un dispositivo cuántico para poder ejecutar estos algoritmos, qué diferencia hay entre los recursos disponibles y las necesidades reales, qué aproximaciones son más eficientes, etc. 

Para ello, primero es necesario explicar los algoritmos clásicos de química cuántica y sus limitaciones. Estas limitaciones se pueden superar con algoritmos cuánticos. Sin embargo, estos algoritmos cuánticos son costosos, especialmente para hardware cuántico actual. Por ello, es necesario analizar el coste en puerta de cada uno de los procesos que realizan los algoritmos, la simulación del Hamiltonio, la codificación, el mapping, etc. 

Este análisis se ha generalizado para un conjunto de algoritmos cuánticos y cualquier posible molécula en una herramienta software llamada TFermion. Esta herramienta analiza en detalle el coste en puertas T de los algoritmos que marcan el estado del arte en química cuántica.

Para mostrar un caso de uso de TFermion en el entorno industrial, se ha escogido el problema del diseño de baterías de litio-ion con el fin de mejorar su autonomía. El problema está centrado en el diseño del cátodo. Para ello se pueden usar algoritmos cuánticos que estiman con mayor precisión el estado fundamental y es interesante estudiar el coste de en puertas cuánticas de esos algoritmos.

\newpage
\thispagestyle{empty}
\tableofcontents

\newpage\null\thispagestyle{empty}\newpage

\newpage
\section*{Introduction}\label{pt:intro}
\thispagestyle{empty}
\addcontentsline{toc}{section}{Introduction}
\markboth{Introduction}{Introduction}

\epigraph{The most beautiful thing we can experience is the mysterious. It is the source of all true art and science. He to whom the emotion is a stranger, who can no longer pause to wonder and stand wrapped in awe, is as good as dead —his eyes are closed.}{Albert Einstein, {\it Living Philosophies}}

\lettrine[lines=1, findent=2pt]{\resizebox{!}{1.2\baselineskip}{T}}{}hese words written by Albert Einstein reflect the reality of researching in quantum mechanics and by extension in quantum computing. This thesis became an exercise of dealing with the mysterious foundations of quantum computing from a different point of view from the current one, from the perspective of the theory of computation and algorithms. This new approach has been applied to the development of hybrid classic-quantum algorithms that solve current industry problems.

Quantum computing entails the application of quantum mechanics principles to the realm of information processing. Quantum mechanics, which ignited a revolution in 20th-century physics, challenged fundamental principles that had stood for centuries. Renowned physicists such as Max Planck, who introduced the concept of quanta through his exploration of black body radiation, Albert Einstein with his work on the photoelectric effect and extensive contributions to quantum physics, as well as notable figures like Louis De Broglie, Niels Bohr, Edwin Schrödinger, Max Born, Heisenberg, Dirac, among others, significantly advanced the field of quantum mechanics. This counter-intuitive and groundbreaking domain found practical utility in the creation of lasers, microprocessors, superconductors, and chemical models.

However, simulating the intricate quantum behaviors of a system using a classical computer proves to be extremely complex. Such simulations necessitate encoding intricate states and managing myriad interactions among them. This challenge led Richard Feynman to propose an unconventional solution: employing the principles of quantum physics themselves, leveraging quantum states to simulate quantum systems. Feynman, driven by the desire to expedite simulations, inadvertently laid the groundwork for what now is recognized as quantum —an anticipated revolution poised to shape the 21st century. Feynman's proposal is only the tip of the iceberg of quantum computing, and he was not the first. In 1985, David Deutsch introduced the concept of a quantum Turing machine \cite{deutsch1985}. It was equivalent to the Turing Machine designed by Alan Turing \cite{turing1936} but it proved that it was possible to execute the same set of operations with quantum hardware. Following the idea of simulating experiments on quantum computers, other scientists have proposed processing information faster and with lower energy costs to develop areas such as quantum chemistry, communications, or artificial intelligence.

If classical computing is conceived as the manipulation of 0 and 1 signals (indicating the presence or absence of electric current) and the implementation of logical gates that integrate these signals (AND, OR, NAND, XOR gates, etc.), Quantum computing shares similarities in terms of logic while diverging significantly in its properties. Quantum computing leverages quantum particles existing in superposition and entanglement states, enabling it to not only represent states but also to perform operations in an optimized manner compared to classical computing. The main similarity between both is that in classical and quantum computing the unit of information is the bit (or the qubit) with states 0 and 1. So, both use binary representation. Although it is important to note that this is just a simplification, it is sufficiently explanatory.

\begin{figure}[t]
\centering
\includegraphics[width=1\textwidth]{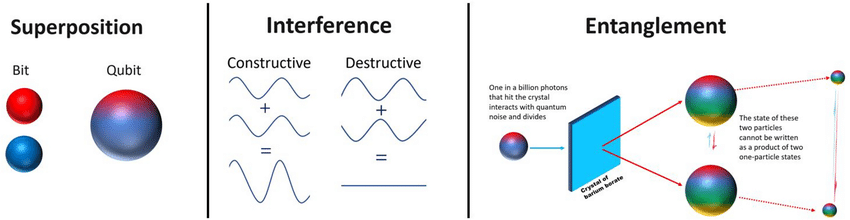}
\caption{\label{fig:superposition} Scheme extracted from \cite{giani2021}. It represents the phenomenom of superposition in a qubit represented with two possible states, red and blue and a qubit that is a combination of both. The interference effect, in the middle, can be constructive if two waves interfer to create a pattern or destructive if both cancel each other. Finally, the entanglement can be undestood as two particles that can not be written as a product.}
\end{figure}

Superposition is the property that allows a quantum state to be in two states simultaneously. This implies that a qubit can exist in a state of 0, 1, or in a linear combination of both possibilities, resulting in a remarkably condensed representation of information. Entanglement, on the other hand, is the phenomenon through which two quantum particles behave as though they are a single entity. In an entangled quantum system, any alteration that impacts the system equally affects both particles. The significance of two or more entangled qubits lies in their collective impact: operations (quantum gate) on the entangled system can affect all its qubits \cite{sutor2019}. In Figure \ref{fig:superposition}, these three concepts are visually represented.

Quantum computing offers numerous advantages over classical computing: superdense encoding of information, transmission through secure channels, reduced energy costs, operations in superposition, etc. This thesis will focus on the ability to operate on a superposition of states, which allows for the reduction of computational complexity in algorithms \cite{nielsen2010}.

All these advantages have strong theoretical underpinnings. There are works in computational complexity analysis that have created new classes like BQP \cite{aaronson2010}. Other works have formulated the quantum Turing machine \cite{deutsch1985} and derived the set of universal gates for quantum computation \cite{chow2012}. There are also studies that demonstrate polynomial or superpolynomial advantages over classical algorithms \cite{deutsch1992}. However, it is crucial to highlight that all these demonstrations are purely theoretical and have not yet been proven in quantum algorithms executed on real hardware. This means that, at this moment, quantum algorithms do not have an advantage over a classical one in solving interesting problems.

At this juncture, it is worth asking why quantum computing has not yet solved real-world problems. There is a simple answer, there are many limiting factors need to be analyzed.

\begin{itemize}
    \item \textbf{Hardware}

        \begin{itemize}
            \item \textit{Quantum HW size (number qubits)}: The number of qubits is often considered the primary metric for evaluating the scale of a quantum computer. However, it is important to emphasize that this metric, while significant, is neither the most crucial nor the most constraining factor. Although some may draw parallels between the number of qubits and the storage capacity, RAM size, or CPU capabilities of a classical computer, such comparisons are not accurate.

            The number of qubits is directly tied to a quantum computer's capacity for representing different states. In simpler terms, a quantum computer with a higher number of qubits can represent a larger number of states, allowing it to address more extensive problem sizes. However, it is essential to recognize that the ability to solve larger problems does not necessarily equate to faster problem-solving.
            
            Instead, the true potential of quantum computing lies in its capability to tackle problems that classical computers may find unbeatable. It is also true that increasing the number of hardware noise qubits is moving towards the possibility of having logical noise free qubits which will speed up the quantum algorithms, but it is still a future scenario.

            \item \textit{Quantum Errors}: The execution of algorithms in a hardware device is never error-free, either classically or quantumly. Classically, these errors have been extensively analyzed and studied with different strategies to correct them. Quantumly, there are inherent errors in devices that make classical error correction techniques inapplicable. These errors are phase errors. This new type of error added to the quantum limitations of not being able to copy the states without altering them means that, at present, quantum errors limit the execution of algorithms. Advances in quantum error correction will mark the ability to run longer and more complex algorithms.
            
            \item \textit{Coherence times}: The duration during which a quantum system retains its quantum attributes, such as superposition and entanglement, is referred to as its coherence time. When a quantum computer has a limited coherence time, it implies that there is only a brief window available for executing quantum gates before the system's qubits lose their quantum properties. Consequently, enhancing the coherence times of quantum chips opens up the potential for executing more intricate and longer algorithms. Coherence times limit the circuit depth, which is the real limiting factor in the present quantum hardware \cite{amy2013}.

        \end{itemize}

    \item \textbf{Software}

        \begin{itemize}
            \item \textit{Algorithms}: The full harnessing of quantum computing's advantages remains a subject of ongoing exploration, and consequently, quantum algorithms have yet to achieve their maximal speed and efficiency. Quantum computing processes information in a manner so distinct than classical computing, that current efforts primarily involve attempting to replicate classical algorithms to ascertain any potential advantages. However, in the long-term trajectory, this strategy must evolve to encompass the development of entirely quantum algorithms that are not reliant on classical inspiration.

            \item \textit{Representations}: A significant constraint of quantum algorithms lies in how information is represented. Many of these algorithms attempt to encode information into a formula, rendering quantum processing similar to quantum simulation rather than information processing. It is imperative to recognize that in the term ``quantum computing'', the emphasis is more on ``computing'' than on ``quantum''. In other words, the main focus should be on information processing.

            \item \textit{Operators}: A quantum operator is composed of a set of gates, and it is an extra layer of abstraction in the process of implementing a quantum algorithm. An operator, in the context of quantum computing, serves as a programmer's tool for executing specific operations in a manner that leverages the capabilities of a quantum computer. However, a significant challenge arises from the relatively limited number of available operators. In fact, many algorithms that theoretically exhibit a quantum advantage are derived from or heavily influenced by two operators: the Grover diffusion operator \cite{grover1996} and the amplitude estimation operator, famously employed by Shor in the realm of prime number factorization \cite{shor1994}.
            
        \end{itemize}

    \item \textbf{Inherent}

        \begin{itemize}
            \item \textit{No cloning theorem (information storage)}: While quantum computing presents transformative quantum properties, it also introduces challenging features that complicate its manipulation. Perhaps the most crucial of these challenges is the fact that observing a quantum system inherently disrupts it. In practical terms, this means that quantum states cannot be duplicated without undergoing destruction because the very act of examining the state unavoidably alters it. This principle is encapsulated by the non-cloning theorem and affects, currently prohibits, some crucial operations in computer science, copying and storing information. However, this problem can be avoided since quantum states can be stored in a quantum memory and retrieved by quantum teleportation, which does not violate the non-cloning theorem. On the other hand, this limitation becomes an advantage in the context of quantum cryptography, where not being able to copy a state guarantees that it is secret.
        \end{itemize}

\end{itemize}

All these factors listed above are just an enumeration of the key points guiding the development of this new technology, the faster they evolve, the sooner quantum computing will become a reality over or alongside classical computing.

Currently, the development of quantum computing is a global and combined effort between academia and companies in which everybody is focus on simulation, hardware and software. This collaboration is almost complete and results in both companies and universities publishing their advances. It should be noted that companies are willing to collaborate for several reasons. First, there is no direct application of quantum computing to industry, so they do not yet have a clear and defined business. Second, there are still many challenges to solve and work on, so the relationships between the various players are still symbiotic. Finally, the volume of business offered by quantum computing is so large that a few companies are unable to cover it, and it is more profitable for them to collaborate than to compete. This also translates into a high shortage of workers in quantum computing.

Looking at the profile of scientists working in quantum computing today, most of them are physicists specialized in quantum physics who have made the step to quantum information for both quantum algorithms and quantum many-body systems. There are also many experimental physicists working on the development of quantum computers. Looking for other profiles working in quantum computing, most of them are engineers working on the adjacent circuitry in quantum chips. 

However, in the field of algorithms and software development, the number of computer scientists is minimal. This is understandable because the first steps of quantum computing had to be developed by those who understood quantum theory with great precision, and that required extra effort from computer scientists. Today, that situation has changed; It is already feasible to devise and implement algorithms with a certain level of complexity, and the simulation-oriented approach traditionally employed by physicists in algorithm design is proving to be inefficient. For this reason, companies and universities are increasingly demanding physicists who are capable of programming and designing algorithms, or even computer scientists with a background in quantum physics.

This thesis, among its various contributions, endeavors to introduce a distinct perspective to the development of quantum algorithms. Rooted in the experiences of someone accustomed to working with data processing algorithms and classical artificial intelligence, it advocates an approach centered on information processing rather than the mere simulation of physical phenomena. Although this statement may appear somewhat ambiguous, subsequent sections will elucidate the development of algorithms that emphasize the representation of information through variables, with diverse values giving rise to a multitude of states. Subsequently, an effort will be made to combine these variables with specific problem rules to derive solutions. This approach aligns more closely with Grover's philosophy of state processing than with Shor's emphasis on phase estimation or adiabatic computation grounded in Hamiltonian evolution.

Following this approach of creating quantum algorithms with a stronger focus on solving industrial problems from a more computer-centric perspective, this doctoral research program has been developed. It has earned the distinction of an industrial doctorate. This recognition acknowledges the ability of this work to provide solutions originating in academia to the industrial domain, bridging the gap that often exists between these two realms. This distinction is particularly apt, considering the past decade has witnessed an explosion of startups following the spinoff model. In this model, expert scientists from various universities have developed prototypes that are both marketable and valuable to industry.

In the specific case of this work, the Spanish company Quasar identified quantum computing as a promising line of business for its interests. Quasar is an aerospace company that develops artificial intelligence and optimization algorithms applied to data analysis and spatial solutions. One of Quasar's main customers is the European Space Agency, for which it has developed products such as autonomous analysis of satellite images for the observation and monitoring of the Earth or an autonomous robotic telescope with the capacity to manage by itself the observations of a night in different hemispheres.

Quasar understood that the complexity and volume of space data increasingly required more computing power, especially when the space sector has just entered the so-called new space era, with private companies licensed to launch and operate satellites \cite{kodheli2020}. This led the company to look for new solutions to overcome the bottleneck of data processing and optimization algorithms. To this end, an industrial doctorate was proposed in collaboration with the GICC group of the Complutense University of Madrid and was awarded the Industrial Doctorate scholarship of the Community of Madrid (IND2019/TIC17146).

In order to carry out this industrial doctorate and to be able to create interesting algorithms for industry and collaboration with leading companies, during the development of this thesis, the main lines of research being pursued by other companies and universities have been analyzed in detail.

As discussed above, quantum computing offers state processing capabilities far superior to classical capabilities. However, it has been analyzed that quantum computing still has numerous problems to be solved. Some of these problems are particularly well solved by classical computing. For this reason, some years ago scientist started to discuss about hybrid classical-quantum algorithms. These algorithms try to combine the advantages of both worlds, the large data handling and storage capacity of classical computers and the superposition processing of quantum computers.

\begin{figure}[t]
\centering
\includegraphics[width=1\textwidth]{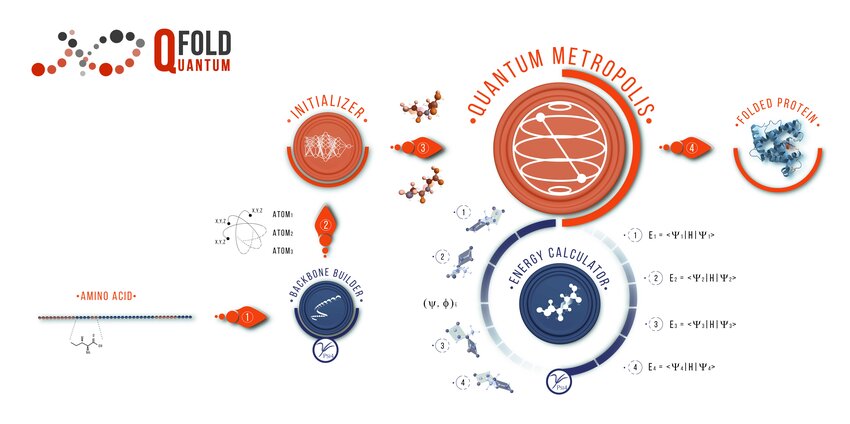}
\caption{\label{fig:hybrid} Scheme extracted from \cite{casares2022} representing a hybrid algorithm. Red modules are quantum and classical ones are in blue. Quantum and classical modules cooperates to get a solution. The system started with a chain of aminoacids with all angles initialized to 0. Then, an initial guess of the angles is done in the initializer and this initialization is refined in the quantum metropolis module. The energy calculator in blue is a classical module to calculate the energies of each possible configuration. Finally, the folded protein structure is obtained.}
\end{figure}

Hybrid algorithms consist of architectures that include modules from both technologies. Typically, a first module for classical data preparation and processing, then a quantum analysis module, and finally a classical module to analyze the output of the quantum module as it can be seen in Figure \ref{fig:hybrid}.

Following this architecture, hybrid algorithms offer numerous advantages in the short term.

\begin{itemize}

    \item \textbf{Good points of both worlds}: Hybrid algorithms combine the fast analysis and preprocessing of data from classical computing with the ability to perform superposition processing and quantum operations from quantum computing.

    \item \textbf{Speed up against classical}: Hybrid algorithms, which incorporate a quantum module, perform intricate operations of classical algorithms with a notable speedup compared to purely classical algorithms.

    \item \textbf{Execution on real hardware}: The limitations of quantum hardware mentioned earlier may prevent the execution of a complete quantum algorithm. However, contemporary quantum chips are sufficiently capable of executing certain components or segments of these algorithms. This capability allows for the execution of hybrid algorithms on real quantum hardware.

\end{itemize}

These hybrid algorithms fit perfectly with the current approach to quantum computing. QC development has already been discussed to help overcome the limitations of classical computing. However, classical computing works at many levels, from personal computers, office systems, to computing centers, and high performance computing (HPC) \cite{severance2010}.

The fundamental goal of QC development is to assist or replace those supercomputers used in HPC. This is due to several reasons, first because finding quantum advantages with current devices requires problems of high complexity, something that a personal computer will never solve, second because current classical computing is both effective and efficient at solving simple problems, third, quantum computers must be controlled by experts, something that makes their mass production and maintenance difficult, and finally, because the cost of a quantum computer must be borne by a large corporation.

The current short- and medium-term approach may be modified in the long term. It would not be surprising to find applications of quantum algorithms for personal computers over time. For example, the scientists who operated the first room-sized computers never thought of mobile devices, virtualization, or social networks, but over time these were developed. However, it is clear that, should this development take place, it will be in the long term.

Research in quantum computing endeavors to construct quantum algorithms capable of addressing problems that are computationally intractable for classical algorithms. Many of these algorithms designed to tackle such challenging problems are currently running on supercomputers \cite{humble2021}. Consequently, in the near term, the development of quantum computing is geared toward either supplementing or potentially supplanting supercomputers in specific computational tasks. Nevertheless, a fundamental characteristic of supercomputing is its ability to handle large volumes of data, often referred to as big data. Quantum computers, on the other hand, face limitations when dealing with such extensive datasets. Therefore, in the present scenario, the integration of both technologies becomes imperative, and hybrid algorithms emerge as a sensible approach to bridge the gap between these two realms.

Besides, different works \cite{endo2021} consider hybrid algorithms as a bridge that ends in pure quantum algorithms. Currently, it is not well known how to quantize some operations or how to get advantages from quantum computing in areas like quantum artificial intelligence (QAI) or quantum chemistry, and the only solution is to apply quantum computing to some modules of the classical pipeline. This hybrid solutions will be an useful intermediate step to learn more about quantum information processing and to develop pure quantum algorithms.

\begin{figure}[t]
\centering
\includegraphics[width=0.6\textwidth]{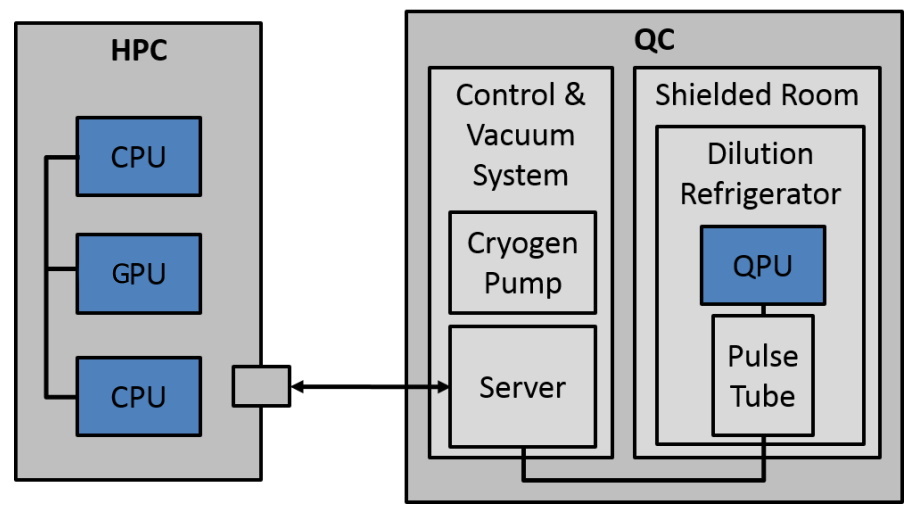}
\caption{\label{fig:qpu} Scheme extracted from \cite{nvidia_qpu} representing a hybrid architecture integrating a QPU in a HPC system. It represents a possible architecture in a HPC pipeline, in which the quantum computing is integrated as one extra module to execute specific algorithms. Following the proposed architecture, QC module would be equivalent to GPU or CPU modules.}
\end{figure}

On the other hand, there are authors who claim that quantum computing will be just an additional processing module within a large classical architecture \cite{britt2017}. This concept would be to have a quantum processing unit (QPU) inside a classical computer equivalent to current graphics processing units (GPU) as shown in figure \ref{fig:qpu}.

Whichever view of hybrid algorithms prevails, it will be a battle that will be fought over the medium to long term. In the short term, hybrid algorithms will continue to develop because it is the fastest way to the concept of quantum advantage, understood as an algorithm that uses quantum computation and is faster than a classical one. Gradually, better interfaces between classical and quantum modules will be developed, optimizing the encoding of information and optimizing its execution.

\begin{figure}[t]
\centering
\includegraphics[width=0.3\textwidth]{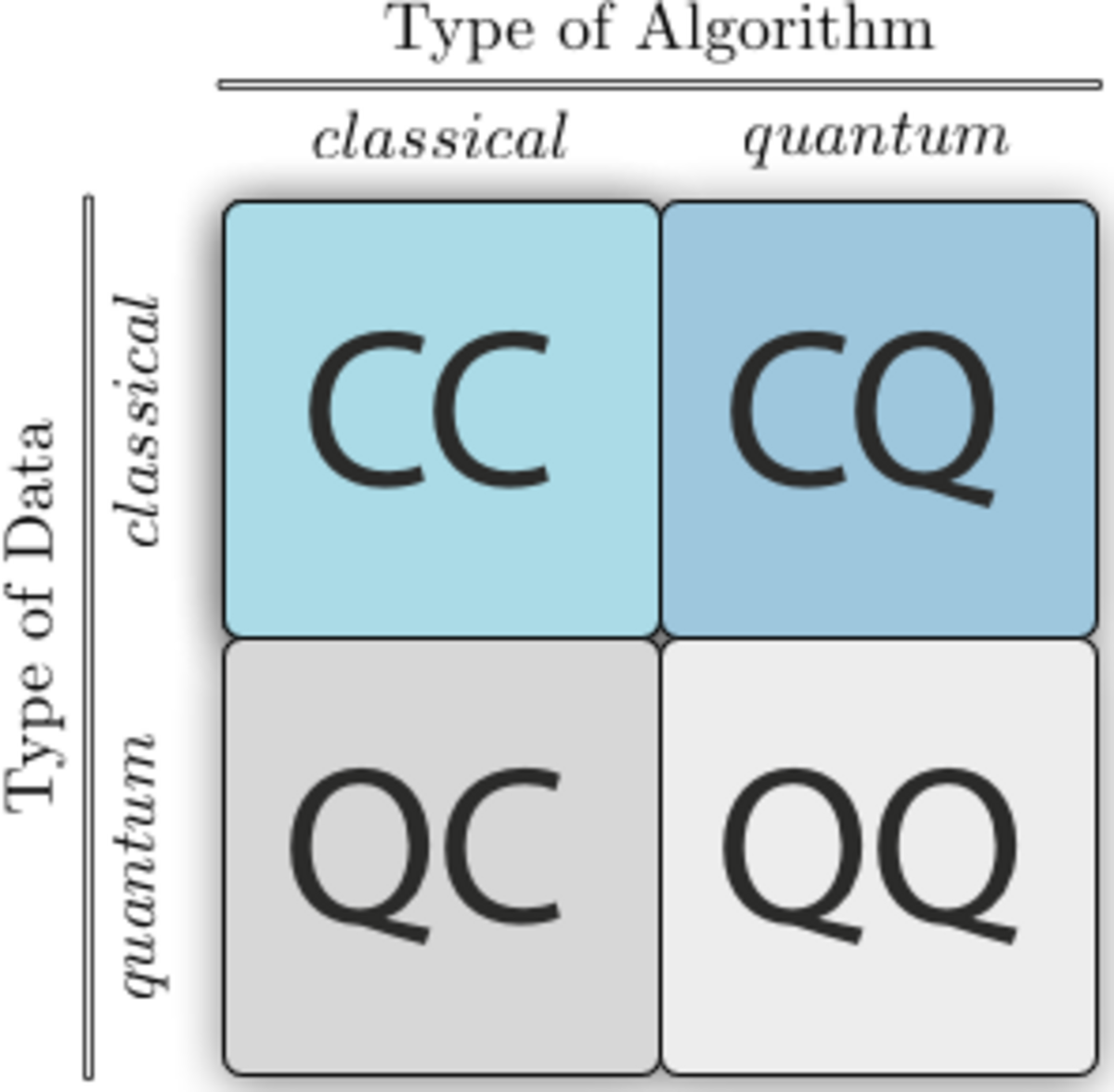}
\caption{\label{fig:cc} Scheme extracted from \protect\url{https://en.wikipedia.org/wiki/Quantum_machine_learning\#/media/File:Qml_approaches.tif}. In this figure is possible to observe four different paradigm for algorithms attending to the type of algorithm and type of data. If the algorithm is classical and also the data, the algorithm is similar to any existing algorithm in classical computing. The option of a quantum algorithm working with classical data is similar to the quantum algorithm proposed in this chapter applied to optimization problems. If the algorithm is classical, but the data is quantum could be the classical simulation for quantum physics. Finally, if both, data and algorithm, are quantum, would be similar to a quantum algorithm working on quantum chemistry. This paradigm will be explored in part \ref{pt:qchem}.}
\end{figure}

Hybrid algorithms also hold significance due to their capacity to handle varying data types, encompassing both classical and quantum data. As mentioned earlier, quantum computing was initially proposed primarily for quantum simulation, yet it swiftly emerged as a valuable tool for a broader spectrum of algorithms. Hence, based on the nature of the information involved, a classification can be drawn, distinguishing between classical or quantum data and classical or quantum algorithms, resulting in four possible categories, as shown in figure \ref{fig:cc}. 

Hybrid algorithms are capable of working in three of these four categories, such as classical algorithms with quantum data \cite{giordano2022}, quantum algorithms with classical data \cite{escrig2023} or quantum algorithms with quantum data \cite{low2019}.

The ability to run quantum algorithms on real hardware represents one of the significant advancements in the field of quantum computing in recent years. This milestone owes its thanks to the advent of noisy intermediate-scale quantum (NISQ) computers. These NISQ computers are compact devices comprising 10 to 1000 qubits, characterized by considerable noise levels, but capable of executing small-scale algorithms. They serve a dual purpose as a test-bed for algorithm development and as an intermediate step towards achieving fault-tolerant algorithms, where the anticipated quantum advantage is expected to shine truly \cite{preskill2018}.

NISQ hardware development has created an industrial opportunity for many companies. Big companies like IBM, Google and Microsoft started the research in this technology and have made great strides in quantum computing in both hardware and software, strongly supported by research conducted at universities. This juncture was the breeding ground for the emergence around 2015 of small startups led by scientists coming from large universities and financially supported by rounds of funding and venture capital. These startups have continued the development of all QC research lines following a more academic than industrial model.

This diversity of actors and their collaboration is mainly due to the lack of competition and the early stage of the development of the technology. A paradigmatic example of this situation is that the three main proposals that have demonstrated quantum advantage over an artificial problem:

\begin{itemize}
    \item \textbf{Google (2019)}: Using Sycamore chip, they solved a sampling random circuits problem faster than a classical computer \cite{arute2019}.

    \item \textbf{Chinese University of Science and Technology (2021)}: Using the Jiuzhang chip, they solved a problem of gaussian amplitude sample faster than a classical computer \cite{wu2021}.

    \item \textbf{Xanadu (2022)}: Using the Borealis chip, they solved a Gaussian boson sampling problem faster than a classical computer \cite{madsen2022}.
    
\end{itemize}

These three experiments were useful in demonstrating that quantum computing can solve certain problems faster than classical algorithms. However, they have no real applications. The interesting lines with real applications in which quantum algorithms are expected to be developed in the future are as follows:

\begin{itemize}
    \item \textbf{Optimization}: Problems such as how to deliver packages in a city while minimizing the number of trucks or the fuel used are of enormous relevance and can be interesting for QC because they encompass many possible combinations to check.

    \item \textbf{QAI}: Problems such as classification, speech recognition, image generation, or prediction are presently addressed by classical computers, although with certain inherent limitations. To tackle these problems at scale, a technology like quantum computing, which leverages operations in superposition, becomes essential.

    \item \textbf{Quantum chemistry}: Classical computers often expend significant resources when encoding a chemistry simulation. In contrast, quantum computers inherently represent chemistry quantum systems, enabling data processing with far greater efficiency and precision.

    \item \textbf{Quantum simulation}: Unlike quantum chemistry, quantum simulations will have substantial benefits from quantum computing. This is because executing operators that are inherently present in the quantum circuit enhances both the precision and the scale of systems that can be effectively simulated.

    \item \textbf{Communications}: The paradigm of quantum mechanics opens opportunities for improving classical communication methods by providing secure channels, superdense encoding, and speedy communication.

\end{itemize}

However, all of these lines of research must be accompanied by a strong development of quantum hardware. As seen before, the limitations of noise, coherence, and size of quantum computers affect the algorithms to the point of rendering them useless. Currently, NISQ computers are adequate as proofs of concept, but if quantum computing is to be used in real environments, a transition to a more robust quantum chip paradigm must be made. Fault-tolerant computers, capable of correcting their own errors and increasing coherence times, have emerged. Moreover, even these robust quantum computers seem to be assisted by classical computers, i.e. hybrid algorithms \cite{andreasson2019}, which is a further motivation for this thesis. At least this is how research is developing at the moment.


Throughout this thesis, hybrid algorithms that address some of the main problems of interest to the industry have been explored. The interest was not only to develop algorithms to solve some of the problems in these areas, but also to better understand the concept of information processing in quantum computing and to create algorithms that were not only theoretical but had code implementation and the ability to be executed in simulators or quantum hardware. Therefore, each of the works explained has an associated code that can be executed. The topics that this thesis has delved into are optimization, quantum artificial intelligence, and quantum chemistry, which are three of the main ones listed above.

This thesis is organized as follows. The material has been divided into two main parts: (I) Part \ref{pt:qss} on the work on search \& sample algorithms very used in optimization and quantum artificial intelligence; (II) Part \ref{pt:qchem} on the work done analyzing algorithms of quantum chemistry and their relevance to the battery sector. Each part is divided into chapters.

Chapter \ref{ch:introQSS} provides a concise introduction to the concept of search \& sample techniques in quantum algorithms.

Chapter \ref{ch:cSS} provides a review of the classical algorithms already implemented for search \& sample.

Chapter \ref{ch:qSS} provides a review of the quantum algorithms already proposed for search \& sample.

Chapter \ref{ch:qms} provides a review of an algorithm for applying search \& sample to optimization problems. This chapter is based on the publication \ref{pub:p1}.

Chapter \ref{ch:qaiSS} provides an explanation of how search \& sample can be applied to quantum artificial intelligence. 

Chapter \ref{ch:SSspace} provides an explanation of how search \& sample can be applied to solve problems in the space industry. This chapter is based on the publications \cite{escrig2023, escrig2024}.

Chapter \ref{ch:SSchem} provides an explanation of how search \& sample can be applied to solve problems related to biology, in particular, the protein folding problem. This chapter is based on the publication \ref{pub:p2}.

Chapter \ref{ch:tfermion} provides a brief introduction to Ground State Preparation (GSP) techniques. It also includes a detailed explanation about the TFermion software tool. This chapter is based on the publication \ref{pub:p3}.

Chapter \ref{ch:cars} provides an explanation of how GSP techniques can help in battery design. This chapter is based on the publication \cite{delgado2022}.

\newpage
\vspace*{\fill}
\part[Quantum Search and Sample (QSS)]{\centering\Huge Part I\\Quantum Search and Sample (QSS)}\label{pt:qss}
\thispagestyle{empty}
\vspace*{\fill}
\newpage

\newpage\null\thispagestyle{empty}\newpage

\pagenumbering{arabic}

    \section{Introduction to S\&S}\label{ch:introQSS}
    \thispagestyle{empty}

        \lettrine[lines=1, findent=2pt]{\resizebox{!}{1.2\baselineskip}{S}}{}earching and sampling techniques encompass a family of algorithms that can be described through various definitions. In order to present these concepts in simple terms:
    
        \begin{itemize}
            \item \textbf{Sampling} is a technique to approximate an unknown distribution defined by a function. This can be likened to a plumber who needs to understand the shape and path of a pipe hidden inside a wall. To achieve this, the plumber makes small openings in the wall and examines the pipe at various points, gradually gaining a sufficiently accurate understanding of its shape to perform the necessary work.

            \item \textbf{Searching} is a technique that involves finding a specific element within a distribution, typically the maximum or minimum, by making queries to the underlying function. Comparatively, this is similar to the plumber's task of locating a hole or issue in the pipe by tracing the flow of water within the pipe. Effective searching benefits from prior sampling, as it provides information about the distribution being searched.
        \end{itemize}
    
        These analogies illustrate the fundamental concepts of search \& sample techniques in a more accessible way. In the realm of quantum computing, algorithms based on search \& sample are developed to address optimization problems efficiently, often surpassing the capabilities of classical methods.

        In this thesis, a search \& sample technique is defined as an iterative method that allows to perform a search to find a reduced subspace of the complete search space and to sample that subspace to know its distribution. In the next step, using the information from the previous sampling, make a new search for another interesting subspace. By using this iterative methodology, it will be possible to know the global distribution of the problem to search in some cases for a minimum or maximum (pure search) or to sample the space.

        In this work, the set of search \& sample (S\&S) algorithms are considered as a distinct category within the broader realm of techniques to solve optimization problems. While this perspective might be considered somewhat innovative, it is important to note that focusing specifically on search \& sample algorithms, in reality, does not introduce anything fundamentally novel. Instead, this new category simplifies the algorithms, allowing to abstract certain features of optimization algorithms that are not essential for these problems.

        The reason why these two types of algorithms can constitute a category of their own within optimization problem techniques is that they have many similarities and applications. 
        
        \begin{itemize}
            \item \textbf{Numeric methods}: Search \& sample techniques are numeric methods that calculate the solution with repeated queries to a function that returns a value. Both techniques refine the solution as more iterations of the algorithm are executed. Besides, it is important to notice that each iteration of a S\&S algorithm is very simple to execute, so a problem can be solved efficiently executing many iterations in a row, as opposed to more precise and complex methods. \cite{donald1999}

            \item \textbf{Complex functions}: Search \& sample techniques are specially useful with complex distributions that are difficult to analyze with other techniques. For example, functions that are costly to derive. The objective of S\&S is to reduce the number of times that the function is executed, in particular if exhaustive search is the only known solution. \cite{russell2010}

            \item \textbf{Approximated methods}: Since they are approximate methods, in most of the cases, S\&S return a solution that is not optimal, just a suboptimal approximation. For example, sampling a function will require to sample the same number of states that the function have to return an exact solution, which is not efficient. In the case of searching, it could return a maximum that is not the absolute minimum, but it is close enough.

            \item \textbf{Scalability}: S\&S methods should work on large state space in order to get an advantage over other methods complex methods. Classically, this type of problem with large state space suffers from the curse of dimensionality \cite{bellman1966}, which results in a problem for classical algorithms due to its poor scalability. The number of times that a S\&S algorithm needs to call the function tend to scale exponentially when the problem grows linearly in the number of states \cite{verleysen2005}.

            \item \textbf{Closely related}: Search \& sample algorithms are closely related because combining them could be very useful in some problems. For example, performing a search task is much simpler if a good sample was taken prior to the problem \cite{aaronson2014}. 
            
        \end{itemize}

    
        During the development of this thesis, it has been observed that search \& sample algorithms are especially effective in optimization problems that meet certain characteristics. In this work, optimization algorithms are understood as those in which it is necessary to find a solution that meets certain criteria and is of maximum reward or minimum cost, depending on whether one is working on a maximization or minimization problem, respectively.

        The main characteristics that an optimization problem must meet for S\&S algorithms to gain any quantum advantage is that it requires an almost exhaustive search in the state space to find the solution, and that the state space is large or grows almost exponentially. Although these may seem like very detailed or specific features, they are common in optimization problems \cite{krentel1986}.

        If solving an optimization problem requires the use of algorithms and computers, it means that it involves so many variables, that the relationships between them are considerably large and that the evaluation function of the problem is complex to evaluate. Therefore, every time the problem changes or grows, the amount of resources to solve it will also increase. In addition, optimization algorithms, since they require finding the best solution from among many possible solutions, it is necessary to try almost all combinations between nodes until the optimal or at least a suboptimal one that is close to the solution is found. Therefore, in many optimization algorithms, the best technique for finding a solution is to search for possible combinations. Moreover, this search is almost exhaustive because it is very difficult to say which combinations of nodes should not be tried because they will not improve the quality of the solution.

        The complexity of an algorithm that must do an exhaustive search on a problem grows exponentially. This is because each time a new variable is added, the number of new states grows by combining that new variable with all existing ones. This implies that the scalability of the algorithm is bad, because by slightly increasing the size of the problem, it becomes intractable \cite{maltese2016}. This poor scalability is mainly due to the fact that classical algorithms evaluate states individually. Quantum, on the other hand, evaluates a superposition. Thus, the quantum algorithm will increase its complexity logarithmically in the representation of the states, but it will still always act on a superposition of the state space and its scalability will be better than that of the classical algorithm, always remembering that this applies to exhaustive search algorithms \cite{ambainis2004}. For example, in the quantum S\&S algorithm is based on Grover, the advantage will up to polynomial.

        It is at this point of poor scalability of classical algorithms, that quantum S\&S algorithms appear, with an expected advantage in their scalability. The strength of these algorithms comes from the theoretical advantages obtained by Grover's operators and their different variations \cite{galindo2000}. The main concept for understanding why quantum algorithms can accelerate the solution of optimization problems using QS\&S algorithms is the application of operators in superposition. 

        To understand this concept, it is necessary to think that classically, the difficulty of the processing consists of evolving the states until finding the one with the minimum cost. Quantumly, the difficulty lies in increasing the probability of the minimum cost state. To evolve the classical state, it is necessary to apply the operators individually and sequentially. However, the modification of the quantum probabilities can be done over the entire superposition, with a single operator simultaneously affecting all possible states. If the number of operations to increase the probability of the quantum state is less than the number of operations to evolve the classical state, then there will be a quantum advantage in the algorithm.

        Finding optimization problems in which an almost exhaustive search must be performed and, therefore, are likely to have a quantum advantage is not complicated. Classically, computer scientists have identified some of these problems as very representative and have tested different algorithms on them. Then, when they wanted to solve other problems efficiently, they adapted them to be as similar as possible to the benchmark problem and thus applied the same algorithm. Even going a step further, they defined certain toy problems that were a simplification of real problems, but faithfully reproduced the complexity of the problem, making it very easy to go from the theoretical problem to the real problem. Some of the theoretical problems that were proposed according to the category of problems they solved were as follows:

        \begin{itemize}
            \item \textbf{Theoretical routing problem:} Problem of finding the best route between two cities. The test bed problem proposed is the Travel Salesman Problem (TSP) where a traveler starts in one city, visits a series of cities once, and ends up in the same city he started in \cite{junger1995, hoffman2013}.

            \item \textbf{Theoretical configuration problem:} Problem of finding the best load balance for a small space. The proposed example problem was the knapsack problem. In this problem, the objective is to fill a knapsack that supports a given weight with a configuration of objects that maximizes a certain reward. \cite{salkin1975, bretthauer2002}

            \item \textbf{Theoretical artificial Intelligence:} In machine learning, there are many processes that need to be optimized, such as hypothesis search or hyperparameter optimization. Along these lines, problems such as N-Queen have been used to test algorithms that were then used in more complex techniques. The N-Queen problem consists of distributing $n$ queens on a chessboard $n$ by $n$ so that they do not attack each other \cite{bowtell2021}.

            \item \textbf{Theoretical sampling:} Problem to characterize a function by taking only some samples of it. For example, to find the value of pi, it is possible to take a square and a circle such that the diameter of the circle is equal to the side of the square. By taking random samples and seeing which ones are inside the square and which ones are outside, it is possible to estimate a fairly accurate value of pi \cite{kroese2012}, as seen in Figure \ref{fig:pi_montecarlo}.
            
            \begin{figure}[t]
                \centering
                \includegraphics[width=0.6\textwidth]{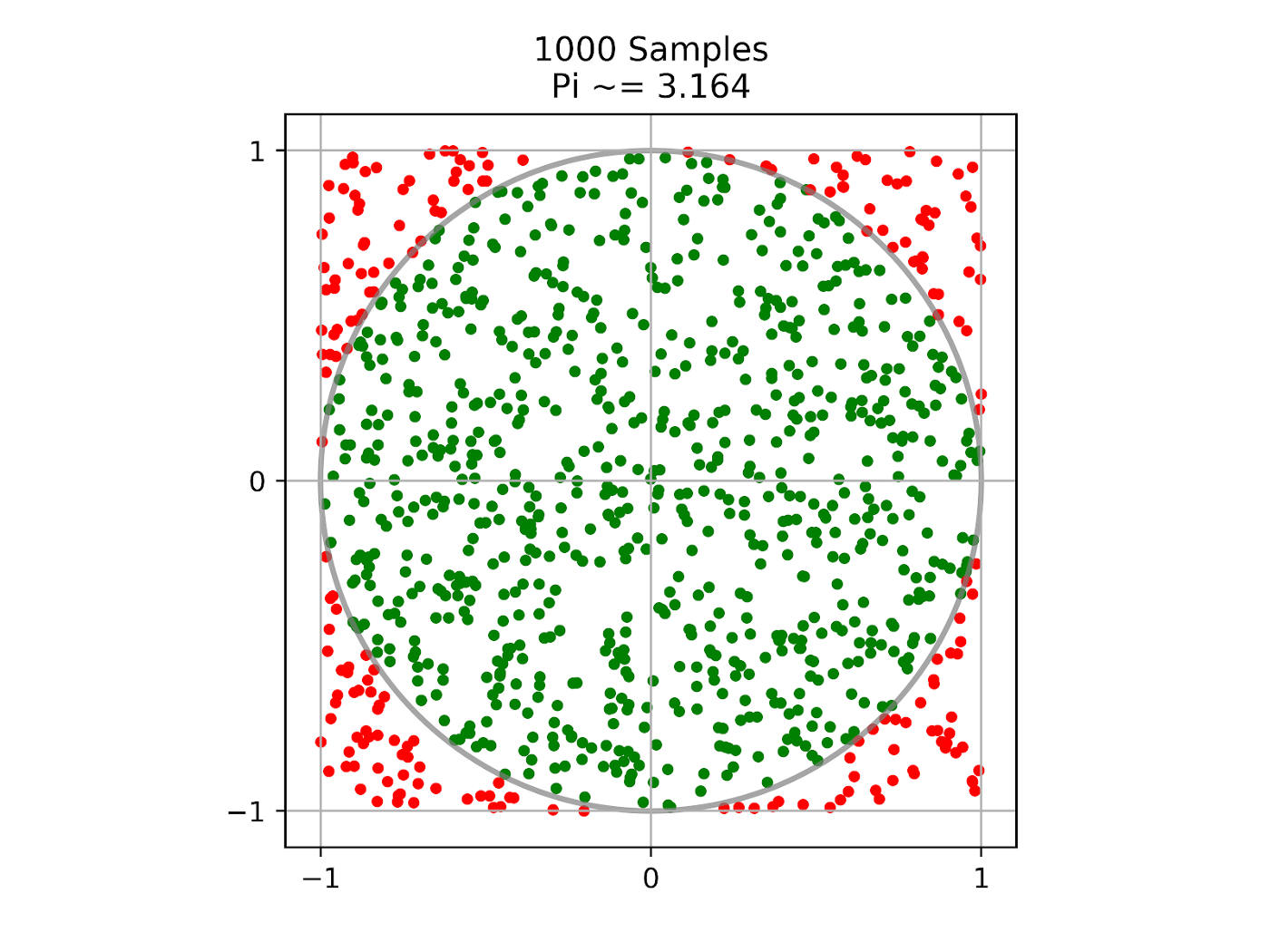}
                \caption{\label{fig:pi_montecarlo} Pi is estimated using the number of green points that fall into the circle and the red points that fall out of the circle. This is a typical example to explain Monte Carlo simulation.}
            \end{figure}

        \end{itemize}

        The important thing about these theoretical problems is their simple applicability to practical problems.

        \begin{itemize}
            \item \textbf{Routing problem:} The problem of finding a route between two or more points through a series of intermediate points is the problem solved by any package, courier or transport company to minimize the number of vehicles or the distance traveled. The larger the problem solved, the greater the number of points, vehicles, people or packages transported, etc., the greater the company's profit. There are numerous variants such as minimizing distance, time, fuel, etc.

            \item \textbf{Configuration problem:} The problem of delivering a number of items in a set of containers is a common one for packaging companies, international freight forwarders or almost any company in the industry known as last mile delivery. It is about cramming more products into a smaller space. Solving this problem more efficiently helps to reduce the number of containers or the size of containers, to give a few examples.

            \item \textbf{Artificial intelligence:} Optimization problems in artificial intelligence can range from selection of the hypothesis that best explains a data set, greater generalization and coverage with fewer clauses, to selection of hyperparameters that improve the performance of a neural network on a particular problem. For example, the unsupervised learning algorithm, k-means requires, in each iteration, to distribute centroids and calculate distances in such a way that the points are as well distributed as possible.

            \item \textbf{Sampling:} Sampling problems are widely used at the industrial level when it is difficult to know a certain distribution. For example, in the financial sector, to know the trend of an asset, or to improve risk analysis. In these cases, the ability to solve larger problems in shorter times allows obtaining more accurate short-term results.

        \end{itemize}

        So far the interest in using quantum S\&S algorithms has been generally motivated, the next step is to study both techniques in more detail and explain the relevance of this type of algorithms as they fit into optimization problems.

        \subsection{Search}\label{apt:search}

            \begin{figure}[t]
                \centering
                \includegraphics[width=0.5\textwidth]{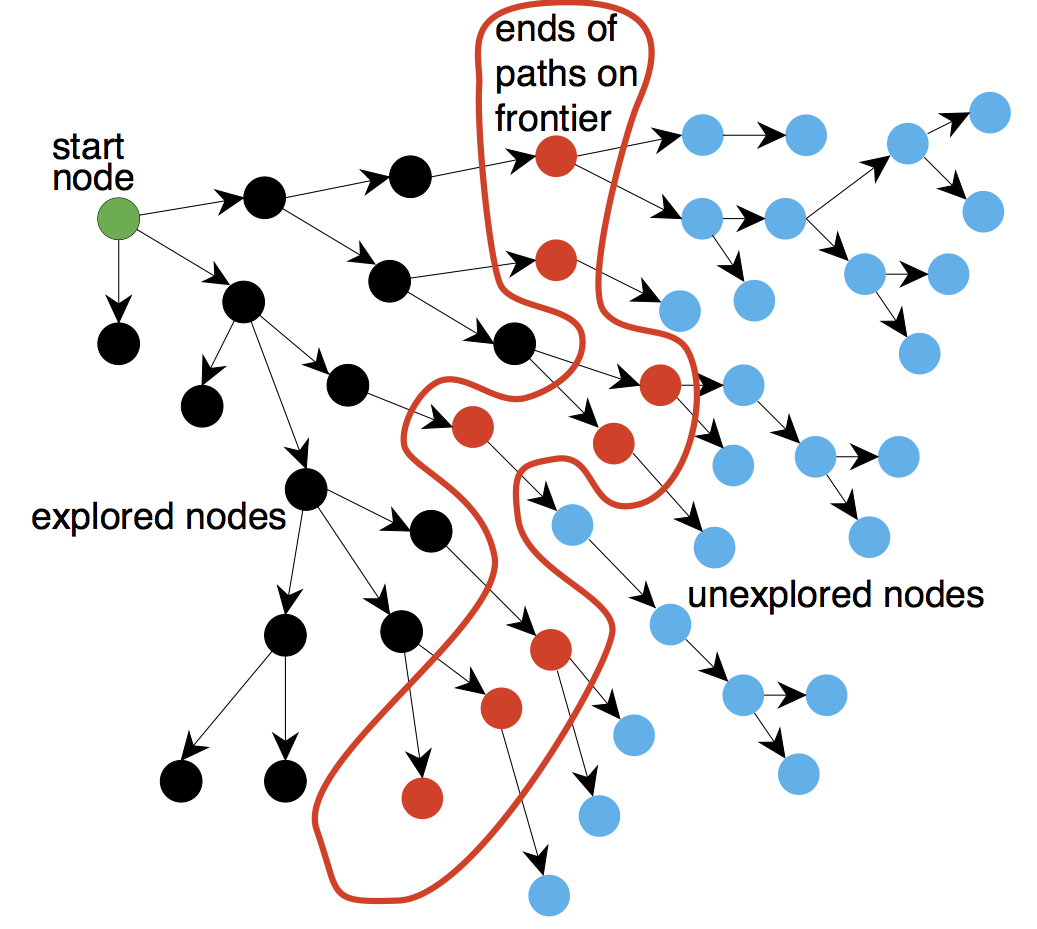}
                \caption{\label{fig:search} Scheme extracted from \protect\url{https://artint.info/2e/html2e/ArtInt2e.Ch3.S4.html}. It shows the elements of a search from a start node. During the search, there are different categories of the nodes. In black, the already explored nodes which have been already evaluated and successor have been calculated. Then, in red the frontier nodes that are under evaluation in this step. Finally, the blue nodes have not been explored yet, the solution is one of these nodes.}
            \end{figure}

            Search algorithms are techniques that visit nodes of a problem with a large search space until the state of minimum cost or maximum reward is found, as shown in figure \ref{fig:search}. The main objective of these algorithms is to evaluate the minimum number of possible states until the solution is found. As with sampling algorithms, the search will always take a set of problem states and try to find a solution that is extrapolable to the entire problem.

            Applying a search algorithm to a problem is interesting if the problem is large enough that evaluating all possible states or combinations is too costly. Thus, the fundamental objective of a search algorithm is either to evaluate a subset of states that contain the solution to the problem (minimum cost or maximum reward state) or to evaluate the fundamental nodes that give it information on where the solution node is.

            To simplify the search task, the problem can be represented in the most efficient way possible for the algorithm. In this way, there are representations of problems such as graphs, trees, constraints to satisfy, Stanford Research Institute Problem Solver language, STRIPS \cite{villegas2009}, etc. If search is understood as a problem in which an agent must find a solution state, an efficient representation may be to represent the problem as the union of states, actions and rewards. Thus, there would be a set of states S, a set of actions A, and a set of rewards R.

            In this context, whether the search algorithm is understood as an agent or not, it must visit the states sequentially, one after the other, to evaluate each one and generate the successors. This process will then be repeated iteratively until the solution or a stopping criterion is found. This sequential analysis of the states is one of the most limiting factors for the feasibility of a search algorithm. Each time the problem grows, a new variable is added to the state space, the state space grows exponentially and, therefore, the number of states that need to be evaluated also increases.

            Among the various techniques for speeding up search algorithms, search parallelization stands out. This optimization aims to ensure that the states are processed in parallel by $n$ threads. While it is true that this is an advantage and is used, the computational cost remains high and scalability limited. \cite{greenlaw1995, cook2023}. Other techniques to avoid sequential search are bidirectional search and random search, but they are still algorithms with low accuracy and some computational cost.

            There are even problems in which the search to find the solution is practically exhaustive, i.e. it is necessary to visit all the possible states of the problem until the solution is found. To give a simple example, to find the correct combination of a lock, it is necessary to try all the possible solutions until the correct one is found, and by evaluating a state it is impossible to know how close or far it is from the solution. In cases where an almost exhaustive search is required, the simultaneous evaluation of nodes has the greatest impact.

            Search algorithms can be divided into two main categories according to their choice of successors, informed or uninformed search. According to this classification, if the algorithm chooses the next node to be evaluated without having an estimate up to the solution node, it is called an uninformed search. If, on the other hand, among the possible nodes to be expanded, an estimate is made of the distance of each one to the goal and the one that is closest to the goal is chosen, it is called informed search.

            Within an uninformed search, systematic search algorithms such as breadth-first, depth-first, or bidirectional algorithms, or stochastic algorithms such as random walks can be distinguished. Uninformed search is fast because it does not need to evaluate the distance from the state to the goal, but it either has low precision, like a random search, or high cost, like a breadth-first search.

            Informed search is much more accurate, but requires some domain knowledge to develop some kind of heuristics to guide the algorithm to the solution. In this category are heuristic search algorithms such as A* or alpha-beta pruning. In this type of algorithm, a balance must be struck between a sufficiently informative evaluation function and one that is fast to compute because it is to be evaluated at each node.

            Search algorithms have numerous applications for problems where it is necessary to find a solution that meets certain constraints. These constraints can be optimality constraints, maxima or minima, restrictions, etc. For this reason, search algorithms are closely related to optimization problems. These problems involve numerous constraints, and it is necessary to try different configurations of parameters until one is found that satisfies all of them.

            Some examples where search problems are useful are routing problems to find the best path among possible paths, configuration problems to find the combination that maximizes reward, or constraint satisfaction problems to find a particular configuration. In this context, artificial intelligence problems can be easily adapted to problems where the best technique to solve is search \cite{russell2010}.

        \subsection{Sampling}\label{apt:sampling}

            \begin{figure}[t]
                \centering
                \includegraphics[width=0.8\textwidth]{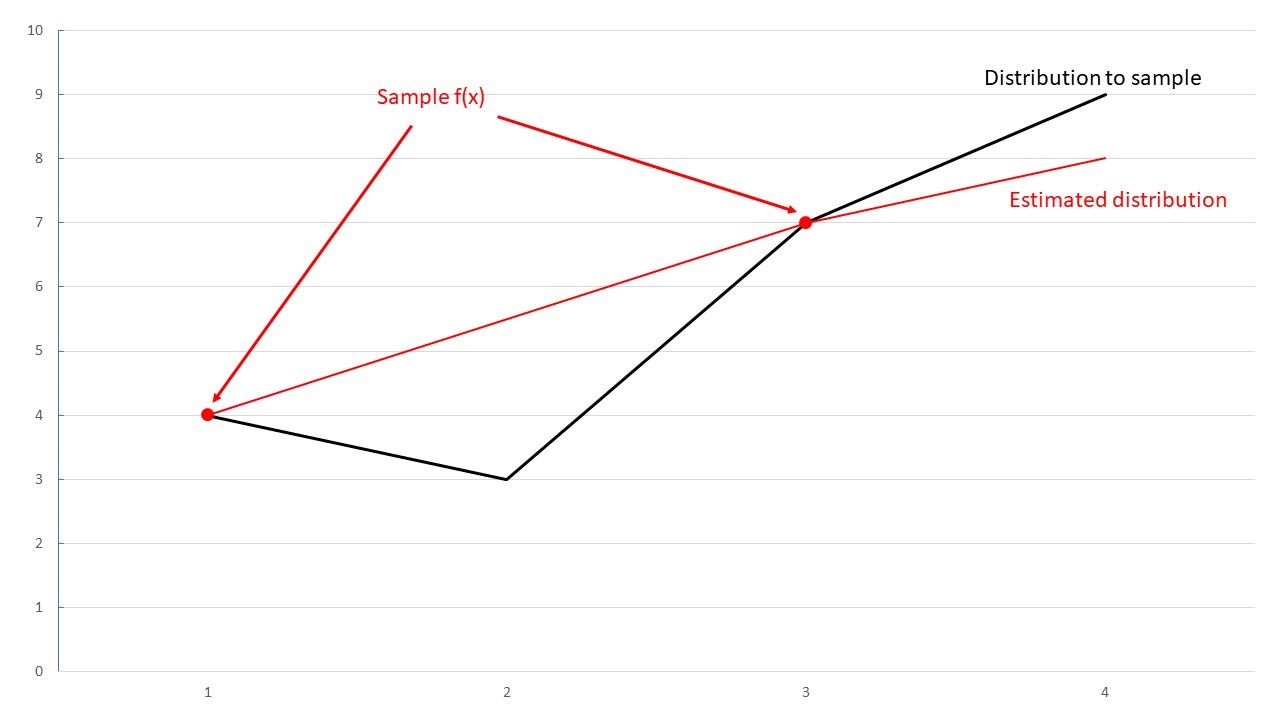}
                \caption{\label{fig:sample} Scheme that shows how a distribution (black) can be estimated (red line) just getting some points (red circles) from it. The red line, the estimated line, is far from the real distribution because there are only two points sampled. Increasing the samples points, the estimated distribution will be closer to the real distribution to sample.}
            \end{figure}

            Statistical sampling of a function is a technique that, from a set of data called population, selects a smaller subset called a sample. This sample is expected to be sufficiently representative to be able to extrapolate any analysis made on it to the population. In this way, working with the reduced sample will allow one to know a general and approximate trend of the population, but at a lower computational cost, as shown in figure \ref{fig:sample}.

            Sampling algorithms are especially interesting for problems where it is difficult to obtain probability distribution information by other methods. If it were easy to use analytical methods, such as the derivative of the function, to learn key aspects of the function, sampling would not be necessary. Likewise, if it were a very simple function to sample because the function $f(x)$, which returns information about the function at a specific point x, is executed very quickly, it would not be necessary to use a subset of the population, the entire population would be sampled and thus a greater amount of information would be obtained with greater precision.

            Therefore, sampling algorithms are used in those problems in which analyzing the set of states is too complicated due to its size or the characteristics of the problem. Once again, large exponentially scaling state spaces reappear. This makes analyzing a small subset of data very useful but not the solution, because it has to be sampled individually. This legitimizes quantum techniques that allow for sampling over the superposition of all states.

            Another fundamental characteristic of the problems on which sampling techniques are applied is the relationship between the elements that make up the population. It has been previously mentioned that the population must be large enough to be very slow to analyze all elements. In addition, there must be certain relationships between the elements so that when a subset of these is chosen, the properties and relationships that exist in the entire population are reflected with a certain probability in the reduced subset to be analyzed. Thus, general conclusions can be extrapolated to the whole problem.

            Depending on the relationships between the data and the type of problem to be solved, one sampling algorithm or another can be applied within the wide range of possibilities for sampling a problem. Later, in chapter \ref{ch:classical_sampling}, classical sampling algorithms will be studied in detail. However, in order to have a global view, these algorithms can be classified as probability and non-probability sampling.

            In probability sampling, the elements that make up the subset of the total population to be analyzed are chosen following a more or less random method. For this reason, any state of the population has a non-zero probability of being in the sample subset. This helps in choosing some confidence elements of the population that have general characteristics. Within this category are the methods of simple random sampling, stratified sampling, systematic sampling and clustering sampling.

            Non probability sampling techniques use non-random sample selection. By eliminating the stochasticity of the process of sampling elements of the population, the selection is made by a criterion of convenience or explicit knowledge of the problem. This type of alternative selection makes the sample less representative of the population, but helps to avoid false data, outliers, or unbalanced samples.\cite{singh2014}

            Using any of the techniques that belong to one of the above categories, a sampling error occurs. This error is due to the fact that a certain sample subset of a population may not contain all the characteristics of the population. In order to know a particular characteristic of the population, the sampling error is the difference between the estimated value of that parameter using the sample and the actual value of the parameter calculated using the entire population. To dampen this possible error, a confidence margin is used, which allows an estimate to be made within a certain confidence interval.

            The error during sampling can be of different types and there are techniques to mitigate it depending on its nature and the algorithm used. This error causes the results obtained on the sample to differ from the real results that would be obtained on the population. This difference can be understood as bias because it gives information about the problem that is related and closely connected to the samples that have been chosen from the population. This bias is unavoidable during sampling, but it is not necessarily detrimental.

            Biases can be voluntary or involuntary. Unintentional biases should be avoided because they reduce the generality of the results obtained and lead to erroneous results, precisely because of the ignorance of these biases. Voluntary biases, however, help to focus on a particular characteristic of the data set. This helps to avoid evaluating population data that lack relevance and to obtain more accurate information on certain parameters. For example, these voluntary biases are crucial in any machine learning algorithm because biases are the element used to generate a knowledge model of the problem. Having a knowledge model without biases means that the model is too general that it is not useful due to unspecific results.

            Sampling algorithms are widely used to learn functions that are difficult to analyze with other methods. For example, for signal analysis, they allow to approximate the signal and reproduce it without having to evaluate it at all possible points. This is especially useful for continuous signals that need to be discretized. Another example is in the financial sector, the analysis of the trend of an asset or the risk of a certain portfolio.

            Another field of great application for sampling is machine learning. When running an intelligent algorithm that is trained on a set of input data, it is necessary to analyze these data to generate a set of hypotheses that explain them. In this hypothesis space, a sampling is made indicating how the data are being generalized, coverage of the hypothesis, and how complex they are, and length or number of hypothesis clauses. This is applicable to both supervised learning, for example, certain types of neural network, and unsupervised learning, for example, the k-means algorithm \cite{ahmed2020}.

            In the case of this work, sampling algorithms will be especially used to sample a large space of states, seeking to understand where a maximum or minimum of the probability distribution that defines a certain problem may be. In this way, sampling will serve as a first step to guide the search and make it over a much smaller state space. Thus, the better the sampling, the more accurate and faster the search will be.

        \subsection{Relevance of S\&S} \label{apt:relevance}

            As explained above, the S\&S algorithms share certain characteristics that allow a joint study of both techniques. This joint study can also be extended to the types of problems that can be solved with S\&S algorithms.  Some characteristics that have to be present in the problems are:

            \begin{itemize}
                \item \textbf{Defined initial state:} The initial state of the problem must be a defined state. It can be the initial state of the search or an initial point of the function to start sampling.

                \item \textbf{Transition function:} A definition of how one transits from one state to another. If it is a network, the connections and their probabilities would be defined; if it is a tree, the generation of successors, etc.

                \item \textbf{Evaluation function:} A function that can be applied in any state and provides information about the global or local state of the problem. For example, that calculates the cost of that state or of the path to reach it.

                \item \textbf{Well-defined Goal:} It must be clear how to solve the problem, whether the goal is to minimize the expenditure of a resource, maximize a certain reward, reach a particular state, etc.
                
            \end{itemize}

        Another common feature of S\&S algorithms is the representation of the problem to be solved. Although there are different techniques \cite{villegas2009}, one of the most common is the Markov Decision Process (MDPs). MDPs are based on Markov chains determining that the probability of an event occurring depends only on the immediately preceding event \cite{puterman1990}. This type of causality hypothesis is very convenient because it ensures that the decision of the next state to be evaluated depends on the state of the system at the previous instant, facilitating the adaptation to changes and minimizing the possibility of getting stuck at a local level.

        The MDPs define the problem as a tuple of (S, A, P, R) where:

        \begin{itemize}
            \item \textbf{S, States:} Set of actions called state space. It is the set that contains all possible values of the problem variables.

            \item \textbf{A, Actions:} Set of actions called action space. It is the set that contains all possible actions that the agent can perform in the state space.

            \item \textbf{P, Probability transition model:} It is the model that assigns a probability of each action that the agent performs in any state. The probability determines the new state after performing an action.

            \item \textbf{R, Rewards:} It is the expected return to the agent after performing an action in a state.
        \end{itemize}

        There are numerous variants of MDPs and they are specially designed to model the behavior of an agent in an environment, but they are suitable for modeling both sample and, especially, search. For sampling algorithms, MDPs are adapted in such a way that the agent has only one possible action to do, sampling.

        Another important point to be emphasized in this paper is that S\&S algorithms can act in concert. It is possible to iteratively combine first a sampling algorithm and then a search algorithm to reduce the search space in each iteration. It has been explained before that one of the difficulties of the problems in which S\&S are applied is the size of the problem. Therefore, it seems logical to divide or reduce the size of the problem to make it manageable.

        Especially in quantum computing, where hardware development is so limited, making an iterative algorithm that combines two techniques, search \& sample, allows one to be quantum and the other classical. In this way, one technique will be able to process the state space in superposition and the other will be able to handle the large volume data set. This paradigm especially motivates the focus of this first part \ref{pt:qss} of this thesis.

        Some of the limitations of the S\&S algorithms have already been succinctly described. Specifically addressing the complexity of these algorithms and the problems they solve, it is possible to define the complexity classes P and NP. These complexity classes serve to classify problems and algorithms into categories in which they share certain characteristics useful for finding a solution. The complexity determines whether a problem can be solved faster or slower, or how much memory it will consume. The computational complexity classes are not invariant elements, but are open to constant revision and discussion \cite{goldreich2008}.

        Although the analysis of computational complexity is very extensive and there are numerous classes beyond P and NP, for the study of S\&S algorithms, it is interesting to focus on these two.

        P complexity class encompasses all those problems that can be solved in polynomial time. This means that a polynomial formula can be established that, given the number of variables in the problem, returns the execution time of the algorithm. A problem belonging to complexity class P is considered tractable because a classical computer can execute with relative simplicity a number of operations that grows polynomially. Therefore, problems belonging to P are said to have good scalability, since it is not expected that there are so many variables that the algorithm cannot be executed on a classical computer. An example of a problem belonging to P is addition; the number of operations is polynomial proportional to the number of variables to be added.

        The NP complexity class is used to classify decision problems. The solution to these problems can be verified in polynomial time, but to find it a non-polynomial time is needed. Within this class, there are other classes such as NP-complete and also the NP-hard class \cite{wigderson2006}. The important thing about this class is that each time a new variable is added to the problem, the complexity or time needed to find the solution grows in a non-polynomial way. This implies that the scalability of this type of problem is poor and if the problem grows it is possible that it cannot be solved with a classical computer in a reasonable time \cite{beame1995}.

        The relationship between the complexity classes P and NP is somewhat unclear and is the subject of constant study. There are currently two currents among scientists, although one is assumed to be the most probable.

        \begin{itemize}
            \item \textbf{P = NP}: If classes P and NP are equal, it means that there are efficient algorithms for NP class problems that have not yet been discovered. This is the least accepted idea among computer scientists because it would mean that there is a way of doing algorithms that is completely unknown so far.

            \item \textbf{P $\ne$ NP}: If both classes are different, it means that there are problems belonging to NP that cannot be solved by a classical computer efficiently. This opens the door to the use of new computing paradigms that can work on this type of problem. In this thesis, hybrid quantum-classical computing algorithms are explored to tackle these NP problems.

        \end{itemize}

\begin{figure}[t]
            \centering
            \includegraphics[width=0.6\textwidth]{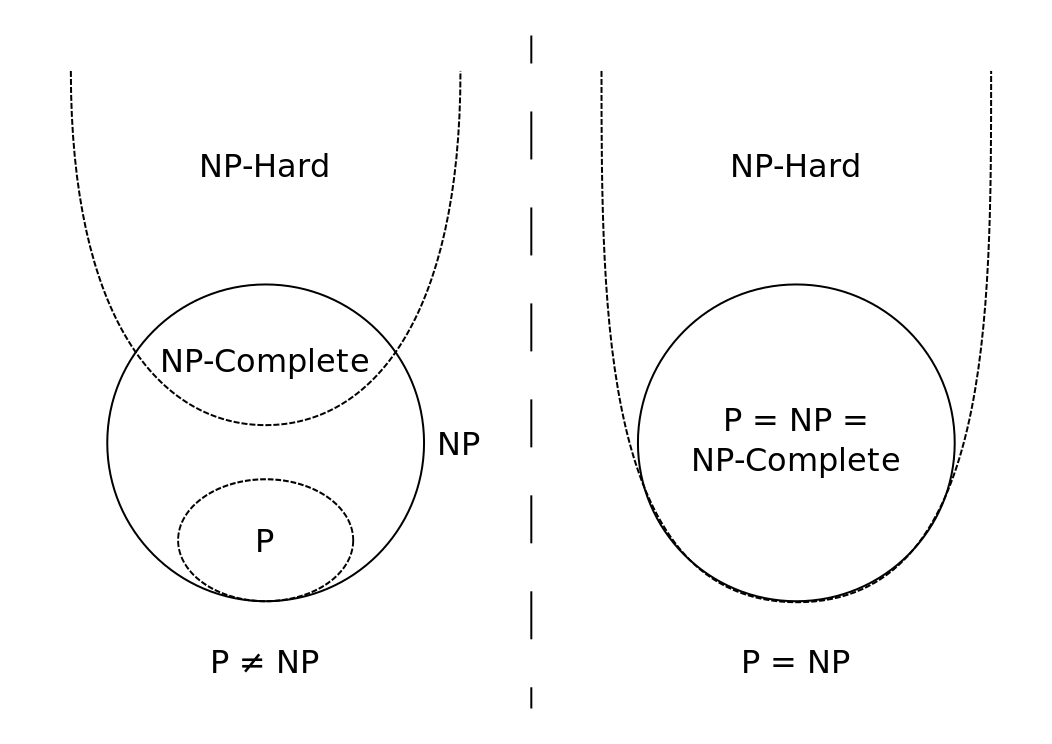}
            \caption{\label{fig:p_np} The complexity classes P and NP play a pivotal role in elucidating the computational complexity of existing algorithms. While the relationship between these classes remains elusive, prevailing evidence suggests that P and NP are not equivalent, as depicted in the left-hand scenario.}
        \end{figure}

        Following this distinction between P and NP, it is easy to assume that P and NP are different, because although in theory they are not, right now, there are no algorithms to convert NP problems to P. Therefore, many of the problems to which S\&S algorithms are applied are NP problems that cannot be solved classically faster. Some examples already mentioned are the TSP, the knapsack problem or the portfolio optimization problem. Figure \ref{fig:p_np} shows a scheme of these possible relationships between P and NP.

        At this point, close to a dead end for classical computing, the option arises to think of new computing paradigms that approach the problem from another perspective. Quantum computing proposes to work with the entire state space in superposition. This new approach seeks the simultaneous modification of states by applying only one operation, as opposed to sequentially that would exist classically and that would require an operation for each state to be generated. This difference translates into the fact that quantum search is done exhaustively with the application of a single operator, while classically it is necessary for the operators to be much more precise in order not to have to do this exhaustive search. 

        While it is true that this superposition of states is a great advantage over classical computing, it also brings with it certain differences in information processing. While classically, the tricky thing is to generate states closer and closer to the goal state until it is reached, quantum computing is not about creating a superposition of states; the real tricky step is to increase the probability of those states that belong to the solution.
        
        To be able to solve problems using quantum computing requires a new way of designing algorithms, new ways of processing information, and designing specific operators. In this part \ref{pt:qss}, it will be discussed how to design quantum S\&S algorithms with advantages over the classical ones and how to apply them to industrial problems. For this purpose, a hybrid tool will be created that will combine the application of operators in quantum superposition with the processing of large volumes of classical data.
        
        One of the main objectives of this work is to design and apply these new quantum computing techniques to create practical algorithms at an industrial level, including their implementations. This approach is part of a general trend in recent years that advocates to start creating algorithms that can be executed on quantum computers and not only theoretical proposals. In this chapter, how to apply S\&S quantum algorithms to optimization problems will be discussed. 

        \subsection{Optimization problems}\label{apt:opt_prob}

        There are many ways to define an optimization problem. In this thesis, a definition of optimization problems has been selected that is advantageous for the S\&S algorithms. The problem definition consists of a successor generation function (probability matrix of transitions) and an evaluation function (reward function). In this way, any optimization problem can be defined as the probability of transitioning between states and the function that gives a value to that state or to the state of the world at that point. This definition of the optimization problem may certainly seem profitable for the interests of S\&S algorithms, but the reality is that it is the definition that is common to most optimization problems at the industrial level.

        The problems to be addressed in this thesis are the following:

        \begin{itemize}
            \item \textbf{Artificial intelligence:} In artificial intelligence, models with multiple variables that require a constant adjustment are used. This adjustment can be understood as an optimization of their weights to maximize a reward or reduce the distance between the state of the learning algorithm and the real world. \cite{gambella2021}. Chapter \ref{ch:qaiSS} presents a study on how to apply a quantum S\&S algorithm to the N-Queen problem. This problem is a common test-bed for search algorithms that optimize a certain learning function.

            \item \textbf{Space exploration:} In space exploration, it is common to collect large volumes of data. These data are often associated with problems that are studied using complex models. Applying a complex model to a large amount of data is costly in itself. However, in addition, these data often require some optimization. For example, in the detection of gravitational waves, it is necessary to make a parameter estimation that can be seen as fitting a certain function to the received wave by modifying its parameters. It will be discussed in more detail in chapter \ref{ch:SSspace}.

            \item \textbf{Quantum chemistry:} Quantum chemistry problems are especially interesting from the perspective of quantum computing. If, in addition, it is a problem that requires optimizing a particular configuration of variables, then quantum computation is more likely to work well. Chapter \ref{ch:SSchem} addresses the protein folding problem, in which it is necessary to optimize the configuration of the protein to find the one with the lowest energy and thus avoid its denaturation inside the body. 
            
        \end{itemize}

        The use cases discussed above share a common characteristic; all of their variables have a discrete domain. These types of problem are known as combinatorial optimization problems where having discrete variables, they tend to have a finite problem domain. Thus, the state space is large but bounded \cite{korte2011}.

        In the following sections, the algorithms currently applied to solve these problems and the quantum algorithms designed during this thesis will be detailed. In addition, it will be explained how to apply these quantum algorithms to the aforementioned use cases.

    \newpage
    \section{Classical Algorithms for S\&S}\label{ch:cSS}
    \thispagestyle{empty}

        \lettrine[lines=1, findent=2pt]{\resizebox{!}{1.2\baselineskip}{S}}{}earch and sampling algorithms have been extensively studied and evolved in classical computer science since the 1980s \cite{russell2010}. It is interesting to know how this evolution has been, to understand that the initial performance limitations have always existed and have been avoided in different ways as the theory evolved.

        Whenever a new search or sampling algorithm or paradigm has emerged, it has evolved until a solution was found that faced an insurmountable performance limit. Once that limit has been reached, an attempt has been made to move forward by changing paradigm. For example, in heuristic search, different algorithms were created until stabilizing in the A* algorithm, which has remained the fastest optimal heuristic search algorithm for almost 50 years. In that time, other algorithms have emerged but with suboptimal search, other heuristics, etc. But the A* algorithm remains the most widely used in practice because it perfectly combines a fast and informative method.

        The following is an overview of the main classical search \& sample algorithms. After this review, the different limitations they currently have are analyzed.

        \subsection{Classical Search} \label{apt:classical_search}

\begin{figure}[t]
            \centering
            \includegraphics[width=1\textwidth]{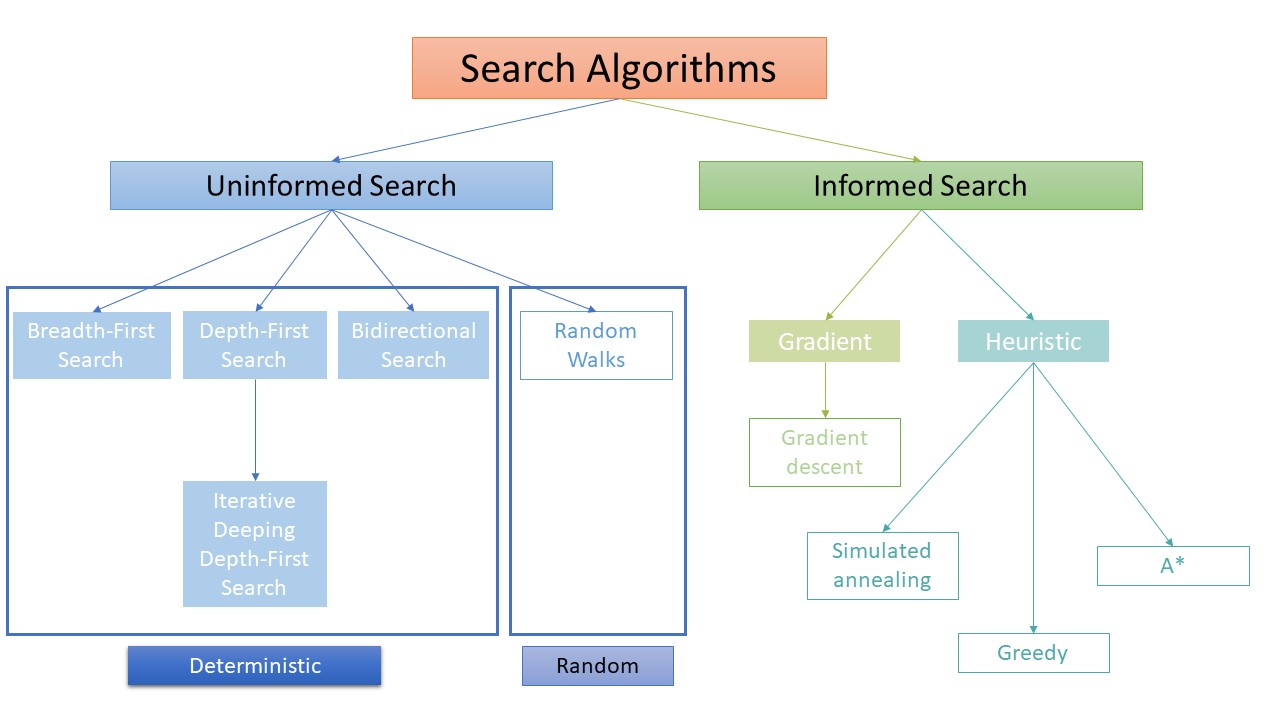}
            \caption{\label{fig:search_algorithms} Scheme with the main search algorithms divided into categories attending to the type of search. Depending if the search uses any information of the problem to be guided, the search can be uninformed or informed. Inside of the uninformed category, there are deterministic searches or random algorithms. Among informed search algorithms, the information can be given using a gradient or an heuristic.}
        \end{figure}

        Search algorithms are abstract techniques to find a state that meets certain conditions from a set. Depending on the characteristics of the problem, one algorithm or others can be used. The main types of search are shown in figure \ref{fig:search_algorithms}. 
        
        Due to this definition can be somewhat abstract and difficult to visualize, initial search algorithms were illustrated as trees. This metaphor serves both didactic and practical purposes by being able to easily represent the evolution of the search algorithm.

        In this way, the state in which the search starts can be seen as a node called root. This root node is evaluated to decide whether it is a node that represents a solution state. If not, it is expanded to obtain its successors. When a node is expanded, it is passed to a list of visited nodes. The successors go to a list of expanded nodes, not visited nodes, from which one is chosen to continue the search. Thus, the tree search is a process of visiting a node, expanding successors, and choosing a successor to repeat the process.

        These search trees can have many shapes and characteristics. It may be that the same node can be reached from several paths, there may be loops between nodes, there may be nodes that have no children and are known as leaf nodes, and so on. What is important is that during the search, the algorithm is able to avoid two fundamental problems, getting stuck in a loop and redundant paths. To do this, many search algorithms check which nodes have already expanded in the list of expanded nodes and avoid entering them in the list of nodes to expand, the list of successor nodes.

        These algorithms are classified according to their ability to achieve the concepts of optimality and completeness. Assuming that each expansion of node, successor generation, has a cost of 1.

        \begin{itemize}
            \item \textbf{Optimality:} Optimality means that the algorithm guarantees to have found the minimum cost solution.

            \item \textbf{Completeness:} The completeness of an algorithm guarantees that it is able to find the solution no matter which node of the state space it is in.
        \end{itemize}

        To review the most important search algorithms that exist and are relevant for the development of quantum search algorithms, a division will be made between informed and uninformed search algorithms. Within this category, the concepts of completeness and optimality will be analyzed.

        Uninformed search algorithms are characterized by having no information about how close or far away the goal state is. This type of algorithm simply expands the nodes until one of them is found to be the goal. Returning to the tree scheme, the quality of the solution will be given by the distance from the goal node to the root, closer implies better quality. The cost of the algorithm will be given by the number of nodes needed to expand until the solution is found.

        The main uninformed search algorithms are divided into sequential or random search and with different algorithms in each category.

        \begin{itemize}
            \item \textbf{Sequential:}
                \begin{itemize}
                \item Breadth-First Search
                \item Depth-first Search
                \item Iterative deepening Depth-limited search
                \item Bidirectional search
                \end{itemize}
            \item \textbf{Random:}
                \begin{itemize}
                    \item Random walks
                \end{itemize}
        \end{itemize}

\begin{figure}[t]
            \centering
            \includegraphics[width=0.65\textwidth]{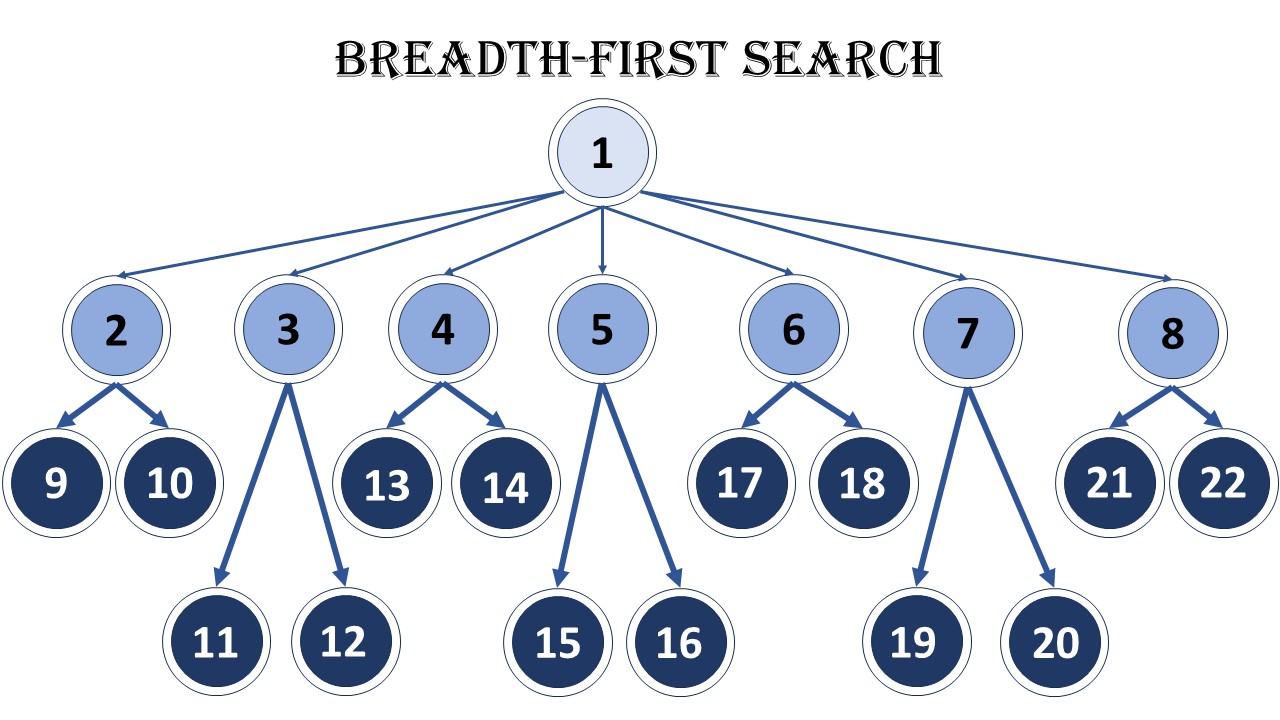}
            \caption{\label{fig:breadth_first} The breadth-first search algorithm systematically explores nodes at each level of a graph before descending to deeper levels. In this figure, nodes are organized into three depth levels, each represented by a distinct shade of blue, ranging from lighter to darker. Prior to traversing nodes at the third depth level, all nodes at the second level are exhaustively visited.}
        \end{figure}

        The breadth-first search was the first attempt to do a search in which, even if it did not have information on what state the goal was in, it might not have to expand all the nodes to find it. This type of search starts at the root node, then visits all expanded nodes from the root, and saves the successors of those nodes for the next round. In the next round, it expands all nodes at a distance of two from the root and saves their successors. In this way, it iteratively traverses the entire breadth-first search tree, visiting all nodes at one level before jumping to the next. An example can be seen in Figure \ref{fig:breadth_first}.

        This type of search was initially proposed because it satisfies the criteria of completeness and optimality. By traversing all nodes at all levels of the tree, it is certain that at some point it will find the solution and it will be optimal. Optimality is ensured because it first visits all nodes of minimum cost, then all nodes of the next cost value, and so on.

        Although this algorithm is perfect from the formal point of view because it meets two very important criteria, it always finds a solution and this solution is of the highest quality, in practice it is not a very useful algorithm. The reason for this lack of usefulness in practice is the number of nodes required to expand. It is a level-by-level exhaustive search algorithm, that is, until the solution is found, it will have expanded all the nodes \cite{nievergelt2000}.

\begin{figure}[t]
            \centering
            \includegraphics[width=0.65\textwidth]{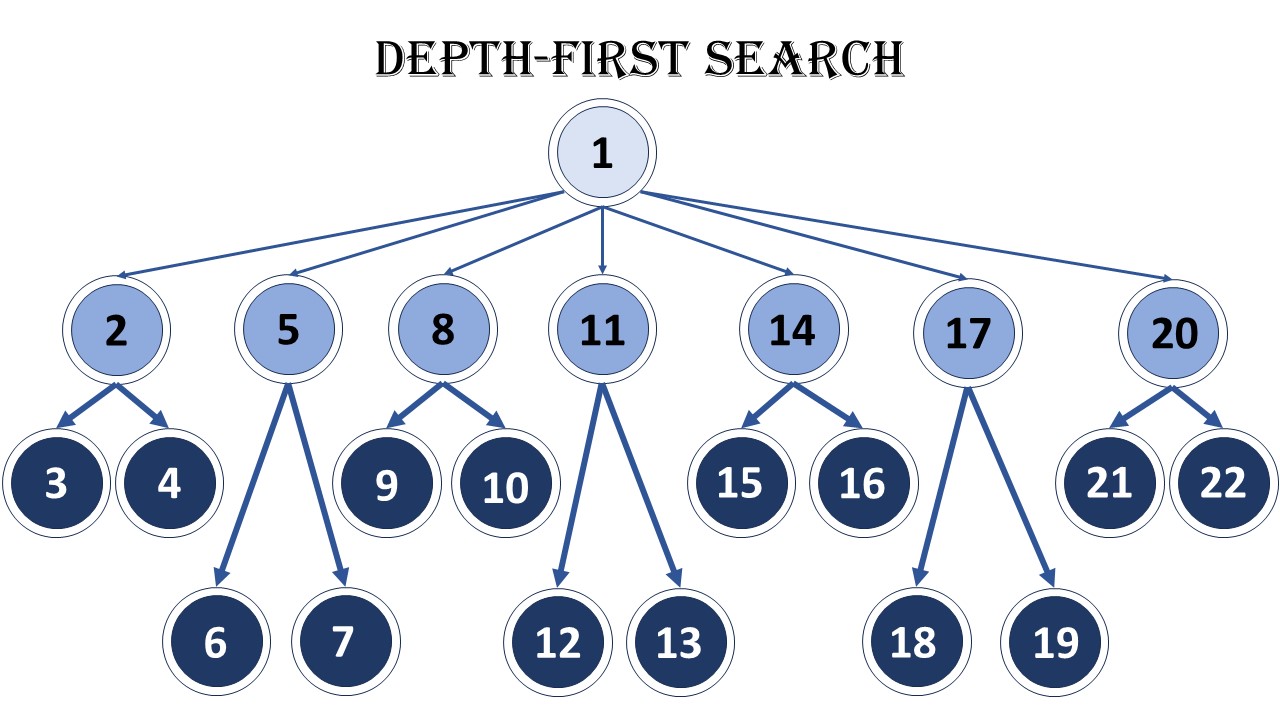}
            \caption{\label{fig:depth_first} The depth-first search algorithm explores nodes along a single branch until reaching a leaf node, then moves to the next branch. Unlike breadth-first search, the depth level is not prioritized. In contrast to the approach depicted in figure \ref{fig:breadth_first}, this algorithm focuses on swiftly reaching a solution rather than exhaustively searching all nodes.}
        \end{figure}

        The depth-first search follows the opposite paradigm to the breadth search. This search always expands the deepest node among the possible successors. Once a node that has no successors, leaf node, is reached, it returns to continue expanding the next deepest nodes. An example can be seen in Figure \ref{fig:depth_first}.

        The search in depth was initially proposed at the same time as the search in breadth, but later forgotten. Its main problem was that it was neither a complete nor an optimal algorithm. In case of trying to find the solution in a very deep tree, the algorithm was lost by expanding nodes even if the search was at shallow depth, which resulted in an unreliable technique. Moreover, even if a solution was found, it was not possible to know if it was at minimum depth, and it was necessary to continue the search almost as if it were in breadth until other solutions were discarded.

\begin{figure}[t]
            \centering
            \includegraphics[width=0.65\textwidth]{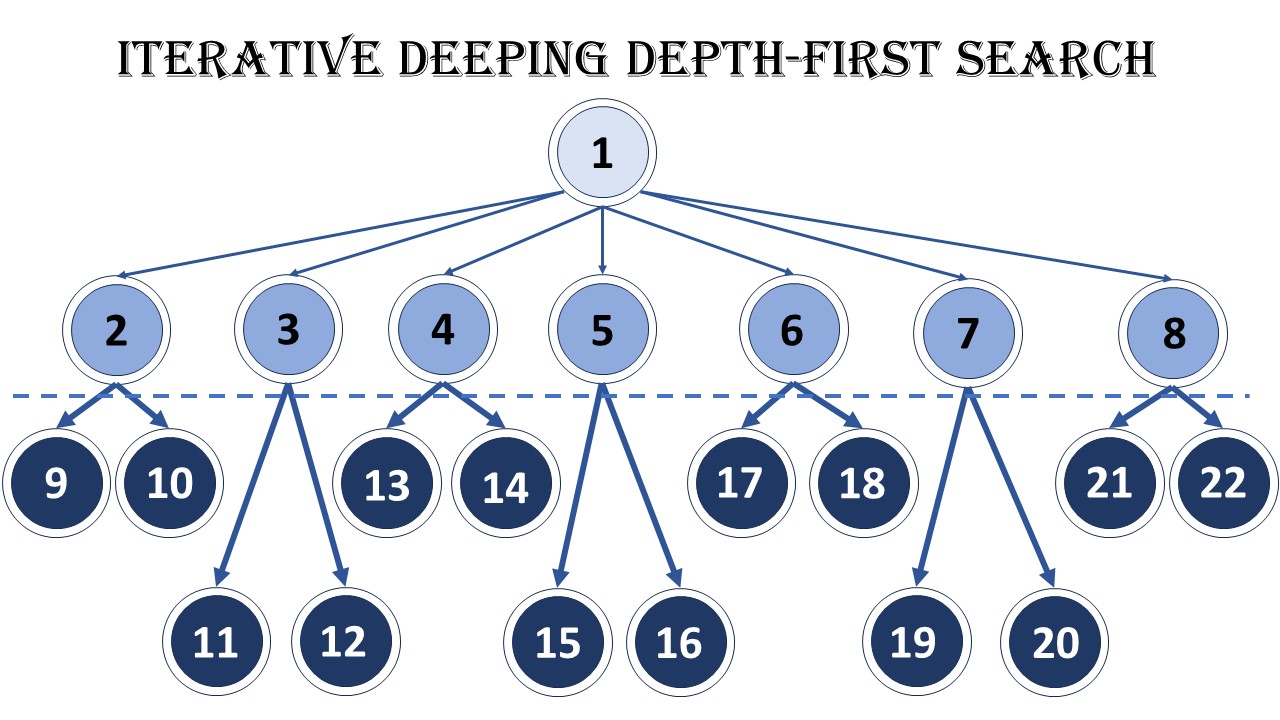}
            \caption{\label{fig:iterative_deeping} This scheme illustrates the order of node visits in the iterative depth-first search. Here, nodes within the same branch are explored until reaching a specified depth limit, denoted by the dashed line (set at 2 in this example). The distinctive feature of the iterative depth-first search, as opposed to previous algorithms, is its initial behavior resembling breadth-first search, as depicted in Figure \ref{fig:depth_first}. Once the limit is reached, however, the search becomes exhaustive, visiting all nodes less depth than the limit.}
        \end{figure}

        To solve the problem of the previous algorithm and try to take advantage of all its benefits, a modification known as iterative deepening depth-limited search was developed. The advantage of this new version of the algorithm was that it searched in depth up to a certain depth. Once that depth was reached. The algorithm went back to the beginning and started again from the root or the previously marked depth. The depth limit was getting bigger and bigger, deeper and deeper, until the solution was found. An example is shown in figure \ref{fig:iterative_deeping}.

        By having a bounded depth at which the algorithm stopped, the problem of infinite search was avoided. This allowed the algorithm to be complete and also, in many cases, to find the solution faster than the breadth-first search algorithm. It makes the iterative deepening depth-first search algorithm the preferred algorithm when an uninformed search algorithm had to be chosen. It is even possible to view this algorithm as a breadth-first search, if at each iteration one adds one to the limiting depth to search. In this case, the algorithm will perform a breadth-first search at each level. In contrast, if the maximum depth allowed is greater, it will find the searched node earlier or, in the worst case, with the same complexity as the breadth-first search, always faster or equal.

        The bidirectional search executes two searches simultaneously. One from the initial state forward and one from the target state backward. In this way, both searches will meet at one point, and a path can be established between the initial state and the goal state.

        This search is more specific because it requires that the problem have specific characteristics. The goal state must be known, and the problem must consist of finding the path to that goal state. In addition, this search algorithm, at each node visited, must not check if it is the goal state; it must check if that node has already been expanded by the other part of the search.

        This algorithm has completeness and, moreover, can be optimal, provided that certain checks are made when both forward and backward searches meet. This bidirectional search algorithm plays with the boundaries of both search processes to try to minimize the number of states visited and reduce the complexity of the algorithm.

        Unlike a sequential search, there are random search algorithms. The advantage of a randomized algorithm is that the successor generation time is reduced to a minimum. The stochasticity allows that, instead of processing which node is the best to visit, one is chosen randomly to continue the search. The drawback of this search is that it may repeat nodes and return low-quality solutions.

        Random searches are a solution to problems where there is little information about the environment or where it is very complex and expensive to evaluate which type of search works best. In addition, random search behavior is very easy to implement.

        There are different random search algorithms. The most important for this work is the random walks algorithm. A random walk is a type of Markov process in which the algorithm explores the state space in a completely random manner. The idea of these algorithms is to make a search in the state space as fast as possible and without regard to the repetition of nodes or the efficiency of the search. The main advantage, in addition to speed and derived from stochasticity, is that a random walk never gets stuck at local minima or maxima and therefore, with some probability, makes a uniform state space coverage. The main drawback is that there is no information about the quality of the solution. It cannot be known whether the solution is optimal or how far it is from the optimal solution. 

        Random paths have different applications. For example, they are very useful to simulate the Brownian motion of particles or to perform fast searches to give a first solution that can be later refined, as in link prediction or semi-supervised learning \cite{xia2019}. Although one of their main applications, and the reason why they are studied in this work, is because they are useful as a base technique for other more complex informed search algorithms.

        If during the search, the expansion of nodes is guided by some knowledge of the problem environment, it is considered a guided search. The main advantage of this type of search is that it generates nodes more intelligently, resulting in fewer nodes being expanded until the solution is reached. In addition, it can allow knowing the quality of the solution, knowing if it is optimal or the distance to the optimal solution. In this work, gradient search and heuristic search are going to be analyzed.

        Informed search is currently the most widely used for solving search, optimization and artificial intelligence problems. This type of search has proven to be useful because, by selecting nodes with certain criteria, it considerably reduces the search space explored. Depending on the information that the algorithm uses to guide the search, a distinction is made between searches with domain-independent and domain-dependent information. 

        Searches using domain-independent information do not need to know specific information about the problem, only about the search process itself. For example, heuristic searches based on landmarks \cite{richter2008}. These domain-independent searches are useful if little domain information is known, but tend to be slow and explore a larger set of state space. In contrast, domain-dependent searches, although they require fairly precise knowledge of the problem being solved, allow very precise searches, which do not explore as many nodes of the state space and, therefore, are faster.

        Gradient descent search analyzes the value differences, from a value function, between the different nodes of the search space to always advance to the point of minimum energy. In this case, value can be understood as an evaluation function or cost function constructed with specific knowledge of the problem. For example, in the case of neural networks, the gradient is derived from the differences between the predictions and the actual data calculated by a function such as the least squares error.

        Gradient descent follows minimization schemes with a convex distribution and one absolute minimum and, in some cases, several local minima. The goal of the search is to find as quickly as possible the node that represents the absolute minimum of the function without getting stuck in local minima. To do this, it performs an analysis of the slope of the function at each node it analyzes that lets it know the direction in which it has to continue the search.     
        
        Instead of using a gradient, a function that returns an estimate can be used to guide the search. This scheme is known as heuristic search.  The heuristic gives information to the algorithm on how it should guide the search based on the distance to the goal or the time of the search. There are different types of heuristic search and numerous algorithms. This search, by expanding a node, calculates the value of the heuristic for each of its successors. Then, taking into account the heuristic value of each node, it chooses the one of greatest interest following a certain criterion.

        The heuristic is a function h(x) that returns a value, where $x$ is the node to be evaluated. To construct such a function h, the constraints of a problem are relaxed and simplified. The key is to make h as informative as possible, but also fast to compute. For example, if the search problem is to find the fastest path to the center of the maze, one possible heuristic would be the Euclidean distance, which is a simplification because it ignores the walls of the maze.

        Greedy search is the simplest type of heuristic search available. In this case, the algorithm will always choose the node with the best heuristic value. In this way, the algorithm will blindly rely on the ability of the heuristic to approach the solution. This strategy has a very clear limitation, since the heuristic is a simplification of the problem, there will be times when the greedy search will fall into local minima and will not be able to get out.

        The simulated annealing algorithm uses heuristics to tell the algorithm when to perform an exploratory search and when to perform a focused search within the explored space. This algorithm, based on metallurgical processes, uses a random walk algorithm. Initially, it imposes a high temperature that makes the selection of the next node to visit completely random. However, as the number of iterations of the algorithm progresses, the temperature decreases and the selection of nodes is less determined by chance and more by a decrease in the energy of the nodes. This algorithm makes it possible to combine a broad initial exploration with an accurate search after a given number of iterations.

        The most important heuristic search algorithm is A*. This algorithm is similar to the greedy search, but instead of selecting the node with the highest heuristic value, it takes into account the cost of the previous nodes expanded so far. Thus, not h(x) but $f(x)$ is used to choose which node to expand. $f(x)$ is equal to the sum of h(x), the heuristic value of the node, and g(x), the actual non-heuristic cost of having reached that node. A* solves the problem of getting stuck in local minima and, in addition, ensures that the search is optimal \cite{candra2020}.

        In this section, two different ways of guiding an informed search have been discussed, one based on gradient and the other based on heuristics. This analysis, which classically has some relevance, quantumly becomes of vital importance. As will be seen in the following, there are hybrid classical-quantum algorithms that use the classical gradient-guided search to optimize quantum circuits. It has been seen that this search suffers from a problem called barren plateaus (BP), which makes it difficult to find the absolute minimum. A possible solution may be to use heuristic-guided and non-gradient guided searches to avoid the BP phenomenon.

        As it has been explained, the classical search, both informed and uninformed, is a broad topic with numerous variants, a small part of which has been analyzed.

        While it is true that these search algorithms have numerous applications, it is also necessary to understand that they have clear limitations. These limitations are so strong that they result in a very limited scalability of the algorithms. As a result, search algorithms make slow progress in solving increasingly complex problems and are very limited in solving optimization problems in industrial environments. In \ref{apt:limitations}, the factors that limit these algorithms, their consequences, and how they are tried to be solved will be analyzed.

        \subsection{Classical Sampling}\label{ch:classical_sampling}

        Classical sampling algorithms are techniques that allow to know a probability distribution P of a population $X$ just analyzing a subset $x$ of this population called sample. The objective of the sampling algorithm is to obtain an estimated probability distribution \^{P}, as close as possible to the real P, by analyzing the smallest sample $x$ possible.

        During the sampling process, a sampling error occurs. This error is the difference between the data estimated by the distribution obtained from the previous samples and the real value of that sample. This error becomes smaller as more samples are taken. A sampling algorithm must always maintain the balance between analyzing a few samples and maintaining a small sampling error.

        In general, sampling can be classified as probabilistic and non-probabilistic. The main difference lies in whether the samples of the population are taken completely randomly following a certain criterion, probabilistic sampling, or whether samples are chosen from a limited subset of the entire population following a specific criterion or bias, non-probabilistic sampling.

        Probability sampling is more complex and slower to perform because there is no a priori information about the distribution being sampled, and it is not simplified in any way. The advantage is that the sampling process does not introduce any bias, so that the function and parameters learned from this type of sampling are much more general. This work focus on probabilistic sampling techniques, specifically on Monte Carlo methods and derivatives.

        The Monte Carlo sampling method can be defined as a probabilistic sampling method used to approximate mathematical functions. This method is integrated within a larger family of algorithms that allows to know and perform certain estimations on physical experiments using random sampling \cite{hammersley2013}. The Monte Carlo method is based on sampling a function completely randomly until there is a sufficient number of samples to know this function in sufficient detail. Thus, the Monte Carlo method calculate an approximate estimation of the real probability distribution.

        This estimation obtained by the Monte Carlo method will have a certain error that will depend on the number of samples used during sampling. This error decreases as \( \frac{1}{\sqrt{N}} \), where $N$ is the number of samples taken from the function.

        The Monte Carlo method is very useful for approximating certain mathematical functions, but if the objective is to approximate the distribution of a particular problem about which there is some a priori knowledge, it is possible to use more accurate methods based on Monte Carlo sampling. One such method is Metropolis-Hastings.

        The Metropolis-Hastings algorithm is a method that combines the above elements of Markov and Monte Carlo processes to construct a Markov Chain Monte Carlo (MCMC). This method uses domain-specific knowledge to guide the sampling method between the proximate values. This type of sampling can also be understood as a search guided by a heuristic, generally based on energy minimization.

\begin{figure}[t]
            \centering
            \includegraphics[width=1\textwidth]{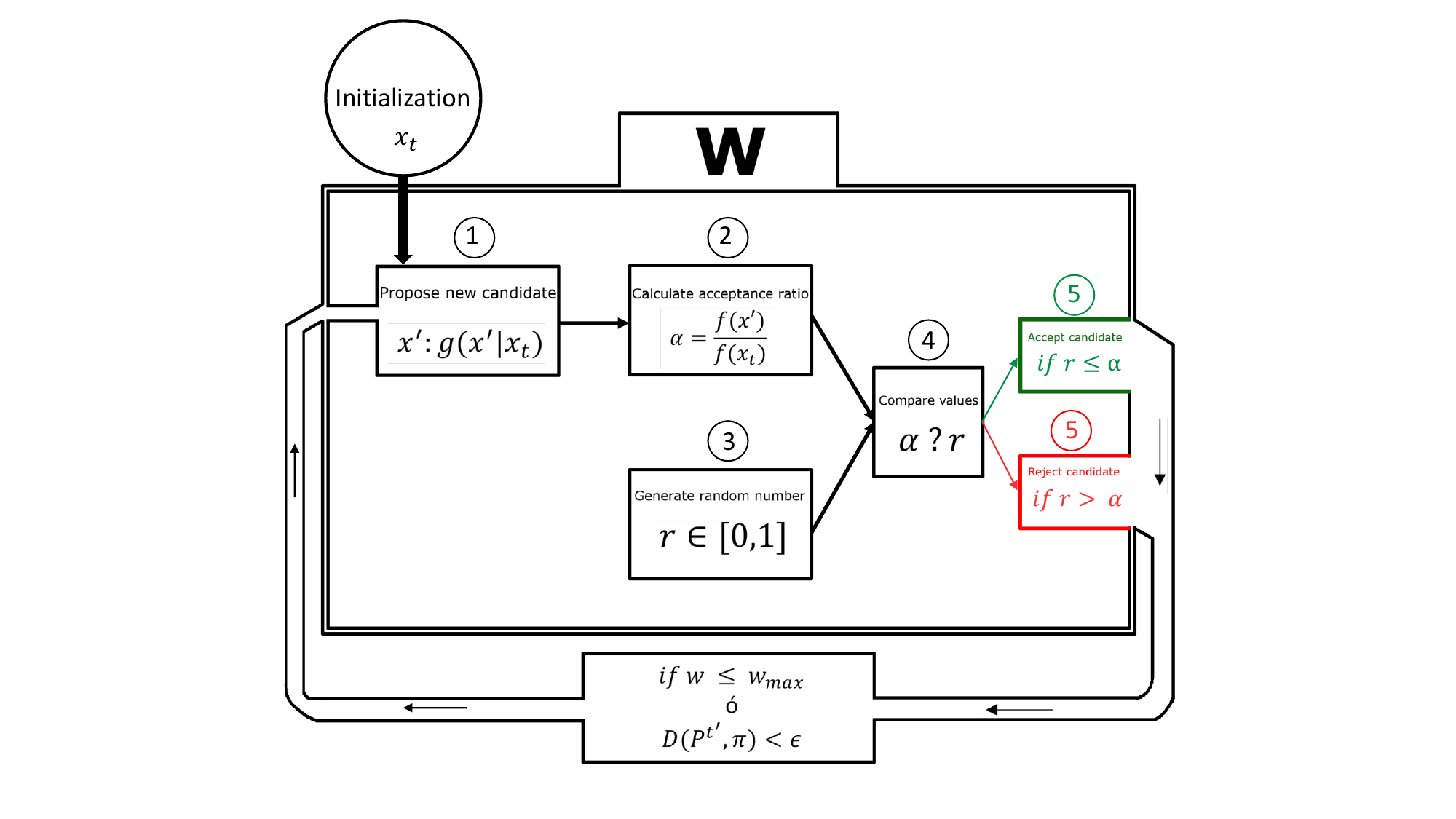}
            \caption{\label{fig:mh_scheme} This scheme shows the Metropolis-Hastings algorithm process to propose a new candidate and reject or accept it. This process is divided in 5 steps. Step 1 consists in a random candidate proposal. In step 2, an acceptance ratio is calculated using the evaluation function for the actual state and the candidate selected in step 1. In step 3, a random value between 0 and 1 is calculated. In step 4, both values, acceptance ratio and random value, are compared to decide if the new value is accepted or not. In step 5, the candidate is rejected or accepted. This iterative process is repeated until a maximum number of steps is reached or the distance between the real and the inferred distribution is below an epsilon value.}
        \end{figure}

        The Metropolis-Hastings method uses random paths to decide which node to evaluate or expand next. Once the next node has been randomly selected, it compares the value of the evaluation function of both nodes. If the problem is a minimization problem and the energy of the new node is lower, the change is accepted, and the algorithm moves to that new node. If, on the other hand, the energy of the new node is higher, the change is initially rejected. However, the energy difference between the two nodes is analyzed and, with a probability proportional to the energy difference between the two nodes, the new node is accepted. This probability of accepting a transition to a node that is not better than the previous one is essential to prevent the algorithm from stagnating in local minima. These steps are shown in figure \ref{fig:mh_scheme}. The Metropolis-Hastings algorithm is repeated until a converge criterion is found. The criterion can be that a maximum number of $W$s has already been executed or the distance $D$ between the calculated distribution and $\pi$ is lower than an epsilon value, $\epsilon$.

        The acceptance ratio of the change is defined as

        \begin{equation}
            A_{ij} = \min \left(1,e^{-\beta(C_j-C_i)}\right),
            \label{eq:Metropolis update probability}
        \end{equation}

        In this equation, $i$ is the current state and $j$ is the candidate state. $C_i$ and $C_j$ are the costs of these new states and $\beta$ is a parameter representing the inverse of the temperature. With more temperature, the algorithm performs a wider exploration, with lower temperature, the algorithm focuses on finding a local minimum.

        The Metropolis-Hastings algorithm is particularly useful in bayesian contexts where one seeks to approximate the posterior distribution. In these cases, the functional form of the posterior distribution is difficult to compute analytically. By sampling in parameter space, a representation of the posterior distribution is obtained that can be used to estimate credible intervals, means, and other statistics of interest. The likelihood of each new sample is decided with a function that must be proportional to the posterior that from where sample are taken.

        As can be seen, the Metropolis-Hastings algorithm aims to sample a function but uses methods very similar to those of the informed search. In fact, it is not unreasonable to think that a pseudo-random search can be performed using Metropolis-Hastings so that each step of the algorithm maintains an appropriate balance between exploring new paths and exploiting paths already visited that are close to the solution.

        \subsection{Classical limitations for S\&S} \label{apt:limitations}

        As seen in the previous two sections, both classical search \& sample methods have been extensively studied and developed. Moreover, they have numerous industrial applications in which they solve problems that, although small, are totally impossible to solve in any other way. However, more complex problems that can only be solved by high-performance computing (HPC) require algorithms with good scalability that are capable of solving larger and larger problems. This type of problems, usually combinatorial optimization problems, have the particularity that their complexity increases exponentially.

        If problem complexity scales exponentially but problem analysis capabilities scale linearly, a limit is soon reached \cite{ferson1996}. Increasing the size of the problem by a single unit takes the problem from being solvable by HPC in days or weeks to being unsolvable because its execution time increases to decades or centuries \cite{milojicic2021}. One of the main reasons for this poor scalability in search \& sample algorithms is the phenomenon known as the curse of dimensionality.
        
        This phenomenon occurs when the dimensionality of the data grows very fast, causing the volume of the data to grow as well. As a result, the data become scattered and difficult to cluster \cite{bellman1966}. The curse of dimensionality causes that in order to find a solution in the data, it is necessary to evaluate many more dimensions as the problem grows because the state space becomes sparse. This problem is difficult to solve classically and, for that reason, this paper proposes a quantum solution for the search \& sample algorithms.

    \newpage

    \newpage\null\thispagestyle{empty}\newpage
    
    \section{Quantum Algorithms for S\&S}\label{ch:qSS}
    \thispagestyle{empty}

        \lettrine[lines=1, findent=2pt]{\resizebox{!}{1.2\baselineskip}{Q}}{}uantum search \& sample algorithms (QS\&S) represent a paradigm shift in the approach to solving problems involving some form of quantum algorithms for search or sampling. It is quite intuitive and easy to apply it to problems where it is necessary to analyze the entire search space and, therefore, use exhaustive algorithms. However, QS\&S algorithms are also useful if to find the solution it is necessary to explore an important part of the state space and this is spatially very large.

        As discussed above, the curse of dimensionality affects classical algorithms because they must perform an exponentially larger number of operations even though the dimension of the data increases to a lesser extent. Faced with this problem, QS\&S algorithms are interesting because their complexity is not so dependent on the data dimension, as will be seen in some of the algorithms explained below. However, this is only an intuition followed by the community that has not yet been proven. In fact, there are some authors that call for caution with this type of algorithms and problems, saying that it is necessary to find better advantages for quantum computing to be really useful \cite{babbush2021}.

        In quantum algorithms, the winning bet seems to be the reduction in the number of operations needed to find a solution. This lower number of operations means that fewer quantum gates than classical ones have to be applied. If this statement proves to be correct, it will ultimately result in fewer operations, which, due to the physical foundations of quantum computing, will also translate into lower energy costs. This saving in the energy spent on executing an algorithm to solve an optimization problem, for example, is not trivial, since classical computers, specifically HPC centers, often have an energy consumption that is unbearable for computing centers \cite{patel2020}.

        Problems solved by QS\&S algorithms have been of great interest since the inception of the concept of quantum algorithms. It was considered that in search, sampling and optimization problems, QC would have a great impact. Therefore, a part of the academic effort was devoted to this topic, giving rise to numerous techniques and algorithms. In this section, main ones are going to be analyzed. First, the Grover operator will be analyzed because it is a technique commonly used by other algorithms. Based on Grover, it is possible to study quantum walks, and their implementation in the quantum Metropolis-Hastings algorithm, and quantum backtracking algorithms. Also, algorithms based on quantum annealing will be analyzed because of their wide use and easy implementation in adiabatic hardware. Finally, an analysis will be made of how these algorithms can be adapted to the hybrid classical-quantum paradigm, with special interest in the quantum Metropolis-Hastings (QM-H) algorithm, which will have special relevance in this work.

        It should be noted that these QS\&S algorithms have a wide range of applications for both theoretical and real problems. Although the problems currently being tested are restricted to an academic rather than an industrial setting due to the limitations of quantum HW, it is expected that in the short-medium term, QS\&S applications will start to be noticed at an industrial level. In this work, practical applications of QS\&S algorithms are going to be shown as use cases with proofs of concept using real data of the problem they are solving, in order to bring a little closer the demonstration of the practical utility of these algorithms in these problems.

        \subsection{Grover Operator}\label{apt:grover}

         The Grover operator did not emerge as such, but was proposed as an algorithm by Lov K. Grover \cite{grover1996}. After being studied in depth by the scientific community \cite{galindo2000}, it was revealed as a basic technique for other algorithms beyond the initial applications intended by its creator.

        The importance of Grover's algorithm/operator lies in the fact that the fundamental operation of classical computers is to store and retrieve information, i.e., to search and sort information in large volumes of data. Search operation has been isolated and studied since the first computers \cite{donald1999}. For example, one of the natural solutions to find information faster are indexes. Classical computers employ all kinds of indexes that allow searches in logarithmic complexities \cite{bohm2001}. However, it is impossible to create indexes for all data sets because it would make the index search even slower than the search itself. Thus, the only solution is to create a reduced set of indexes for certain datasets and perform most of the searches exhaustively.

        This exhaustive search, commonly called ``brute force search'', consists of making a visit to every possible state in the state space, which makes it a really expensive operation. The cost of the algorithm is defined as the number of queries or times that an element is evaluated. If there are $N$ elements, the cost will be $N$. Grover realized that it was possible to perform a similar operation but with a cost of $\sqrt{N}$ using a quantum algorithm.

        To see it with a practical and very simplified example to clarify the concept, if in a library there are 100 million books without any order, and a specific book is required, in the worst case, it is necessary to read the title of 100 million books. This task is practically unfeasible for a single person. Using Grover's algorithm, it would be necessary to make an effort equivalent to reading only 10,000 titles to find the right one, which, although tedious, is perfectly feasible for one person. It is said that the quantum algorithm makes an effort equivalent to reading 10,000 titles because there is no exact equivalence between reading a title classically and having it in quantum superposition, but the analogy of complexity can be established.

        To construct his algorithm, Grover uses an oracle $f(x)$ given $x \in \{0, 1, ..., N-1\}$. This oracle evaluates each $x$, each possible state of the system (a title of a book in the previous example). Then, the oracle can take two values, $f(x)=1$ if the $x$ value is the desired value, $f(x)=0$ in any other case. If this oracle is applied to a superposition of all possible values $x$ defined in Eq. \ref{eq:grover_init}. The superposition is created, applying a Hadamard gate to each qubit representing the state that initially is in the state $\ket{\cdot}$. Being $N = 2^n$, the oracle will only return 1 when it matches the correct value, and the task will end.

        \begin{equation} \label{eq:grover_init}
            H^{\otimes n}\ket{\cdot}^{\otimes n} = \frac{1}{\sqrt{N}} \sum_{x=0}^{N-1}\ket{x}.
        \end{equation}

        It seems that the algorithm is complete, and it will return 1 in the correct state. Nevertheless, quantum computing is not so simple. Although it is true that after executing the oracle once, the probability of the correct state will go up, the difference between the probability of the correct state and the rest of the states will be so small that when measuring the circuit, it will not differentiate which state has the highest probability. This is a common occurrence in quantum computing in many algorithms, and, therefore, it is necessary to repeat a certain set of gates several times until the probability of the correct state is sufficiently large with respect to the others.

        In the case of Grover's algorithm, this increase in the probability of the marked state is done by repeating two rotation operators. These two rotations play with the phase of the states, making an inversion about the mean probability. The first rotation marks the state in which $f(x)$ is 1 by associating with it a phase opposite to that of the other states. The second rotation, known as the diffusion operator, performs a rotation on the initial state, converting the phase difference between the states into probability amplitudes. In this way, the marked state will have a higher probability of being measured.

        The first rotation in Eq. \ref{eq:rot}, which is identified with the oracle, is responsible for marking the state sought.

        \begin{equation}\label{eq:rot}
            R\ket{x} = (-1)^{f(x)}\ket{x}.
        \end{equation}

        The result of applying the operator $R$, oracle, to the superposition of states in Eq. \ref{eq:grover_init} is a 1 in all states, except in the searched state that is -1. Then, the second rotation, the diffusion operator, has to be applied. The rotation operator is
        
        \begin{equation}\label{eq:dif}
            U_D := H^{\otimes n}(1-2\ket{0}\bra{0})H^{\otimes n}.
        \end{equation}

        This diffusion operator $U_D$ takes all possible states in superposition, applying the Hadamard gate $H^{\otimes n}$ to the reflection on the average value that was previously marked with a -1 $((1-2\ket{\cdot}\bra{0}))$. Using a simplified example \cite{yanofsky2008}, given a state $\ket{\psi}$ with 4 possible elements, each of them with a probability of $\frac{1}{4}$ and a coefficient, amplitude of probability, of $\frac{1}{2}$.

        \begin{equation}
            \ket{\psi} = [\frac{1}{2}, \frac{1}{2}, \frac{1}{2}, \frac{1}{2}].
        \end{equation}

        If the marked element is in the second position, after applying the oracle in Eq. \ref{eq:rot}, the resulting state is as follows:

        \begin{equation}
            \ket{\psi} = [\frac{1}{2}, -\frac{1}{2}, \frac{1}{2}, \frac{1}{2}].
        \end{equation}

        The mean of this state $\ket{\psi}$ is $\frac{1}{4}$. Applying equations \ref{eq:rot} and \ref{eq:dif} multiple times to $\psi$ following the Eq. \ref{eq:dif}, the resulting state is:

        \begin{equation}
            \ket{\psi} = [0, 1, 0, 0].
        \end{equation}

        The 1 is in the second position that corresponds to the searched element. It is an explanatory example, very simplified, but it helps to understand the mechanism used by Grover to its algorithm.

        These reflection and diffusion operators went far beyond Grover's algorithm and have become really important in the quantum search algorithm community. Beyond their application in non-indexed key search problems, they have been applied to all those algorithms that needed to amplify the amplitude of a particular state \cite{brassard2002}. 
        
        \subsection{Quantum walks}\label{apt:qws}

        The classical search algorithms explained in \ref{apt:classical_search} had a common problem, the need to search over almost all, if not all, nodes in the state space in order to find the solution state. One of the algorithms explained, the random walks, proposed to make a very fast but low quality exploration, so that they would explore with a certain probability a region of the search space large enough to find a solution of approximate quality to the optimal solution.

        The main problem with random walks was that, although fast in execution, it was very expensive for them to explore a significant part of the search space. Moreover, this non-repetitive exploration was not guaranteed due to the stochastic nature of the algorithm. For that reason, it was necessary for them to execute a high number of repetitions to ensure the correct exploration of the state space.

        In the quantum case, the state space can be represented as a superposition of states, applying each operation over the entire search space at once. This solves the problem of the exploration limitations of random walks by switching to their quantum counterparts, called quantum walks. In the case of quantum walks, the difficulty does not lie in exploring all possible states, but it becomes complicated to differentiate which state should be chosen because the operations are applied over the entire superposition, not to individual states.

        Any type of walk is easy to imagine, thinking of a graph G with vertices (states, S) and edges (transitions between states, T). This graph is traversed by an agent that makes a path by randomly transiting S states. From a state s to another state s', it will transit with the probability defined in the edge t. The agent can take a series of random W steps until it reaches a state that will be considered final.

        Ambainis' initial proposal for quantum walks \cite{ambainis2004}, taking advantage of the Grover operator, demonstrated a quadratic advantage over classical random walks. Another quantum walks model, introduced by Szegedy \cite{szegedy2004}, is based on graphs and represents the state space as a bipartite graph, as explained in \cite{casares2023},

        Szegedy defines the bipartite graph as two Hilbert spaces $\mathcal{H} \otimes \mathcal{H}$, where each $\mathcal{H}$ represents the state space $X$. These are two identical copies of the original space. Then, two operators to transit around the graph are defined:

        Update operator

        \begin{equation}
            U\ket{j}\ket{\cdot}:= \ket{j}\sum_{i \in X}\sqrt{W_{ji}}\ket{i} = \ket{j}\ket{p_j},
        \end{equation}

        and

        \begin{equation}
            V\ket{\cdot}\ket{i}:= \sum_{j \in X}\sqrt{W_{ij}}\ket{j}\ket{i} = \ket{p_i}\ket{i}.
        \end{equation}

        The matrix $W_{ij}$ can be seen as the weights of the edges T in the graph, which represents the probability transition from $\ket{i}$ to $\ket{j}$. $U$ and $V$ are related as:

        \begin{equation}
            SU=VS,
        \end{equation}

        where $S$ is the swap operation, to move from one $\mathcal{H}$ to the other. Then, it is possible to define the subspaces $A$ and $B$.

        \begin{equation}
            A:= span\{\ket{x}\ket{\cdot}: x \in X\},
        \end{equation}

        \begin{equation}
            B:=U^\dag VSA = U^\dag SUA.
        \end{equation}

        The projector applied to both subspaces:

        \begin{equation}
            \Pi_A := (1 \otimes \ket{0}\bra{0}),
        \end{equation}

        \begin{equation}
            \Pi_B := U^\dag V S \Pi_A SV^\dag U = U^\dag SU \Pi_A U^\dag SU.
        \end{equation}

        Using rotations similar to the Grover proposed, the quantum walk operator defined by Szegedy is defined as:

        \begin{equation}
            W=R_B R_A = U^\dag SU R_A U^\dag SU R_A.
        \end{equation}

        The quantum walk just explained is of the time-discrete quantum walk type. This allows to execute each step of the quantum walk independently and to adjust it to the reasoning model of the Markov agents. 
        
        As seen above, classical random walks were slow due to their random exploration that could repeat multiple states, something that does not happen with their quantum counterpart. Quantum walks are especially interesting in those problems where a virtually exhaustive search of the state space has to be performed. Just as Grover talked about an exhaustive search in a phone book or in the library example, if a quantum walk has to be run on a problem that needs almost all the nodes of a state space, it will find an advantage in having the state space in superposition. 

        Yet, practical implementation limitations hinder the realization of theoretical quantum walk proposals, particularly Szegedy's idea of representing the entire state space in a bipartite graph. Consequently, certain operators within $W$ must be modified to achieve a more efficient representation of the state space in the quantum walk.

        Just as classical search serves as a basis for other algorithms, quantum search based on quantum walks is used on numerous occasions, both to accelerate classical algorithms in hybrid schemes and to build larger quantum algorithms. In \cite{magniez2007} is explained how to search for marked elements in a Markov chain using the quantum walks proposed by Ambainis and Szegedy. In \cite{paparo2014}, a quantum search based on quantum walks is used to accelerate the process of exploiting a Reinforcement Learning algorithm.     
        
        As seen above, the difference between search \& sample algorithms in quantum computing is not as clear as it is classically. The same is true for quantum walks. If the previous paragraph explained quantum walks applied to search, they can also be applied to sampling processes. The best example in this line is the work of Lemieux et al. \cite{lemieux2020} who modified the Szegedy operator by replacing the bipartite graph by a coin and converted the $W$ operator of the quantum walk into an operator to construct a quantum Metropolis-Hastings algorithm, which is going to be explained in detail in the next section.

        \subsection{Quantum Metropolis-Hastings} \label{apt:qmh}

        In the work of Lemieux et al., they take the Szegedy quantum walk operator W and replace the bipartite graph representation with a coin that is in a superposition of 0 and 1. This coin is entagled with the state space, so that changing the probability of the coin also changes the probability of the states. This new quantum proposal of the M-H algorithm has three registers:

        \begin{itemize}
            \item $\ket{\cdot}_S$ is the register containing all the qubits needed to represent the entire state space.

            \item $\ket{\cdot}_M$ is the register containing the move qubits.

            \item $\ket{\cdot}_C$ is the register representing the coin, just one qubit.
        \end{itemize}
        
        Using these registers, the proposed operator is as follows:

        \begin{equation} \label{eq:lemieux}
            U_W = RV^\dag B^\dag FBV.
        \end{equation}

    	The operators are:

        \begin{itemize}

            \item Move preparation operator $V$: It creates the superposition in the register $\ket{\cdot}_M$.

            \item Coin operator $B$: Rotates the coin according to the criteria defined in Eq. \ref{eq:Metropolis update probability}, similar to the classical case. It loads the coin with a certain probability and prepares it to the operator $F$.

            \item Coin flip operator $F$: It flips the states $\ket{\cdot}_S$ according to the probabilities in the coin, $\ket{\cdot}_C$.
          
            \item Reflection operator $R$: It is a reflection in the registers $\ket{\cdot}_S\ket{\cdot}_C$, derived from the Grover reflection operator. It is responsible for increasing the probability of states marked by the coin register $\ket{\cdot}_C$.  
            
        \end{itemize}
    
        This operator $U_W$, in combination with the Markov models explained in chapter \ref{ch:introQSS}, is used to construct the quantum version of a Metropolis-Hastings algorithm. Similar to the graph representation, the agent that traverses the graph G can be modeled as a Markov model in which the decision to move to one or another of the possible states depends only on the state in which the agent is. The probability transition of the agent from one state to the next is codified in the rotation of the coin.

        It is important to note that the operator proposed by Szegedy $W$ and the one proposed by Lemieux et al. $U_W$ are not equivalent. The new operator $U_W$ is a very smart implementation because it avoids the need to uncompute the move register $\ket{\cdot}_M$ and the coin register $\ket{\cdot}_C$, which is a very costly problem when the new move proposal is rejected.

        This initial proposal of a quantum Metropolis Hastings was focused on a specific problem, a one dimensional Ising model. This problem can be used as a simplification of other more general optimization problems. However, as demonstrated later, and as will be shown in the use cases, this version of QM-H can be adapted in different ways to solve more complex optimization, search \& sample problems, such as protein folding \cite{casares2022}, the N-Queen problem \cite{campos2023}, parameter estimation in gravitational waves \cite{escrig2023}, etc.

        The main advantage of this QM-H algorithm is the possibility of using it in combination with a heuristic function. This makes it possible to apply the algorithm to almost any optimization problem. As seen above, in cases of informed search, if one has some information about the problem domain it is possible to guide the search to speed it up. In the case of quantum algorithms, if the heuristic can be evaluated over all possible states, it allows the algorithm to have much more information about the problem with fewer operations.
        
        The quantum M-H algorithm can be used for search problems because the agent always looks for nodes that minimize energy, except for some random movement with increasing energy to avoid local minima. In this way, the problem can be encoded in such a way that the algorithm searches for the absolute minimum of an energy function that can represent a minimum number of kilometers traveled, minimum budget spent or minimum amount of fuel consumed.

        The key to the search performance of the quantum Metropolis-Hastings is its ability to sample the state space with high accuracy faster than the classical algorithm. This sampling is based on knowing perfectly the probability of acceptance of a change in the studied distribution, something that inherently adjusts the quantum algorithm to the problem it is solving.

        \subsection{Other quantum S\&S proposals}\label{apt:otherqss}

        In addition to algorithms based on quantum walks, there are other quantum computing proposals for S\&S algorithms. Understanding the limitations of some of these proposals, strengthens the capabilities of quantum walk-based algorithms.

        One of this proposal is quantum backtracking. Taking the example of an agent moving through a maze, or the problem of loading packages into a truck, a different type of search called backtracking is useful. The key point in backtracking algorithms is that they perform an exhaustive search, but in an intelligent way, avoiding exploring options that are already known not to lead to a valid solution. This helps improve efficiency compared to brute force approaches.

\begin{figure}[t]
            \centering
            \includegraphics[width=0.6\textwidth]{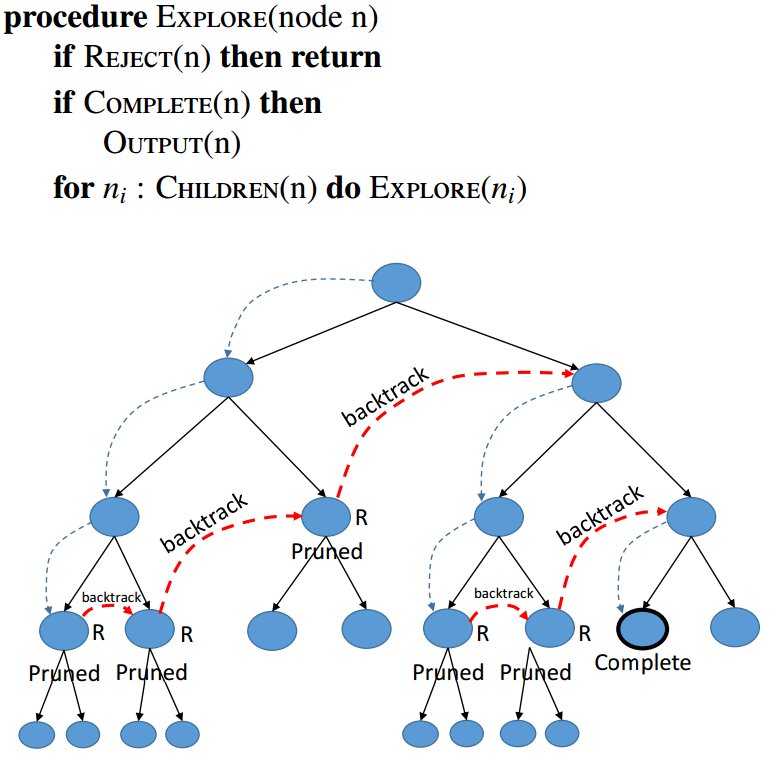}
            \caption{\label{fig:backtracking} Scheme that shows how a backtracking algorithm works. This algorithm visits the nodes following a depth-first-search. If it detects that the solution is not in the explored branch, it prunes the branch and backtracks to a different branch. Extracted from \cite{verenich2017}.}
        \end{figure}

        In this case, Montanaro \cite{montanaro2015} proposed to apply the techniques of quantum walks to speed up backtracking algorithms. The main function of a backtracking algorithm is that it represents the state space as if it were a tree. This algorithm expands and explores branches with different states until a solution state is found. The difference with other algorithms is the ability to evaluate the nodes in such a way that it can know if all the successors, all the branches, generated by that node are going to approach the solution or not. In case it has found an unpromising branch to find the solution, it returns, backtracking, to the root of that branch and takes a different one to continue the search as shown in figure \ref{fig:backtracking}.

        The major disadvantage of quantum backtracking algorithms is intrinsic to their advantage. These algorithms inherit the state space representation limitations explained above for Szegedy quantum walks. They even add additional operators that, while making the advantage greater, also add a high complexity to bring them to a real implementation, either in simulator or hardware, that can solve a practical problem. The limitation of quantum backtracking remains the same as that of many quantum algorithms; it is still difficult to prove its advantage because the very small quantum hardware resources make its implementation impossible until several years from now.

        A different paradigm from quantum walks for implementing QS\&S algorithms is quantum annealing. This proposal, which is almost at the same time an algorithm and device, since the hardware where it is executed is only dedicated to execute this type of algorithms, is inspired by the classical technique of simulated annealing. These algorithms start with an easily achievable quantum configuration and then, by gradually cooling the system, seek to ``anneal'' the optimal solution. The main idea is to configure the system in a quantum state according to the initial conditions of a problem, and then let the system evolve until it reaches a minimum energy configuration that corresponds to the solution of the problem \cite{morita2008}. 

        To encode problems for quantum annealing, various methods can be employed. One of the most common approaches is to formulate them as unconstrained quadratic binary optimization (QUBO) problems. Such representations enable a binary encoding of the variables and constraints of the problem. In this way, the algorithm effectively solves a system of equations where it needs to maximize or minimize certain values while adhering to specified constraints \cite{date2021}.

        QUBO binary encoding can be quite rigid when it comes to representing complex structures that may be necessary in an optimization problem. Therefore, it has been relegated to a method of solving very specific problems. Nevertheless, it remains an alternative to consider for addressing certain QS\&S problems \cite{yarkoni2022}. Scalability is another limitation, closely related to the lack of a formal demonstration of its advantage over classical algorithms in large-scale problems. Although quantum annealing tackles problems of great complexity for classical algorithms, providing an advantage, the scalability of adiabatic quantum hardware devices has proven to be insufficient \cite{hauke2020}. This lack of scalability eliminates the advantage for larger problems, leading to the acknowledgment of quantum annealing's importance within QS\&S algorithms but ruling it out as a research technique for algorithm construction and problem resolution in this work.

        A different perspective for QS\&S started when researchers working on quantum optimization algorithms were aware that executing a purely quantum algorithm to solve an optimization problem faster than a classical one was tremendously complex due to the slow evolution of hardware devices. Therefore, seeking a middle point between complex purely quantum algorithms with significant advantages but with questionable advantages, a new family of algorithms was conceived. These algorithms combining classical and quantum technology are known as quantum-inspired algorithms.

        Within the quantum-inspired category, there are many types of algorithms: quantum algorithms executed on classical hardware, classical algorithms accelerated by small quantum hardware modules, algorithms that use quantum methods but are implemented entirely classically, or algorithms that combine quantum and classical modules, both hardware and software. This work specifically focuses on hybrid algorithms.

        Hybrid algorithms, the central focus of this work, are currently generating high expectations to become the paradigm through which quantum computing enters the industrial sector. Under the umbrella of this concept, there are various algorithmic schemes, but the common thread is their modular nature, which combines both classical and quantum modules.

        Hybrid algorithms have different approaches, but the main ones are twofold. On one hand, hybrid algorithms can be considered an end in themselves. They are constructed following this paradigm because there will be operations where quantum computing will never be faster than classical, and thus, it is always better to combine both. In this approach, quantum processing units (QPUs) are developed as processing modules that are added to classical supercomputers, similar to RAM or GPU processors. On the other hand, hybrid algorithms can be viewed as an intermediate step that must be taken because quantum devices are not yet mature enough to support the entire quantum algorithm. As technology develops, the classical modules of these hybrid algorithms may be phased out.

        This work aligns more with the philosophy of the second approach, which is the development of hybrid algorithms because quantum devices currently lack the capacity to analyze the size of the problem intended for resolution. However, the algorithms are designed with sufficient scalability so that once quantum hardware becomes available, it will be possible to execute the entire algorithm in a quantum manner.

        Taking this intermediate step through classical-quantum hybrid computing is a way to try to gain quantum advantages in optimization problems much earlier. NISQ computers, specifically designed for such algorithms, are expected to be available sooner than other paradigms of quantum computers like early fault-tolerant computers. This implies the possibility of gaining an advantage in certain problems in the medium term. However, large-scale quantum computers capable of executing purely quantum algorithms are not expected until the long term, without specifying a date.
    
        In this first chapter on quantum search \& sample, a hybrid algorithm will be proposed that combines quantum walks technology to create a quantum Metropolis-Hastings algorithm capable of solving optimization problems strongly connected to industrial needs. In the following sections, various use cases where the explained hybrid algorithm has been applied will be discussed.

    \newpage
    \newpage\null\thispagestyle{empty}\newpage

    \section{Quantum Metropolis Solver (QMS)}\label{ch:qms}
    \thispagestyle{empty}

        \lettrine[lines=1, findent=2pt]{\resizebox{!}{1.2\baselineskip}{F}}{}ollowing the philosophy of hybrid algorithms as an intermediate step for future fully quantum algorithms and looking for an algorithm that can be executed in the short term in NISQ, the quantum Metropolis Solver (QMS) algorithm is proposed. This algorithm follows the line of the Metropolis-Hastings quantum algorithm explained in chapter \ref{ch:qSS} and modifies it to create a software tool that can be executed following the universal computing paradigm.

        As explained in the previous sections, optimization problems have always been considered to be of great appeal for quantum algorithms. These problems are complex for classical algorithms and, moreover, require almost exhaustive searches. QMS seeks to solve optimization problems by making use of the advantages of universal quantum computation of circuits. Therefore, the designed tool is fully executable by code in both quantum simulators running on classical hardware and purely quantum hardware following the circuit paradigm.

        \subsection{Motivation}\label{apt:motivation_qms}

        QMS design is strongly influenced by the types of QS\&S algorithms explained in chapter \ref{ch:qSS}. It is an algorithm that combines search \& sample techniques to arrive at a solution, and is fully applicable to both search \& sample problems. To do this, QMS performs two operations iteratively, first sampling the circuit by performing a step, similar to W in the Equation \ref{eq:lemieux}, and then repeating these steps, W, a specified number of times until the probability of measuring the correct state is sufficiently increased.

        Since the core of the QMS tool is the QM-H algorithm, the quantum advantage of this tool derives from this technique. Therefore, the main motivation to build QMS has been to find an algorithm based on QM-H that would allow a real implementation, keeping the same quantum advantage and that could solve a wide range of QS\&S problems, ultimately optimization problems.

        Considering that one of the main motivations of this work and this thesis was to create realistic and implementable QC algorithms, QMS has been implemented and tested in the Python programming language and using the Qiskit quantum computing library. In addition, the QMS code and its modifications to the use cases have been published in public repositories on GitHub. Links can be found in each of the releases.

        By open-sourcing the code, another motivation behind this work has been pursued, to enable any user to have an easy-to-use software tool that solves optimization problems through quantum computing algorithms, providing a straightforward and interpretable result. To run the code, it is only necessary to provide a simple description of the problem, and QMS will return the minimum/maximum energy configuration for their problem.

        Finally, QMS has been designed with a completely modular philosophy, allowing any software component to be modified or adapted to a specific use case. Moreover, this promotes the hybrid model, as any component can be switched from quantum to classical and vice versa. In the following sections, three use cases for QMS are presented: quantum artificial intelligence, space exploration, and quantum chemistry.

        The idea behind these three diverse use cases is to demonstrate that QMS can be applied to each of them and to an even broader range of optimization problems. The only common characteristic among them is the need for their fundamental operations to be quantum search \& sample algorithms. Additionally, there is an intention to open a new line of research, following the hypothesis that delving deeper into these optimization algorithms is necessary through universal computing and not dedicated hardware such as quantum annealing.
    
        \subsection{Methodology} \label{apt:metho_qms}

        QMS operates by providing a description of the problem to be solved and specifying the precision in terms of the number of qubits with which the states should be represented. The problem description consists of defining the state space of the problem and the evaluation function to be applied to each state, guiding the search towards the state of maximum or minimum energy.

        Ideally, this problem description would be an input file that defines two functions: one to generate states from an arbitrary state (inter-state transition function) and one to calculate energies for any state (evaluation function). However, considering the current state of quantum computing, QMS is currently designed so that both functions are implemented directly in the code. Depending on the use case, modifications are required to be made to the QMS code to adapt it to the particular problem. This approach requires a thorough understanding of the problem to be solved. It is also important to realize that defining the problem using an input file with two functions does not require additional research; it would simply imply that the transition and evaluation functions would be executed in the quantum circuit, which implies a high processing and storage capacity for a quantum computer that is not currently available.

        \begin{itemize}
            \item $\ket{\cdot}_S$: State register, representing the entire state space. This register uses a binary encoding of a set of $v$ variables. Each variable has a precision of $q$ qubits and therefore can represent $2^q$ states. In total, the state space size represented by this state register is $2^{v\times q}$
            
            \item $\ket{\cdot}_M$: Movement register, it is composed by two independent registers, move id register $\ket{\cdot}_{move-id}$ that contains information about which variable register is going to be modified and move value register $\ket{\cdot}_{move-value}$ that contains if the modification of the variable is going to be plus or minus 1.
            
            \item $\ket{\cdot}_C$: Coin register, similar to that proposed by \cite{lemieux2020}. One qubit to represent the coin.

            \item $\ket{\cdot}_A$: Probability acceptance register, it is a register in which the value energy associated to each state is loaded. It is used to load the move acceptance probability on the coin.
            
        \end{itemize}

        These registers are used in a modified QM-H operator $U_{QMS}$ to fit the new registers.

        \begin{equation} \label{eq:W}
            U_{QMS} = RV^\dag B^\dag FBV.
        \end{equation}

        The operators that form $U_{QMS}$ are similar to those explained in chapter \ref{ch:qSS}, with differences in the implementations explained in publication \ref{pub:p1}.

        An extra operator $E$, could be added. $E$ is the energy operator that loads or calculate the energies associated to each state represented by $\ket{\cdot}_S$. This operator can be used in two ways. The theoretical approach is for QMS to be a fully quantum algorithm and calculate the energies associated with each state in superposition without the need to measure the states. This first case is the ideal, but as explained earlier, quantum hardware is far from making this a reality. The second way to use this operator follows the philosophy of hybrid algorithms. For this, a module is designed in classical computing capable of calculating the energies of quantum states. These energies are passed to the operator $E$, which reads and loads them into the probability acceptance register $\ket{\cdot}_A$ using a technique called QRAM \cite{miszczak2010, di2020}.

        Another important aspect of QMS is its ability to be configured with different hyperparameters to adapt it to the specific problem being executed. For example, the number of times the operator $U_{QMS}$ is applied, the precision to represent the probability acceptance register, or the beta parameter ($\beta$). Specifically, the $\beta$ parameter is one of the main configuration hyperparameters of QMS, as it regulates the balance between exploration and exploitation in the search. A low value of $\beta$ favors exploration, while a high value favors exploitation. In practice, properly adjusting these hyperparameters is crucial for achieving good results in solving optimization problems.

        $\beta$ is defined as $\frac{1}{T}$ being $T$ the effective (fictitious) temperature of the system. This concept is similar to the simulated annealing explained in subsection \ref{apt:otherqss}. The canonical use of $T$ is high temperature at the beginning to focus on exploration and, progressively, reduce the temperature to find the absolute minima/maxima.

        QMS allows for modification of the temperature $T$ by adjusting $\beta$ in each iteration. This enables the use of various annealing schedules, ranging from a constant schedule with a fixed $\beta$ to an exponential reduction of $\beta$ in each iteration, and including more gradual reductions with linear or geometric schemes.

        Another key aspect of the quantum software tool, QMS, is the comparison with its classical counterpart, the M-H algorithm. For this purpose, a small module has been implemented within QMS that enables the execution of the classical algorithm on the same problem and under the same conditions. This allows benchmarking the results of the quantum algorithm and assessing whether it outperforms the classical one.

        However, this comparison between classical and quantum is not straightforward. The main issue is deciding which metric to compare. Comparing execution times does not make sense because the quantum algorithm runs on a simulator on classical hardware. Counting the number of gates does not make sense either because quantum and classical gates are very different. Therefore, the metric chosen for comparison is Time To Solution (TTS).

        TTS is a metric explained in \cite{lemieux2020}. It is a figure of merit that assigns a numerical value to the cost of executing a specific algorithm. In the case of QMS, it returns a numerical value that allows the comparison of both executions. TTS stands for time; therefore, the higher the TTS, the longer the time to solution and the worse the performance. Thus, in the comparison, the aim is to determine which algorithm has a lower TTS and, therefore, better performance.

\begin{figure}[t]
            \centering
            \includegraphics[width=0.7\textwidth]{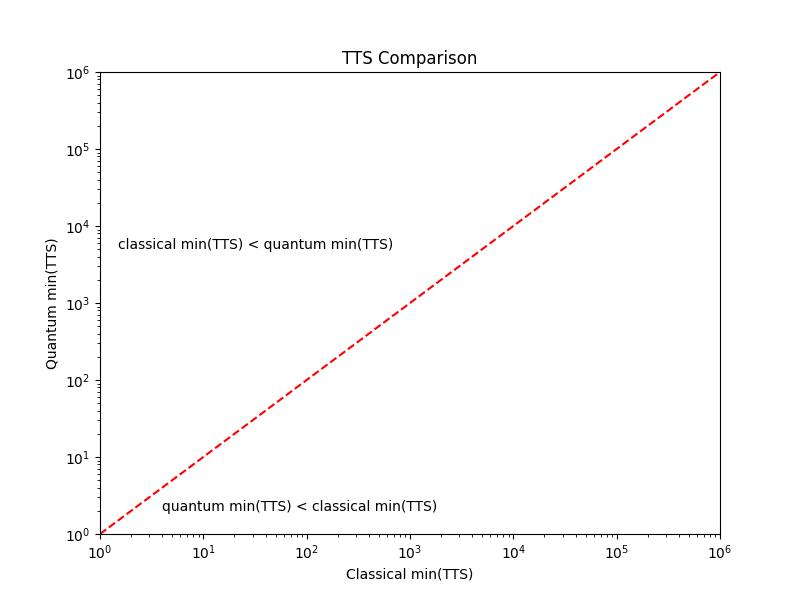}
            \caption{\label{fig:tts_comparison} TTS comparison plot is divided by a red dashed line. This line splits the plot in two triangles: the upper triangle contains points with low classical TTS and high quantum TTS, which means classical advantage region. In contrast, lower triangle contains points with low quantum TTS and high classical TTS, which means quantum advantage region. Therefore, the desired behavior is that all points fall into the lower triangle.}
        \end{figure}

        TTS is calculated following the formula:

        \begin{equation}
            TTS(t):= t \frac{\log (1-\delta)}{\log (1-p(t))},
            \label{eq:TTS}
        \end{equation}

        where $t$ is the number of time steps, equivalent to $W$. $\delta$ is the success probability, and $p(t)$ is the probability of hitting the ground state after $t$ steps. Then the TTS is represented in a plot, shown in figure \ref{fig:tts_comparison}, with quantum TTS on the y-axis and classical TTS on the x-axis. This plot is divided in the diagonal by a dashed line, indicating the frontier between classical and quantum TTS. This dashed line creates two triangles, the upper triangle contains the points with low classical TTS and high quantum TTS, which means classical advantage region. In the opposite, the lower triangle contains values with low quantum TTS and high classical TTS, quantum advantage region. Ideally, all points should be in the quantum advantage region. Nevertheless, the quantum hardware is not ready yet for this situation. For that reason, some points will be in the classical region and others will be in the quantum advantage region. This makes necessary to represent the axis in logarithmic scale and to make a scaling exponent analysis. Using a square mean regression, it is possible to observe the tendency of the points. If the exponent is lower to 1 means that, with more TTS, bigger sizes of problems, all points will tend to be under the dashed line, so in the quantum advantage region. It is important to understand this concept in order to interpret correctly the result of the use cases.

        With the development and testing done on QMS, a behavior has been observed that has been compared with other works. With small optimization problems, it is not possible to gain quantum advantage because the execution of quantum circuits requires a certain computational overhead. The quantum advantage is expected to be observed with larger problem sizes.
        
\begin{figure}[H]
            \centering
            \includegraphics[width=0.6\textwidth]{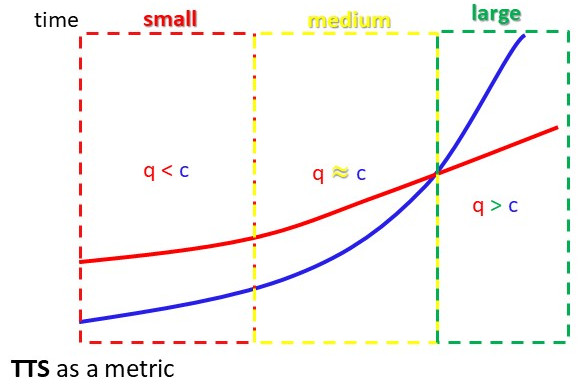}
            \caption{\label{fig:scaling} This figure represents three different sizes of problems solved by QMS: small, medium and large. The boundaries of the three regions are delimited by an inflection point in the ratio between \textbf{\red{quantum}} and \textbf{\blue{classical}} algorithm performance. Between the small and medium size region, the inflection point is that the gap between quantum and classical starts to narrow because the computational overhead is no longer so important. Between the medium and large size region, the tipping point is that both algorithms have similar performance and the quantum starts to be better off.}
        \end{figure}

        In Figure \ref{fig:scaling}, a plot with three different regions can be seen. The x-axis represents the size of the problems, and the y-axis represents the time needed to solve them. The three regions correspond to small, medium, and large problems. This plot is a simplification of the results that have been observed and compared with those of other authors. With small problems, classical algorithms have better performance, lower execution time due to the computational overhead of quantum circuits. This gap remains constant for a certain problem size. When that gap starts to shrink, it is considered a transition to medium-sized problems. In this new region of problems, the constant computational overhead begins to have less impact on performance, and the gap starts to close. In the last region, when equality is achieved between the performance of quantum and classical algorithms, it is clear that the trend for classical algorithms in complex optimization problems is to present a cliff in performance. The time required by classical algorithms increases exponentially. However, the expected performance of quantum algorithms grows more steadily, entering regions of practical quantum advantage. Of course, this figure is only an idealization of reality, but has many similarities with what has been observed in the use cases that will be explained later.

        Regarding the scalability in the third region of Figure \ref{fig:scaling}, which pertains to large problems, two scenarios can be analyzed. First, if the number of states is exceedingly high, and secondly, if the differences in the evaluation function value between states are very low. In such cases, it becomes necessary to apply a large number of $W$ operations.

        In the first scenario, where a large number of states are involved in the problem, thereby demanding a significant number of qubits to represent them, the scalability of QMS is primarily determined by the number of qubits required for the state representation. Other registers, such as those for the coin, move value, and ancillas, either remain constant or increase logarithmically with the number of qubits needed for state representation. Therefore, the focus is primarily on examining how the number of qubits for the state register grows. In QMS, this growth follows a polynomial pattern concerning the number of variables required to represent the states, multiplied by the number of qubits necessary to represent all possible values of these variables.

        If there are 100 variables and each variable is discretized using 20 qubits, QMS will require $100 \times 20 = 2000$ qubits to represent all possible states, resulting in $2^{2000}$ possible states. If a new variable is added, the system will require an additional $1 \times 20 = 20$ qubits to represent the states. Thus, the representation scales polynomially, as the number of variables multiplied by the number of discretization qubits. Although it is difficult to establish a comparison with the classical counterpart of QMS in representation scalability terms, the memory to represent $2^{2000}$ possible states will scale exponentially due to the necessity of representing them sequentially.

        In the second scenario, where the differences in evaluation values between states are minimal, necessitating numerous repetitions of the $W$ operator, it is crucial to examine how the number of $W$ repetitions grows to find the solution. Due to QMS is a heuristic method which its performs depends on the problem, it is not possible to give an exact number, just estimations based on simulations. After the simulations of different use cases that will be explained in the next chapters, it is known that QMS requires to execute operator $W$ up to an exponent of 4 number of the number of qubits, $n_q$. Therefore, while the number of possible states grows as $2^{n_q}$, the upper limit of the number of $W$ repetitions grows as $n_q^4$. The number of states grows much faster than the number $W$ operators. It is important for the comparison of the number evaluations that a classical algorithm needs to search in the whole state space.

        Comparing the number of evaluations that the classical and quantum requires to find a solution, the classical M-H algorithm requieres to explore between 60\% and 80\% of the total number of states \cite{mengersen1996}. Of course, it is a rough number that depends on multiple factors. However, it is easy to see that the classical m-h requires exploring a very large region of space to converge and this scales with the number of states. QMS, on the other hand, requires a number of repetitions of W that scales much more slowly and, by definition, at each repetition explores the entire state space in superposition.

        The scalability conclusions shown here can be applied to all qms use cases because they are all based on the same algorithm derived from quantum walks.

        To obtain the results of the use cases, QMS has been designed to run on both simulators and quantum hardware. All results from the use cases have been obtained using Qiskit's simulators. Additionally, in the use case of quantum chemistry, protein folding, a reduced version of QMS was executed for this specific use case on IBM's quantum hardware. This demonstrated that QMS is entirely valid for running in realistic environments.

        With the ability to run on quantum hardware, coupled with QMS's good scalability in problem representation and algorithm execution, this tool could be considered as part of an industrial pipeline with quantum advantage. To reach that point, the only current limitation is the development of quantum hardware, which, as seen in previous sections, is undergoing rapid progress.

        \subsection{Results}\label{apt:results_qms}

        \begin{itemize}[label=\textcolor{green}{\checkmark}]
            \item A quantum tool has been designed capable of solving real optimization problems and using real data. It is called QMS (Quantum Metropolis Solver)
            \item Implementation of software of a modified version of QM-H with the necessary adaptation of registers and operators.
            \item Implementation of hybrid architecture that can be tested in a early fault-tolerant quantum computer.
            \item Polynomial quantum advantage demonstrated in an important optimization problem.
            \item General software tool that can solve a wide range of optimization problems, just by receiving a description of the problem as input.
            \item Software tool with better scalability than a classical tool for optimization problems.
            \item Modular implementation that allows to modify any component of QMS to adapt it to a different problem. In addition, it was developed under an open source philosophy and published on GitHub.
        \end{itemize}

        \addcontentsline{toc}{subsection}{\numberline{}Publication P1. \textit{Quantum Metropolis Solver: A Quantum Walks Approach to Optimization Problems}}
        \includepdf[pages=-]{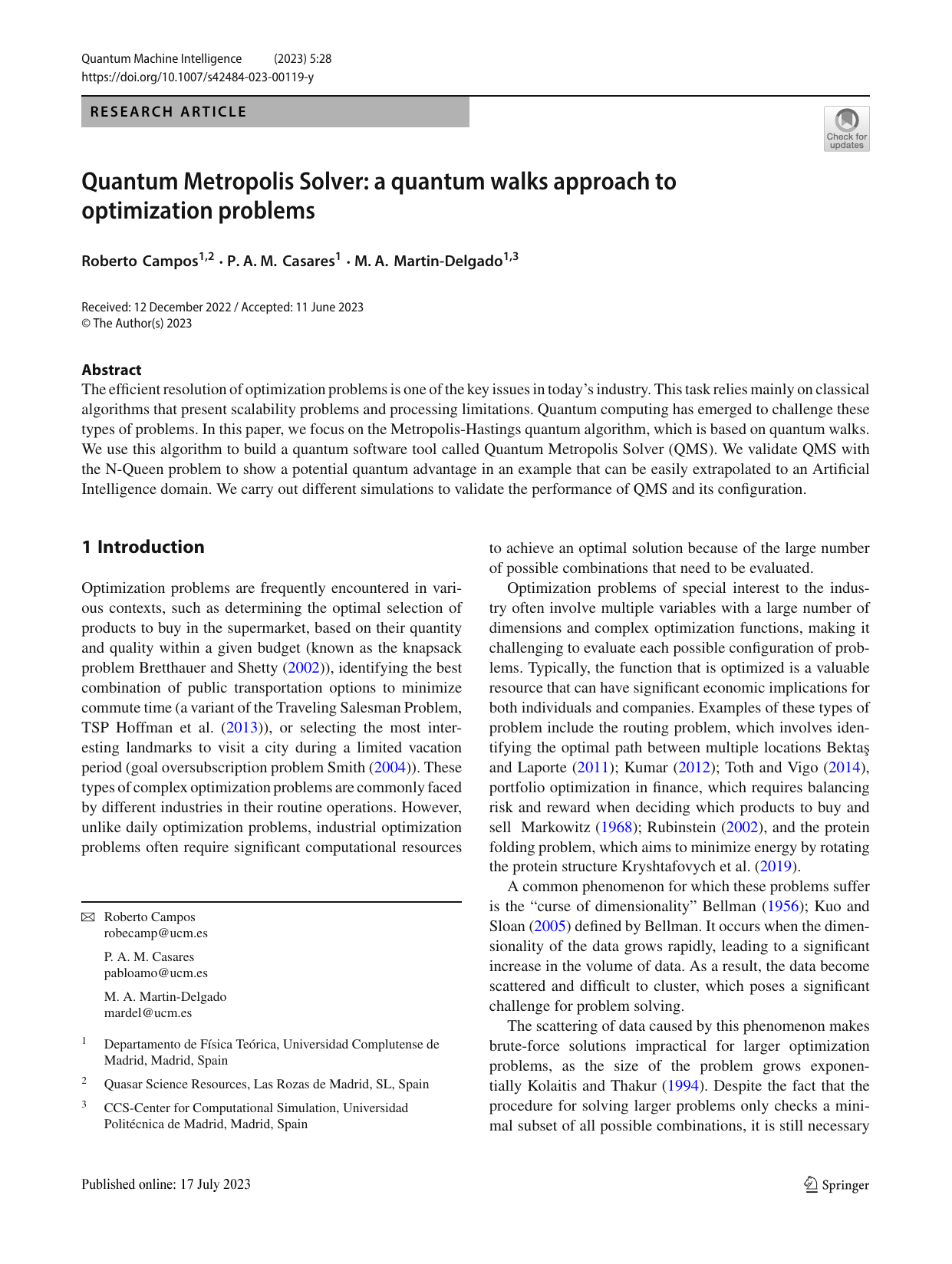}\label{pub:p1}

    \newpage\null\thispagestyle{empty}\newpage

    \newpage
    \section{S\&S applied to quantum artificial intelligence (QAI)}\label{ch:qaiSS}
    \thispagestyle{empty}

        \lettrine[lines=1, findent=2pt]{\resizebox{!}{1.2\baselineskip}{Q}}{}uantum computing promises great advantages over classical algorithms in difficult problems. One of the fields that has classically been characterized by having the most complex problems has been artificial intelligence (AI). In this category included all the problems in which algorithms had to process so much data that they had to be able to make decisions autonomously. Therefore, it is natural that quantum computing has also looked to this field to hybridize with it and create the concept of quantum artificial intelligence (QAI).

\begin{figure}[H]
            \centering
            \includegraphics[width=1\textwidth]{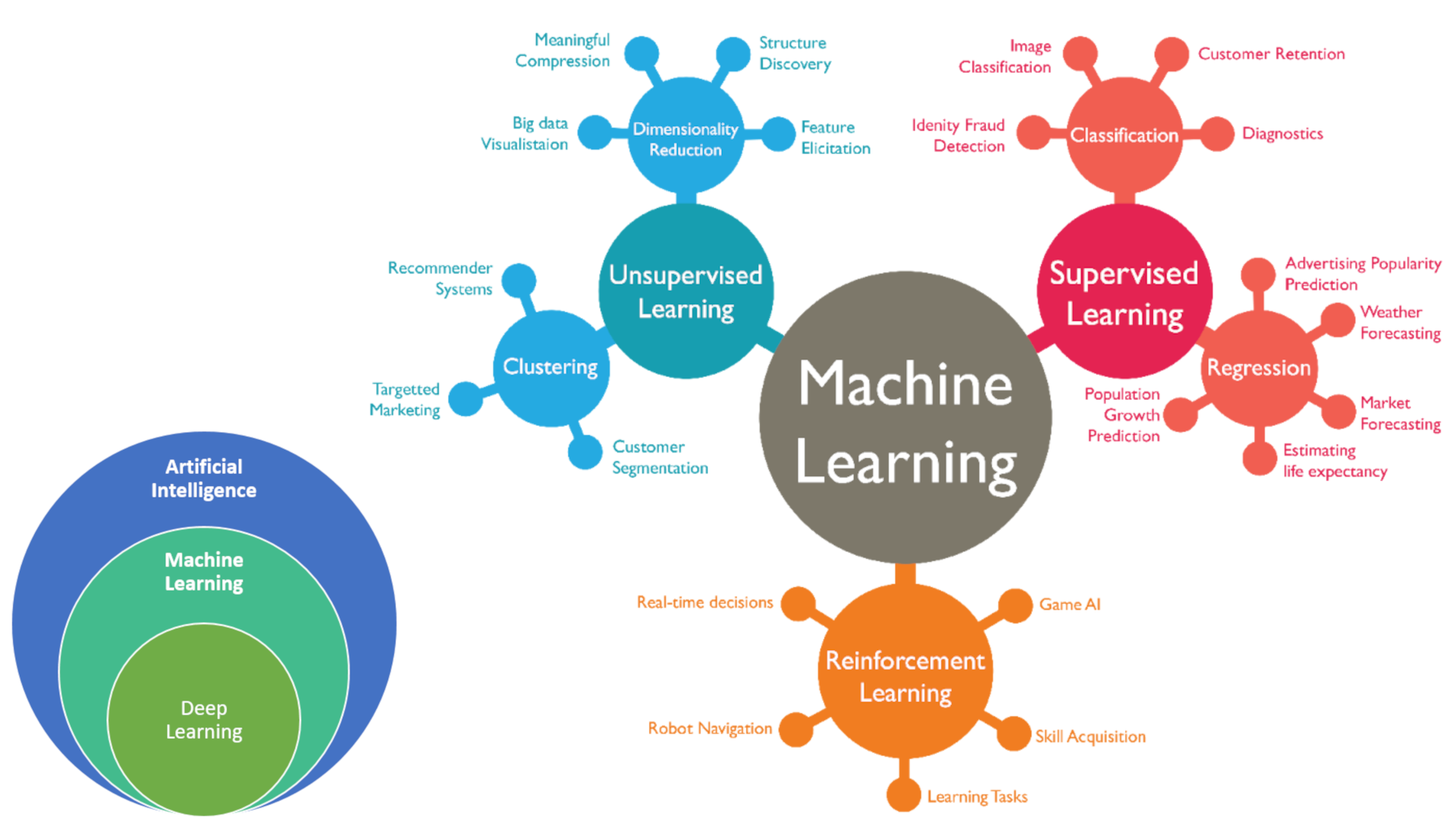}
            \caption{\label{fig:ia_ml} This figure shows the relationship between artificial intelligence, machine learning and deep learning in the left. In the right, there is a scheme of the different algorithms inside the machine learning field (extracted from \cite{ia_ml}).}
        \end{figure}

        The concept of artificial intelligence, whether classical or quantum, encompasses many techniques. In this work, an AI algorithm is defined as one capable of running autonomously under changing conditions, doing some kind of reasoning or searching. Within the family of AI algorithms, three main groups can be distinguished: search algorithms, machine learning algorithms (ML) and multi-agent algorithms. In previous sections, search algorithms have been discussed in more detail, although they were not specifically dedicated to AI problems. In this section, ML algorithms will be discussed in more detail. An AI and ML general scheme is shown in figure \ref{fig:ia_ml}.
    
        ML algorithms take a set of input data and by reasoning learn from it to generate a knowledge model of the problem to find the required solutions. Within ML algorithms there are three other types of algorithms: supervised learning, unsupervised learning and reinforcement learning.
    
        The main motivation for developing the field of QAI in general and QML in particular is to improve classical algorithms. This improvement can occur in different areas, but especially focuses on three, reduction of training or execution times, improvement of the accuracy of the algorithms, and reduction of energy consumption.
    
        Currently, AI is in a moment of high growth and implementation in industrial technologies. However, its development limitations have already begun to become apparent. At the software level, they are beginning to see limitations in data processing that cannot be overcome \cite{chowdhury2012}. Since these limitations cannot be overcome with new algorithms, the solution focuses on creating specific hardware that handles more data at higher speeds, for example, with dedicated AI-only chips \cite{sze2017}. Although this approach also has limitations, the size of the chips cannot be reduced below a certain threshold without quantum effects appearing. In fact, this limit has already been reached and no more transistors can be introduced on a single chip \cite{waldrop2016}. The solution is no longer to introduce more transistors on a chip to double the capacity every two years, as proposed by Moore's law \cite{moore1965}. Now, the solution consists of stacking chips and making computers with more interconnected chips, giving the whole a higher capacity \cite{moore1995}. However, the concept of creating chips linked together in large supercomputers following the High Performance Computing (HPC) philosophy also has limitations and disadvantages. HPC computing centers require increasingly complex architectures for interconnecting the processing chips and, above all, they have very high energy consumption \cite{gupta2021}.
    
        The first step in developing QAI algorithms, especially QML algorithms, has been to look for a source of inspiration. Just as happened with the first scientists who wanted to develop classical AI algorithms but did not know how to do it and started to mimic human brain processes, the development of QML has taken as a starting point classical algorithms that performed well and tried to make quantum versions \cite{schuld2015}.
    
        This strategy is valid because the objective is clear, to have more efficient QML algorithms than the classical ones, but it is unknown how, so it was necessary to start the research at some point. However, unlike what happened in classical ML, which mimicking the human brain and learning proved to be very useful, such as neural networks, evolutionary algorithms or Reinforcement Learning, in the field of QML it seems to have led the research to a dead end.
    
        Mimicking classical algorithms to create their quantum versions has proven to have limitations either because there are processes that cannot be reproduced in the quantum paradigm or because classical algorithms are so well-developed that it is impossible to beat them. Moreover, it does not seem a good solution to try to imitate classical algorithms because then the quantum advantage would fall only on the advantage that the quantum HW can offer, something that, currently, is not guaranteed that running a quantum circuit is faster than doing it classically.
    
        At this point, pessimism seems to prevail regarding the field of QAI and QML in particular. Conceptual and fundamental problems have been encountered that have made it clear that the current trend followed so far in the field is not correct and must be redirected \cite{cerezo2023}. However, the opportunity offered by quantum computing to accelerate algorithms in general remains a fact, and it is clear that classical AI has limitations, so there remains an opportunity for quantum algorithms. Once the target is clear and relevant to research, it is necessary to open up new research lines. In this work, QS\&S algorithms have been developed that can serve as the basis for a new line of QML research that diverges from the limited state of the art.
    
        To explain how QS\&S can be applied to the QAI field, the close relationship between search and artificial intelligence will be explained. Then it will be explained some QAI algorithms and the limitations that have been observed, and finally how QMS has been applied to QAI and what are its future applications will be explained.

        \subsection{Quantum S\&S for QAI}\label{apt:mot_ss}

        There is a close relationship between the field of optimization and AI. Many of the algorithms used for optimization problems are the basis for complex AI algorithms. In fact, many of the basic AI processes can be seen as problems in which a certain parameter needs to be optimized. This situation can be seen in evolutionary algorithms trying to optimize a fitness function, multi-agent algorithms trying to optimize group behavior, or ML algorithms where a hypothesis has to be optimized to achieve a balance between generality and expressiveness or a parameter optimization for deep learning.

        A third element, quantum computing, can be introduced into this relationship. As seen above, optimization problems have large state spaces that need to be traversed almost exhaustively to find the solution. It has also been seen that this is one of the tasks in which quantum computation is superior to classical computation. Therefore, one can speak of a trichotomy or triad that relates the three fields and that, in the center, houses a field of algorithm research with the ability to have a high impact on certain problems.

        Returning to the specific case of QML, it is known that a generic ML algorithm, whether classical or quantum, performs reasoning, usually inductive, on data. It takes concrete examples and tries to generalize to obtain rules common to all examples that allow it to make predictions or classifications. To do this, the algorithm, as it processes input examples, generates hypotheses that explain these examples. Some hypotheses will be more concrete, accurately covering certain cases but being wrong in many others, and, on the contrary, other hypotheses will be more general, covering most of the examples processed, but without accurately describing any of them. Thus, the set of hypotheses generated is called the hypothesis space, which is equivalent to the state space of an optimization problem. On this hypothesis space, it is necessary to look for the one that maximizes a certain value function that will represent a compromise between generality and abstraction as defined in each case.

        This search, which is performed during the training process, is very costly for the algorithms. The greater the volume of examples processed and the greater the expressiveness of the system, the greater the number of hypotheses generated and the greater the similarity between them, the more difficult it will be to guide the search and the longer the training time.
        
        In variational algorithms there is the same concept of searching the hypothesis space. The main difference is that the hypotheses are converted into parameter settings to vary the ansatz circuit. This process is performed by classical gradient descent algorithms, resulting in the barren plateaus (BPs) problem and, moreover, little generalization. It is evident that the same concept that works classically does not work quantumly, and therefore one must look for other models of both parameter configuration search and algorithms, going beyond the variational ones.

        The QS\&S algorithms are different from the gradient descent used so far. QS\&S does not perform a search following the energy landscape and, therefore, cannot end up stuck in a plateau. The key is to leverage some domain knowledge in the form of heuristics to guide the search toward a solution. This heuristic must be run for each of the possible states of the problem being solved, something impossible for classical computation, but feasible for the quantum paradigm.
        
        The key concept of QS\&S for doing heuristic search is to include the computation of the heuristic as an additional operator to the set of operators in the search process. This will compute the heuristic value of the superposition of states, that is, the value of each of the states. Thus, it will be possible to increase the probability of the states with the highest reward by performing an exhaustive search. This concept of QS\&S is not a concrete algorithm; it can be seen as a new philosophy, different from the current one, to develop QML algorithms. Later, this philosophy can be materialized in many types of algorithm, e.g., QMS.

        By applying QMS to the N-Queen problem, a practical implementation following this QS\&S philosophy for QAI has been demonstrated on a problem equivalent to the QAI problems.

        \subsection{Other QML techniques}\label{apt:qml}

        Once the relationship between QS\&S and AI and QML has been analyzed, it possible to explain briefly other techniques. This additional techniques study is interesting to see their problems because the QML field requires a different approach from the current one. 

\begin{figure}[h!]
            \centering
            \includegraphics[width=0.5\textwidth]{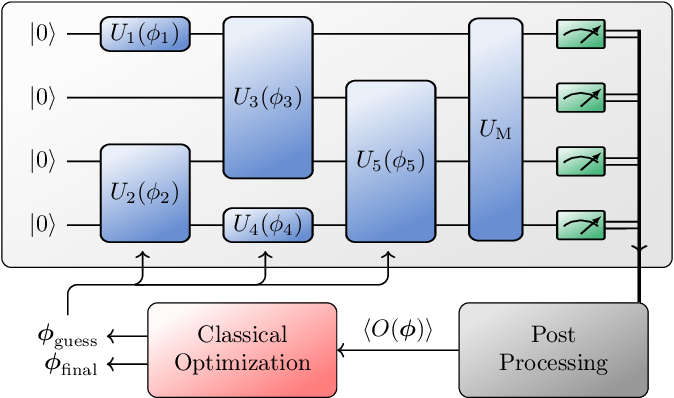}
            \caption{\label{fig:variational} This figure shows a general architecture from a variational algorithm (extracted from \cite{bittel2021}). The variational algorithms is composed by two main modules. The first module is a quantum circuit and it acts as a evaluation function. The parameters in the quantum module are modified following the optimization of a classical module, the optimizer. This process is repeated iteratively.}
        \end{figure}

        The field of QML is currently dominated by variational algorithms. These types of algorithms follow the hybrid paradigm with a quantum and a classical module. The quantum module is a circuit with parameterized quantum gates. Gate parameters are varied until the quantum circuit behaves in a certain way. To vary these parameters, a classical optimization algorithm is used. This algorithm takes the output of the circuit and evaluates it to determine the parameter values in the next run of the quantum circuit. To vary these parameters it follows a gradient descent based method. This system is run iteratively a certain number of times or until convergence is reached. An example of an architecture is in Figure \ref{fig:variational}. However, this paradigm of classic-quantum hybrid programming has not yielded good results.

        This execution scheme closely resembles that of classical neural networks. There is an initial phase known as forward, where the network runs with specific neurons (equivalent to quantum gates) to produce a result (measurement of the circuit). Subsequently, a second phase of backpropagation occurs, where the parameters of the neurons are adjusted based on the output, following a gradient descent algorithm.

        There has been much analysis of the reasons why variational algorithms do not work. The intuition was that variational algorithms would be more expressive than any classical algorithm and therefore learn faster and more accurately. On the contrary, so far the ansatz implementations have not been found to have more expressiveness than other classical algorithms \cite{wiersema2023} and it has been observed that the barren plateaus problem limits the trainability of the quantum circuit \cite{mcclean2018}.

\begin{figure}[h]
            \centering
            \includegraphics[width=0.4\textwidth]{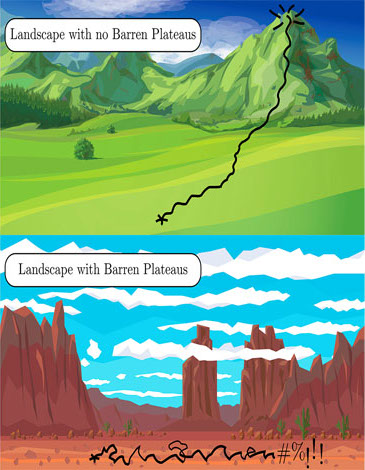}
            \caption{\label{fig:bps} This figure shows an artistic explanation of phenomena of barren plateaus (extracted from \cite{bps_artistic}). The upper picture shows an ideal situation without Barren Plateaus, in which at any point is possible to decide which is the next step to be closer to highest point. The lower picture shows a situation similar to a desert in which is impossible to decide which direction should be taken to be closer to the maximum point.}
        \end{figure}

        The barren plateaus (BPs) phenomenon occurs during the optimization process of the quantum circuit parameters of the variational algorithm. It is expected that each time the circuit is executed, the new parameter configuration of a function value that evaluates it will be closer or farther away from the solution. In this way, it is possible to know if the value of the evaluation function has improved or worsened with respect to the last parameter setting. In other words, the problem is expected to have a distribution in the evaluation function that allows guiding the search towards an absolute minimum. When looking at the BPs phenomenon, the distribution of the evaluation function value is flat in most of the space, except for a very small area around the solution that does have a downward gradient. This means that the search for parameter configurations of the quantum circuit cannot be guided, and it is necessary to carry out a practically exhaustive search, losing all possible quantum advantage \cite{holmes2022}. An artistic conception, and very explicative, of the BP problem is shown in figure \ref{fig:bps}.

        Reusing the most widespread metaphor for this situation, the ideal distribution is similar to a mountainous area, where it is easy to know whether the terrain is going up or down a mountain. In this ideal situation there are several mountains representing local maxima and one mountain higher than the others representing the absolute maximum. In this case, when the peak of a mountain is reached, the local maximum, it is possible to see where there are other higher mountains and guide the search. On the contrary, the barren plateaus scenario is identified with a desert where everything is totally flat and it is almost impossible to guide the search and where there is a small well hidden in the sand where a local minimum is found.
        
        There are multiple reasons for the existence of barren plateaus in problems, but the main ones are: circuit expressivity, local measurements, random initialization, and noise.

        After analyzing the BPs phenomenon in detail, some authors have come to the conclusion that variational algorithms are a dead end for the QML field \cite{cerezo2023}. This situation is problematic because much of the QML community effort was focused on this line.

        After seeing the limitations of variational algorithms, the QML community has tried other fields in which to make relevant contributions. Once is the quantum backpropagation. 
        
        The classical backpropagation method allows adjusting the weights of a neural network to reduce the difference between its prediction of a data label and its actual label. This process is very expensive because the weights of all neurons in a network must be adjusted. In fact, it is the process in which most of the training time of a neural network is spent. Because of its high cost and the possibility of parallelizing the process, something that is already classically done, the quantum computing community has tried to create quantum algorithms for backpropagation. However, this possibility has been discarded due to the need to execute non-linear functions and the impossibility of finding a quantum advantage \cite{abbas2023}.

        The failed attempt to find a quantum algorithm in this field can be understood as further evidence that the community is still looking for a justification for research in the field of QML. It is also necessary to say that, once again, the attempt to find a quantum advantage is using a method based on gradient descent, similar to variational algorithms, and which has proven to be inefficient.
        
        Another field that is being explored in great detail is that of quantum generative models. These algorithms are capable of generating data from noise. Generative artificial intelligence is a machine learning paradigm that differs from the usual discriminative models. The last ones are models used for classification or regression, while the former ones are models that aim to learn or discover relationships or patterns that exist between the different points in the input set, and then use this learned information to generate new points that mimic the characteristics of the input data set.

        Quantum computing provides a perfect fit for this type of AI model for several reasons. First, quantum computing is inherently random, so it is able to generate samples from an ensemble stochastically without introducing any bias. Second, it allows one to compute gradients faster than its classical counterpart. And third, work has already been done that tests these algorithms on current quantum computers, proving that this is a viable solution.

\begin{figure}[h!]
            \centering
            \includegraphics[width=0.75\textwidth]{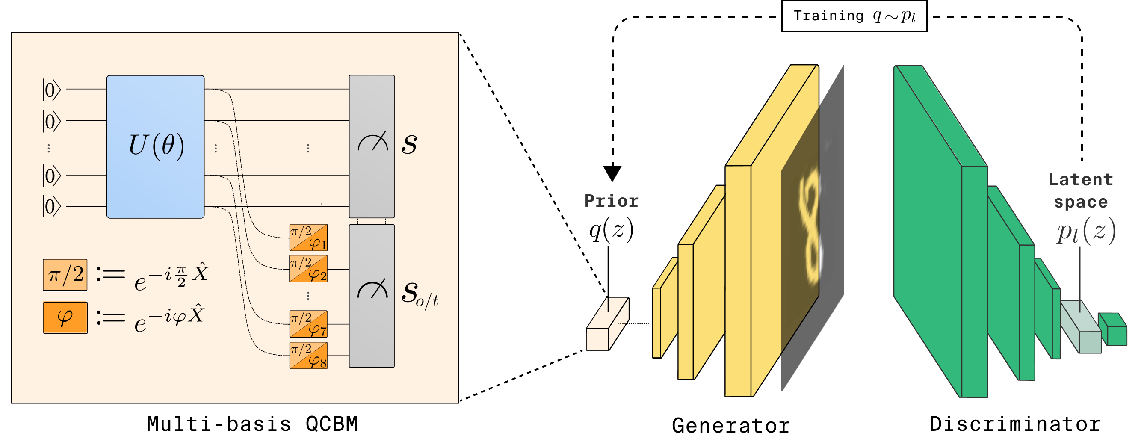}
            \caption{\label{fig:generative} Architecture of a hybrid quantum-classical generative algorithm. It has a quantum prior module and the generator and discriminator are classical modules. (extracted from \cite{rudolph2022}). The quantum prior samples from noise, then, the classical generator and discriminator are trained to transform the sample into data.}
        \end{figure}

        Quantum generative models have been proposed as one of the strongest candidates to take advantage of quantum computing. To this end, metrics have been created that measure the generality of these models and compare them with classical models \cite{gili2022, rudolph2021}. Some examples of this quantum generative model are quantum GANs \cite{dallaire2018, casares2020} and parameterized quantum circuits \cite{leyton2021} and its variations \cite{lloyd2018, zoufal2019}. In addition, small examples of these architectures have been implemented in quantum hardware \cite{rudolph2022}. There are also practical applications that make use of these quantum generative models, such as the discovery of new drugs \cite{li2021}. An example of the architecture of a hybrid generative algorithm is shown in figure \ref{fig:generative}.

        However, although they show great promise and are good candidates for quantum advantage, many of the current generative models are still based on gradient descent techniques, which causes them to have problems similar to those of other variational algorithms. Therefore, in the field of QML, these algorithms are beginning to be seen as useful in the medium to long term.

        The limitations observed in the QML field are due to two main factors that are closely related to each other, the short field run and the imitation of classical techniques. The field of QML is barely less than 10 years old, the first papers on quantum ML algorithms are from the mid-2010s. Just as happened with brain mimicry as the starting point for classical ML, mimicking classical algorithms has been the starting point for QML. Therefore, the current algorithms are, practically, the first attempts of algorithms in this field.

        However, it has been shown that this attempt to copy classical algorithms has serious limitations because gradient descent techniques are useful in classical algorithms but not in quantum algorithms. The future lies in designing new algorithms that, although based on the well-known concepts of classical ML, increasingly move away from the classical part and begin to exploit more purely quantum techniques.
       
        In this thesis, it is proposed that the field of QML be brought closer to QS\&S based techniques. This involves replacing gradient descent techniques with more informed searches. In classical ML, it is not possible to do an informed search in an algorithm like Neural Networks because it would take too much time to adjust the weight of each neuron. However, quantum computing can do exhaustive informed search with advantage over classical computing. The key to this advantage is that the evaluation function can be a quantum operator running on the superposition of states, \cite{miyamoto2022} and that the search for the marked element is quadratically faster.

        To demonstrate that the QS\&S concepts seen in the previous sections can help, it is explained below how QMS has been applied to the N-Queen problem. This problem has been used many times as a benchmark for testing search algorithms that then have an impact on other ML techniques.
 
        \subsection{The N-Queen problem}\label{apt:nqueen}

        The N-Queen problem consists of distributing $n$ queens over a chessboard of dimension $n \times n$ without attacks between the queens, i.e., none of the queens coincide in the same vertical, horizontal or diagonal line. This is a well-known problem in the classical computing world because it is very easy to test different algorithms to find the solution.

        In the Publication \ref{pub:p1} of this thesis, this problem is used to test an algorithm that evaluates all combinations and keeps those that meet the constraints. The solution has been proposed using a heuristic that evaluates the quality of the solution based on the number of attacks between queens until a state is found in which the number of attacks is 0.

        The N-Queen problem has been used as a benchmark for ML algorithms because the search necessary to find the solution is equivalent to the search that an ML algorithm performs in the hypothesis space to find the one that maximizes a certain function.

        Thus, in order to prove that QMS works well with the N-Queen problem following the QS\&S methodology explained above, it is necessary to have one more argument to continue this line of research.

        To take the step of using QMS for QML, it is required to change the evaluation function of the N-Queen problem to the evaluation function needed in QML. For example, having a classifier that can discriminate the examples by 3 parameters, it is possible to create an evaluation function that determines the quality of a hypothesis according to the value of those three parameters. Then, it should apply that evaluation function to all possible states or configurations of the system until the maximum reward is found.

        The results of applying QMS on N-Queen show that there is an advantage over classical algorithms. This consolidates the research line and prepares it for future steps focused on building the generalization process around QMS.

        A final consideration regarding the N-Queen problem and QMS is the execution in quantum hardware. The results detailed in publication \ref{pub:p1} were obtained in a simulator using logical qubits fully connected between them and without noise. However, as it was explained in previous chapters, quantum hardware still has some limitations to be able to execute circuits like a quantum simulator running in a classical hardware does. In a quantum hardware, the quality of these qubits is lower, so it is necessary to increase the number of qubits or reduce the problem. The number of qubits required for QMS is  $PQ + \lceil \log_2 P \rceil + a + 2$, being $P$ the number of parameters, $Q$ the number of qubits for the discretization and $a$ the number of qubits for the ancilla. Using two parameters with two qubits and three ancillas ($P=2$, $Q=2$ and $a=3$), 10 qubits are required. It is a reduced version of the problem but it is useful to understand the requirements. Using this configuration, around 1250 quantum gates are necessary.

    \subsection{Results}\label{apt:res_qai}

        \begin{itemize}[label=\textcolor{green}{\checkmark}]
        
            \item QMS and the QS\&S paradigm have been successfully applied to a problem equivalent to any QAI and QML problem, the N-Queen problem.
            
            \item A new approach to QML problems has been studied, the informed search in state or hypothesis space, as opposed to the mostly used uninformed search, e.g., classical gradient descent optimization.
            
            \item The alternative of using QMS offers the advantage of a search without the phenomenon known as barren plateaus.
            
            \item Scaling advantage results have been obtained using QMS which has a polynomial scalability in the number of qubits to represent the states.

            \item By applying QMS to the N-Queen problem and looking at its results, a tool has been proposed that can serve as a basis for other QML algorithms.
            
        \end{itemize}

    \newpage\null\thispagestyle{empty}\newpage

    \newpage
    \section{S\&S applied to Space Exploration} \label{ch:SSspace}
    \thispagestyle{empty}

        \lettrine[lines=1, findent=2pt]{\resizebox{!}{1.2\baselineskip}{T}}{}he field of space exploration encompasses any field of research that involves learning more about any process outside our planet. Under the umbrella of the concept of space exploration, there is a heterogeneous group of research lines and techniques ranging from the design of new rockets to the analysis of earth data taken from satellites in low orbit. Within space exploration, there are more theoretical lines, such as models for predicting the movement of the planets, or more applied lines, such as communications satellites. 

        A feature common to many techniques of space exploration is the complexity of the analysis of data obtained by satellites or exploration missions. This complexity derives partly from the volume of data to be processed and partly from the complexity of the data. Of course, this complexity cannot be generalized to any line of research, but it tends to be a constant in many problems to be solved \cite{kumar2018}.

        Examples of space exploration problems include analysis of Earth observation data, optimization of satellite orbits, prediction of solar storms, improvement of communications between satellites and ground stations, or analysis of complex data such as gravitational waves. These are just a few examples of space exploration use cases, but in all of them, quantum computing has been applied in some way or another \cite{khan2018, chang2022, miroszewski2023}.

        In fact, the world's major space agencies are devoting great efforts to developing quantum technologies applied to space exploration. Some examples are NASA with its QuAIL and AMES groups, ESA with its ACT and PhiLab groups, JAXA with the Space Exploration Innovation Hub Center or the Chinese space agency with its Mozi and Micius satellites. But this is not limited to government agencies; around them there are large companies that have planned lines of research that apply quantum computing to space exploration problems to overcome the major limitations of classical algorithms.

        In this use case, QMS can be applied to space exploration problems if they can be converted into optimization problems. In some cases, a conversion is not even necessary, as the problems are already studied as such as optimization problems. Specifically, this work will include as a use case the study of the inference of gravitational wave source parameters through an adaptation of QMS to the problem, known as \textit{qBIRD}, quantum bayesian inference with renormalization and downsampling.
     
        \subsection{Introduction to Gravitational Waves}\label{apt:gws}
        
        A gravitational wave is a perturbation of space-time produced by an accelerated massive body. They travel at the speed of light, contracting and stretching anything in their path. They represent ripples in the fabric of space-time, generated by temporal variations in the quadrupole momentum of the source mass, and induce variations in the length of the objects they pass through. This work is focused on gravitational waves produced by two black holes that rotate about themselves until a merge event occurs. 

        Gravitational waves are really difficult to measure and were theoretically proposed by Einstein, albeit with moments of disbelief and doubts encompassing a long and winding road. Today, gravitational waves produced in a merge event are measured because they are of such intensity that they can be measured because of the deformation of the detectors; the rest of the gravitational waves are invisible to the measuring devices currently available.

        In 2015, the LIGO collaboration was able to make the first direct measurement of gravitational waves thanks to large laser interferometers. The detection of gravitational waves is an important new validation of the theory of general relativity.
        
        The LIGO scientific collaboration is the union of different universities and research centers for the study and detection of gravitational waves. Members of the collaboration have access to the advanced LIGO detectors located in Hanford, Richland, and Livingston, Louisiana, both in the USA and to the VIRGO detector located in Pisa, Italy. This collaboration was initially involved in building the detectors with laser interferometer technology and is now involved in upgrading the interferometers and studying all aspects of gravitational waves. The founders of LIGO received the Nobel Prize in Physics for their work in detecting gravitational waves in 2017.

        Studying gravitational waves in detail has a great impact on the study of the evolution of the universe and of phenomena that are still a mystery, such as black holes. A fundamental characteristic of the gravitational waves that are measured is that they have a validity period. This validity period is a problem because the volume of detected events is constantly increasing due to improvements in detector accuracy. If the usual trend of slow processing rate and high detection rate continues, a point of collapse will be reached. In addition, once the phenomenon disappears, it is not possible to observe it, the only possibility is to analyze the static and past already recorded data. 
        
        Gravitational waves make possible the opening of a new window through which to look at the universe, complementary to the observation of the electromagnetic spectrum. This new method promises the observation of astrophysical events that have never been seen before.

        One of the possible analyses that can be done on gravitational waves is to obtain some of the parameters of the event that generated them. It is possible to obtain the distance they were from, the masses of the objects, their spins, etc. This analysis is done with one of the techniques with more results as parameter estimation, giving rise to the field of parameter estimation gravitational waves (PEGW).   
        
        Currently, the PEGW analysis process is slow and requires a high processing capacity and computational power. This is because it requires analyzing many possible states and combinations of variables. PEGW can be viewed as an optimization problem in which the value of certain parameters must be adjusted to find a synthetic distribution that is as close as possible to the actual distribution. 

\begin{figure}[t]
            \centering
            \includegraphics[width=0.6\textwidth]{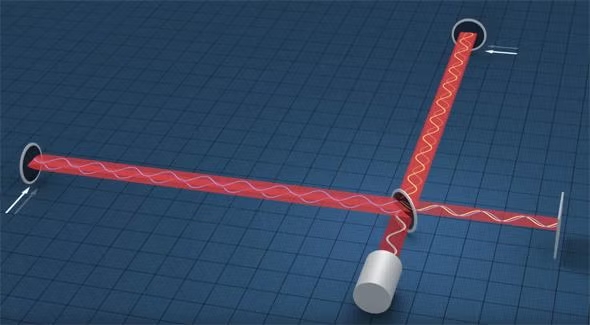}
            \caption{\label{fig:interferometer} Scheme extracted from \cite{interferometers_scheme}. It shows how the interferometers built by LIGO works. Both interferometers are in a perpendicular position. If a gravitational wave event is detected, a interference parameter is received in the detector.}
        \end{figure}

\begin{figure}[t]
            \centering
            \includegraphics[width=0.7\textwidth]{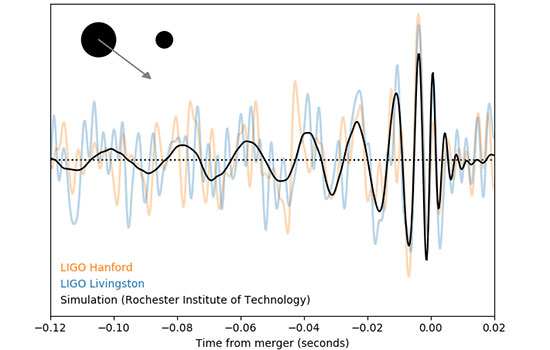}
            \caption{\label{fig:pegw} Scheme extracted from \cite{pegw}. It shows the gravitational waves detected by LIGO Handford (yellow) and LIGO Livingston (blue) and the simulation of the parameter estimation process (black). The estimation process tries to adjust the inferred wave as much as possible to the received wave.}
        \end{figure}

        In order to better understand PEGW, it is necessary to explain the analysis process. The interferometer that detects a gravitational wave is composed of two large lasers. These lasers are joined at a perpendicular intersection and when they detect a gravitational wave, they generate an interference pattern, as shown in figure \ref{fig:interferometer}. This pattern contains all the information about the event that generated the gravitational wave. Therefore, if, by adjusting the value of certain parameters, it is possible to create a sufficiently similar numerical wave, the value of the parameters that form or create that interference pattern can be found. In this way, a state space is created with a number of dimensions equivalent to the number of parameters in which it is necessary to try all possible combinations to find which one generates a synthetic interference pattern as similar as possible to the measured one, as explained in Figure \ref{fig:pegw}.

        Following this PEGW scheme and in order to analyze in detail the gravitational waves as a whole, both the LIGO collaboration and other universities have implemented different gravitational wave analysis software. Among them, two stand out, PyCBC \cite{nitz2024} and Bilby \cite{ashton2019}. Both allow inference, plotting, etc. In this work, the aim is to create a hybrid algorithm that includes quantum computation modules for PEGW and that can be included with some of the existing tools or even be a tool in itself, in order to speed up the PEGW process.

        The classical PEGW algorithms are based on the classical Metropolis-Hastings algorithm seen in the chapter \ref{ch:cSS}. These algorithms encounter computational constraints similar to those experienced by classical optimization algorithms. In addition, in this case, an evaluation function, likelihood, slow to run, and a large volume of data are added. All this makes the number of gravitational wave events to be analyzed grow faster than their analysis capacity, leading to different scenarios of collapse of the gravitational wave analysis \cite{qi2021}.

        The ultimate goal of introducing quantum computing into the field of gravitational waves is to create a pipeline that fully utilizes quantum technologies, both in detection and in processing and analysis. However, this goal is too general and ambitious. Therefore, it is necessary to go step by step, and creating a quantum PEGW algorithm that gains an advantage over classical algorithms is already a breakthrough.
        
        To carry out the development of a PEGW algorithm with quantum computing, after understanding the problem as an optimization, QMS has been adapted to the problem with certain variations. This new version has been called qBIRD.

        \subsection{Methodology}\label{apt:metho_gws}

        Understanding the PEGW problem as an optimization problem in which a search is required, it is necessary to define an evaluation function that determines the quality of each state. In this case, loglikelihood is used as the evaluation function.

        The likelihood function is the conditional probability of the Bayesian theorem $\mathcal{L}(d| \bm{\theta}) = p(d| \bm{\theta})$. Being $d$ the data observed by the interferometer, $\theta$ the set of parameters to infer that explains $d$, $\mathcal{L}$ the likelihood function and $p$ the posteriors of the data and parameters. For this purpose, it is common in the literature of gravitational wave astronomy to assume a Gaussian noise in the detectors \cite{thrane2019}, so that the Gaussian-noise likelihood function has the form:

        \begin{equation}  \label{eqn:likelihood1}
            \mathcal{L}(d| \boldsymbol{\theta}) \propto \exp \left(-\frac{1}{2} \int_{0}^{\infty} \frac{|\tilde{d}(f)-\tilde{\mu}(\boldsymbol{\theta}, f) | ^2}{\sigma^2(f)}df \right),
        \end{equation}

        All these functions are computed in the Fourier domain, where $f$ represents frequencies. Here, $\tilde{d}$ denotes the data set acquired conventionally by the detector, $\tilde{\mu}$ signifies the waveform derived from a specific relativistic numerical model calculated using the parameter set $\theta$, and $\sigma$ indicates the detector noise, expressed as spectral density. Notably, although the integral spans all possible frequencies, in practice, this interval is confined to frequencies compatible with ground-based detectors while excluding certain frequencies like harmonics of domestic electricity. This restriction ensures that $\sigma$ remains non-zero across these frequencies, thereby preventing any divergence in equation \ref{eqn:likelihood1}.
        
        Classically, the M-H algorithm is used to perform PEGW inference. The idea is to run the M-H algorithm many times; it can be run up to 40,000 times. In each run, the M-H algorithm will traverse the state space and stop at a point. All of these points are then taken together to perform a statistical study of the most probable set of values.

        The concept of \textit{qBIRD} is to use a quantum M-H algorithm, seeking greater precision in each execution, and therefore to perform a smaller number of executions. In addition, \textit{qBIRD} is designed so that, in the future, with a more developed quantum HW, the likelihood calculation will be an additional operator in the quantum circuit, which makes the calculation of each iteration faster.

        Due to the limitations of current quantum hardware, the likelihood calculation of each state must be classically preprocessed and introduced into the quantum circuit via a QRAM. However, some authors have already proposed the calculation of quantum likelihood \cite{miyamoto2022}. By including the likelihood calculation as an additional operator, it can be applied over the entire overlay and saves both time and resources by avoiding the inclusion of the QRAM technique, which adds a lot of depth to the circuit.

        When comparing the classical PEGW algorithm with its quantum version, it is necessary to use some figure of merit. Since \textit{qBIRD} is an adaptation of QMS, it is straightforward to use the TTS explained in subsection \ref{apt:metho_qms}. Analyzing the TTS of classical and quantum PEGW methods shows that there is an even greater advantage over the use of other PEGW use cases. Moreover, this advantage is expected to be even higher when larger problems can be run.
        
        Another advantage of \textit{qBIRD} over purely classical algorithms is its scalability in qubits inherited from QMS. \textit{qBIRD} uses exactly the same problem representation as QMS and scales polynomially with the number of parameters being analyzed.

        Currently, quantum simulators running on classical hardware can simulate up to 4 parameters with a discretization of 5 or 6 qubits each. For this, the rest of the parameters are fixed to complete the 17 that are usually inferred. As the capabilities of quantum computers increase, this number of inferred parameters will increase to 17 and be an inference equivalent to the state-of-the-art.
        
\begin{figure}[H]
            \centering
            \includegraphics[width=0.5\textwidth]{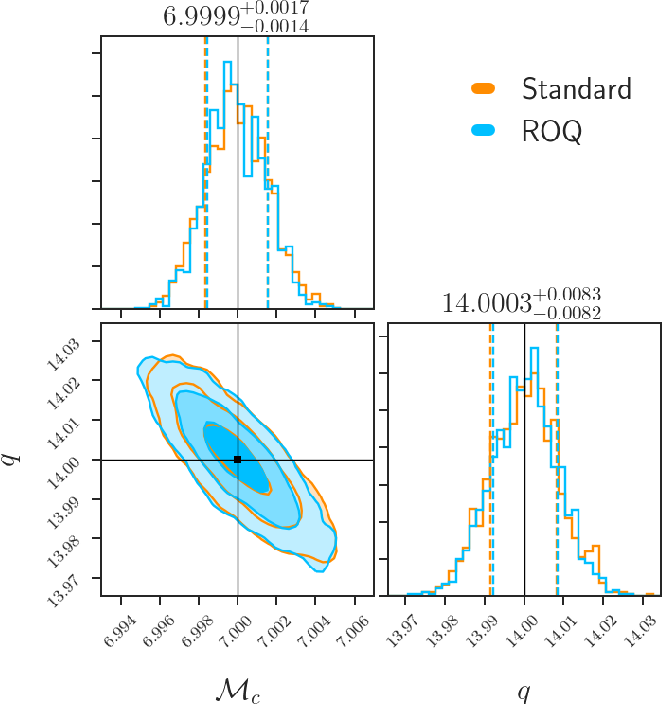}
            \caption{\label{fig:corner_plot} Corner plot extracted from \cite{qi2021}. It shows a combination of the result of estimating two parameters of a gravitational waves. It is a general format to show the results in PEGW.}
        \end{figure}

        Different tools are used to represent the results obtained in the PEGW process. The main one to be seen in this work is the corner plot, shown in figure \ref{fig:corner_plot}. 

        A corner plot, like the shown in figure \ref{fig:corner_plot}, is a commonly used graphical representation in parameter estimation in the field of gravitational waves. This type of plot effectively visualizes the joint distribution of multiple parameters of the model being analyzed. Each axis of the plot represents a pair of parameters under investigation. Along the margins of a corner plot, histograms are included showing the marginal distribution of each parameter individually. The points or contours in the plot represent samples drawn from a Markov Chain Monte Carlo (MCMC) chain or another algorithm used to explore the parameter space of the model. Summary statistics such as the mean or median of the distribution of each parameter and confidence intervals are included to highlight important features of the distribution.

        Corner plots are useful tools for understanding the relationships between multiple parameters in a model and for identifying possible correlations between them and it is especially interesting when there are few parameters, 2 or 4, as in this case, because it allows to see the posterior probability distributions (PDFs) of the parameters 2 to 2.

        As discussed above, the aim of \textit{qBIRD} is to develop a PEGW algorithm that will speed up this process and overcome current bottlenecks. \textit{qBIRD} may end up being included in one of the currently discussed libraries, PyCBC or Bilby, or even being a new library in its own right. To this end, in this work, there has been a close collaboration with a scientist from the LIGO collaboration.

        \subsection{qBIRD}\label{apt:qbird}

        The simplest way to understand \textit{qBIRD} is as an optimization algorithm capable of obtaining good results on the PEGW problem. This algorithm uses sample and search techniques to find the region of the state space in which the likelihood value is highest. It then samples that region in order to find the probability distributions of the posteriors. The search is guided using the likelihood function explained in equation \ref{eqn:likelihood1} with the objetive of maximizing it.

        The QS\&S technique used by \textit{qBIRD} is an adaptation of the one used in QMS, with the objective of adjusting it to the PEGW problem. For this purpose, the concepts of renormalization and downsampling are used.

        Renormalization is a technique represented by a semigroup of transformations that make up an iterative process of elimination (called truncation) of degrees of freedom of a quantum system according to a certain criterion in order to obtain an effective description of the retained degrees of freedom \cite{wilson1975}. The criterion can be of low energy, maximum likelihood, etc. It gives rise to approximate computational methods of variational type. This technique can be applied to QS\&S.

        QS\&S techniques are primarily required for exploring large state spaces. When dealing with quantum computing in the context of a significant state space, the challenge lies in the probability distribution. In smaller state spaces, the probability of finding states with lower energy functions is relatively high compared to other states. For instance, in a search involving 8 states, the normalized probability of lower energy states might be around 0.3, contrasting with other states at 0.001. However, in a state space of $2^{32}$ states, the maximum probability state could still be 0.001, with the next maximum probability state potentially as low as 0.000001. Despite the three-order difference in magnitude, both states become indistinguishable upon circuit measurement. Therefore, it becomes necessary to execute iteratively the quantum circuit, eliminating the least probable states in each iteration until a noticeable difference in probability is achieved after measurement. This iterative process is based on renormalization.

        This renormalization-inspired technique that is an adaptation of the QS\&S to PEGW problem, called qBIRD, can be seen from a different perspective. In computer science, specially in the algorithms of digital signal processing, there is a set of algorithms known as downsampling techniques. These techniques are also widely applied to convolutional neural networks \cite{zhou2020}.

        Downsampling refers to the comprehensive process of reducing both bandwidth (filtering) and sample rate. When applied to a sequence of signal samples or a continuous function, downsampling yields an approximation of the sequence that would have been obtained by sampling the signal at a lower rate (or density, as seen in photograph resolution reduction). In the context of qBIRD, downsampling is employed similarly to approximate or reduce the state space to get the likelihood distribution that requires sampling.

        \textit{qBIRD} receives as an input a set of parameters to infer and it returns the posterior values of these parameters according to a likelihood function. These parameters are initialized to a random values according to some prior knowledge. The prior is composed by an interval and the distribution of this interval. If the distribution is unknown, a Gaussian function is used. The first step of the algorithm is the proposal of new values, \textit{Start} in Figure \ref{fig:qbird_scheme}. Then, steps 1 to 3 are executed.

        \textbf{Step 0.} \textbf{Parameter initialization:} The algorithm begins by initializing values for each parameter drawn from a prior distribution, which is defined by the lower and upper bounds of the prior function. These values are then stored in the state register.

        \textbf{Step 1.} \textbf{Renormalization \& Downsampling: } The first module implements the quantum Metropolis-Hastings algorithm adapted from publication \ref{pub:p1}, and incorporates a renormalization technique defining the \textit{qBIRD} algorithm. During this phase, the quantum walk is iteratively applied to narrow down the parameter space, aiming to identify the values for each parameter that maximize the likelihood.
        
        \textbf{a) Quantum Metropolis:} Iteratively apply the walk operator $W$, defined in Equation \ref{eq:W}, a number of times on the initial state:
                \begin{equation} \label{eqn:ev_state}
                    \ket{\psi} := W_L ... W_2 W_1 \ket{\phi^{(0)}}.
                \end{equation}
        
        \textbf{b) Sampling:} Sample the state register.
        
        \textbf{c) Threshold condition:} If the number of elements is the same than the number of parameters to infer, each parameter is represented by 1 qubit, jump to Step 2, otherwise calculate the number of elements.
                            
        \textbf{d) Qubit reductor:} Reduce the number of qubits in the state register by 1 unit and go to Step 1a). This condition allows eliminating one qubit for each parameter, ending up with a minimum of one qubit per parameter.

    The genesis of this second module stems from the challenge of employing the quantum Metropolis algorithm to pinpoint the state with the highest probability. Drawing inspiration from the renormalization techniques of quantum lattice models \cite{wilson1975}, it is addressed the issue arising from the vastness of the state space. In quantum systems, the normalization factor leads to minute probability discrepancies between the most and least likely states. Despite spanning several orders of magnitude, these disparities fail to attain significance without an impractical number of measurements. 
    
    By progressively eliminating states with significantly lower probabilities through the reduction of problem size and qubits, these discrepancies become increasingly discernible. Employing this technique enables the identification of the state with the maximum likelihood among all proposed values.
    
    \textbf{Step 2.} \textbf{Mean \& Std Deviation calculator:} The third module consists of a classical processing that takes the results obtained in the first two modules to generate PDFs for each parameter.
    
    \textbf{Step 3.} \textbf{Search interval calculator:} To gradually narrow down the search area, a new interval for each parameter is proposed.
    
    \noindent    \textbf{End.} 
    After a given number of iterations of Steps $0 - 3$, PDFs of each of the parameters are constructed from the values obtained in each iteration. This algorithm is vastly detailed in \cite{escrig2024}.

\begin{figure}[H]
        \centering
        \includegraphics[width=0.6\textwidth]{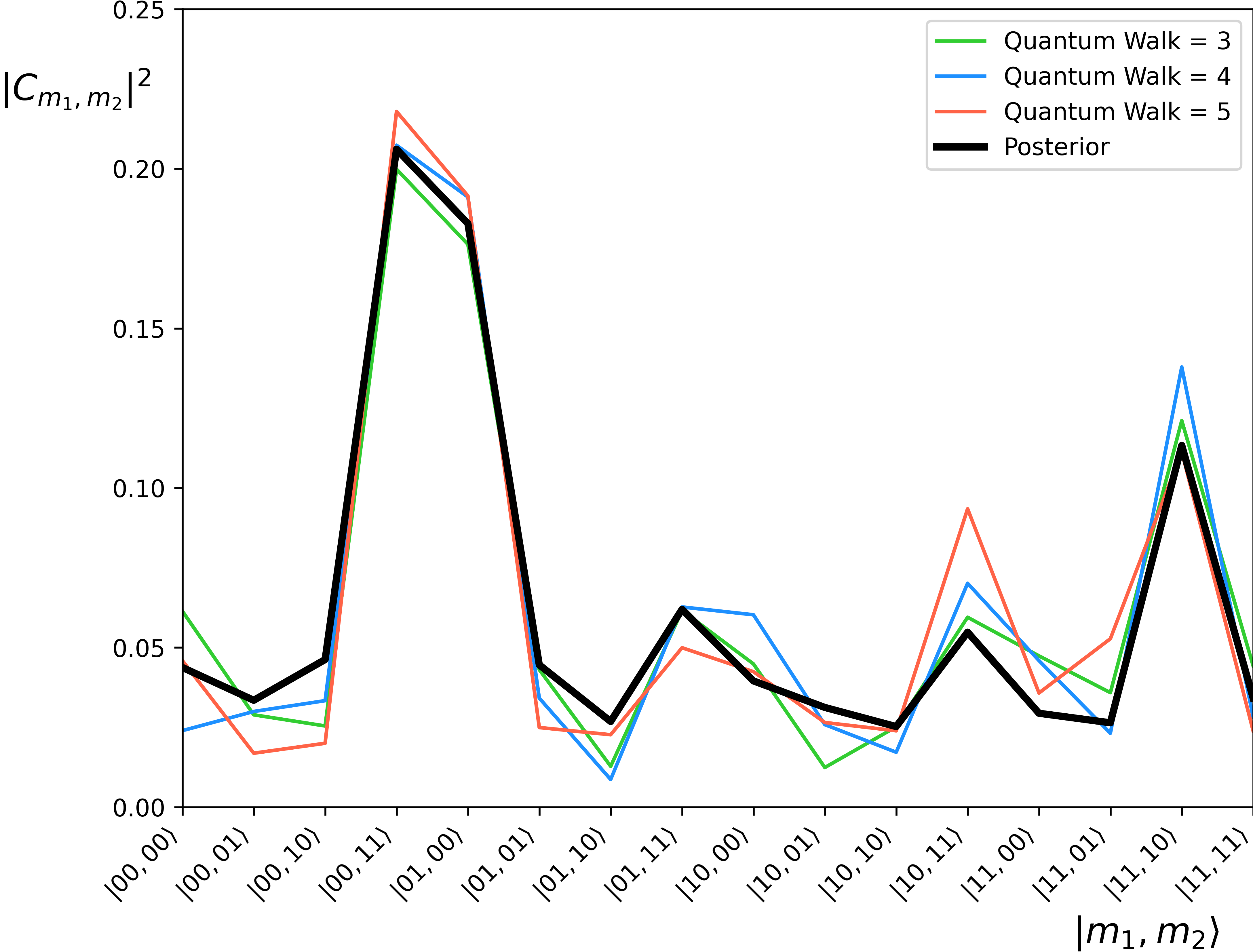}
            \caption{ Colour lines represent the state register probabilities for different number of applications of the quantum walk operator. It is a two source masses inference, using $Q = 2$ discretization qubits, for the event GW150914. X-axis is composed by the 16 combinations that represents the binary encoding of a value for the first mass and another value for the second mass. The black line represents the posterior probability function of each of these combinations.} \label{fig:likelihood_sampling}
    \end{figure}

    The strength of \textit{qBIRD} is its ability to sample a distribution with a low number of iterations. As it is possible to observe in figure \ref{fig:likelihood_sampling}, using a different number of $W$ it is possible to perform almost a perfect match with the distribution to guess (black line). While the classical sampler would need almost an exhaustive sampling, the quantum algorithm can do with a number of $W$ between 3 and 5. It is just a small example, but very illustrative.

    Once \textit{qBIRD} was designed, it was implemented in a circuit. Subsequently, this circuit was executed in a quantum simulator, which ran on a classical computer utilizing the Qiskit library \cite{qiskit}. The results obtained from the simulator are invaluable, as they validate the functionality of the algorithm. The next step, executing \textit{qBIRD} on quantum hardware, does not require any special implementation. This implies that qBIRD's limitations are solely dictated by the constraints of quantum hardware, but it can scale as much as needed. 

    \textit{qBIRD} has been tested using two benchmarks, speed and accuracy. Both metrics are relevant for any quantum algorithm called to replace a classical one. As explained above, the role of \textit{qBIRD} is to replace the classical samplers integrated in tools such as \textit{Bilby} and \textit{PyCBC}. \textit{qBIRD} must be fast because current PEGW algorithms take unacceptable times for the speed at which events are detected. In addition, its results must be as accurate as or more accurate than classical algorithms to be accepted by the gravitational wave community.

\begin{figure}[H]
            \centering
            \includegraphics[width=0.8\textwidth]{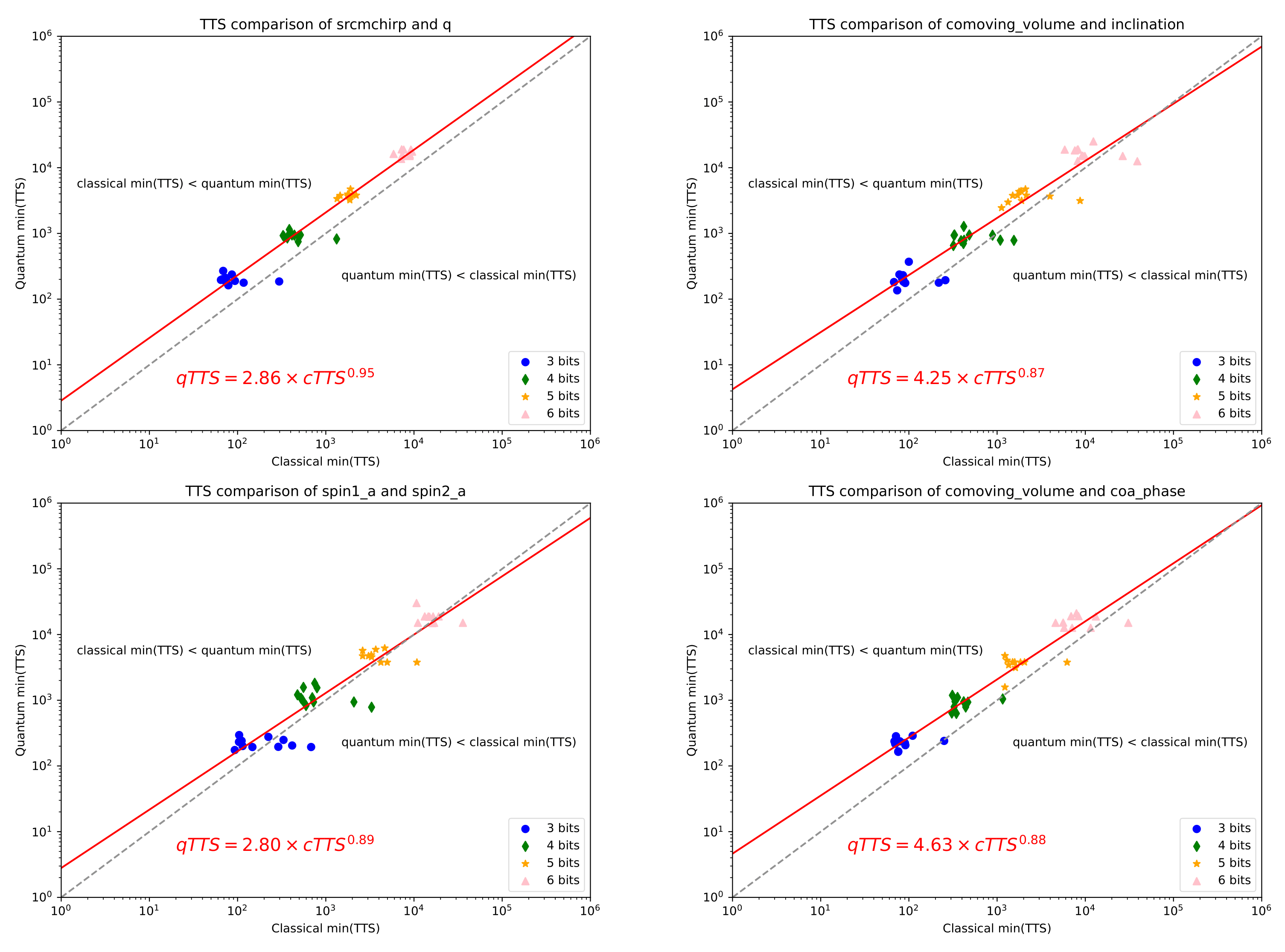}
            \caption{Comparison of the minimum value of the classical and quantum TTS achieved for the 2-parameter inference simulations. Top left the source-frame chirp mass and the mass ratio have been inferred. Top right the comoving volume and the inclination have been inferred. Below left the dimensionless spin-magnitude of the larger and smaller object have been inferred. Below right the comoving volume and the coalescence phase have been inferred. All results have been obtained for 3-6 bits of precision with a constant value of $\beta$.} \label{fig:2_p}
        \end{figure}

\begin{figure}[H]
            \centering
            \includegraphics[width=0.8\textwidth]{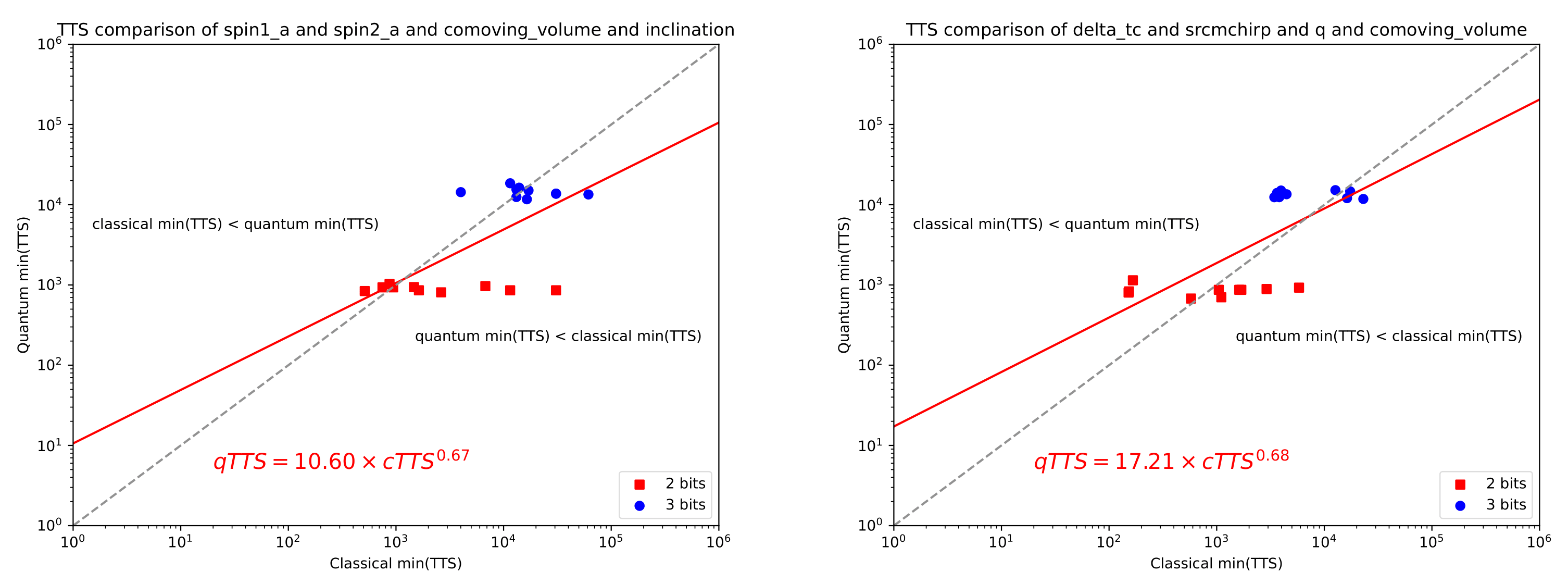}
             \caption{Comparison of the minimum value of the classical and quantum TTS achieved for the 4-parameter inference simulations. On the left the dimensionless spin-magnitude of the larger and smaller object, the comoving volume and the inclination have been inferred. On the right the source-frame chirp mass, the mass ratio, the coalescence time and the comoving volume have been inferred. All results have been obtained for 2-3 bits of precision with a constant value of $\beta$.}\label{fig:4_p}
        \end{figure}

    The core of \textit{qBIRD} is the quantum Metropolis-Hastings algorithm based on equation \ref{eq:W}. To test the execution time of \textit{qBIRD} applied to PEGW, it is necessary to prove the performance of $W$ applied to the problem. In the work \cite{escrig2023} is explained how the quantum Metropolis-Hastings algorithm is able to get a scaling quantum advantage as shown in figures \ref{fig:2_p} and \ref{fig:4_p}. It is important to notice that in both figures, the results are obtained using real LIGO data.

    In both figures, TTS metric is used compare classical and quantum algorithms. As has been shown in other results, the simulator limitations do not allow executing problems with a size equivalent to the solved by classical computers. However, it is possible to analyze the scaling tendency and the quantum advantage is appreciable. In the table \ref{tab:exponents} a summary of the scaling exponent is shown in each inference case.

    This table \ref{tab:exponents} presents the same information as figures \ref{fig:2_p} and \ref{fig:4_p} in a more concise format. The key information conveyed by the figures is the value of the exponent, where a value lower than 1 indicates quantum advantage. Therefore, a lower exponent signifies a better quantum advantage, with 0.5 being the maximum due to the quadratic advantage defined by Grover. Additionally, it is observed that the exponent improves, approaching 0.5, with an increasing number of parameters to infer. This observation is linked with figure \ref{fig:scaling} and the problem size: with two parameters to infer, the problem is considered small, whereas with four parameters, it falls closer to the medium-sized category, where the gap between classical and quantum performance diminishes.

   \newpage

    \begin{table*}[h]
        \centering
        \begin{tabular}{cc}
        \multicolumn{2}{c}{2 PARAMETERS INFERENCE}                                    \\ \hline
        \multicolumn{1}{|c|}{\textbf{Parameters inferred}} & \multicolumn{1}{c|}{\textbf{Fit Exponents}} \\ \hline
        \multicolumn{1}{|c|}{Chirp mass and mass ratio}                  & \multicolumn{1}{c|}{0.95}             \\ \hline
        \multicolumn{1}{|c|}{Dimensionless spin$_1$ and spin$_2$}                  & \multicolumn{1}{c|}{0.89}             \\ \hline
        \multicolumn{1}{|c|}{Comoving volume and inclination}                  & \multicolumn{1}{c|}{0.87}             \\ \hline
        \multicolumn{1}{|c|}{Comoving Volume and coalesence phase}                  & \multicolumn{1}{c|}{0.88}             \\ \hline
        \multicolumn{2}{c}{4 PARAMETERS INFERENCE}                                    \\ \hline
        \multicolumn{1}{|c|}{\textbf{Parameters inferred}} & \multicolumn{1}{c|}{\textbf{Fit Exponents}} \\ \hline
        \multicolumn{1}{|c|}{Coalesence time, mases and comoving volume}                  & \multicolumn{1}{c|}{0.68}            \\ \hline
        \multicolumn{1}{|c|}{Dimensionless spin$_1$, spin$_2$, comoving volume and inclination}                  & \multicolumn{1}{c|}{0.67}            \\ \hline
        \end{tabular}
        \caption{Fitted exponents for different sets of variable parameters.}\label{tab:exponents}
    \end{table*}

    Just as important as the speed of execution is the precision and, derived from this, the visualization of the data. This is the motivation of the second publication that supports this work \cite{escrig2024}. In this continuation of the first work, qBIRD is used to get corner plots with real data similar to those used by GWs community.

\begin{figure}[t]
        \centering
        \includegraphics[width=0.6\textwidth]{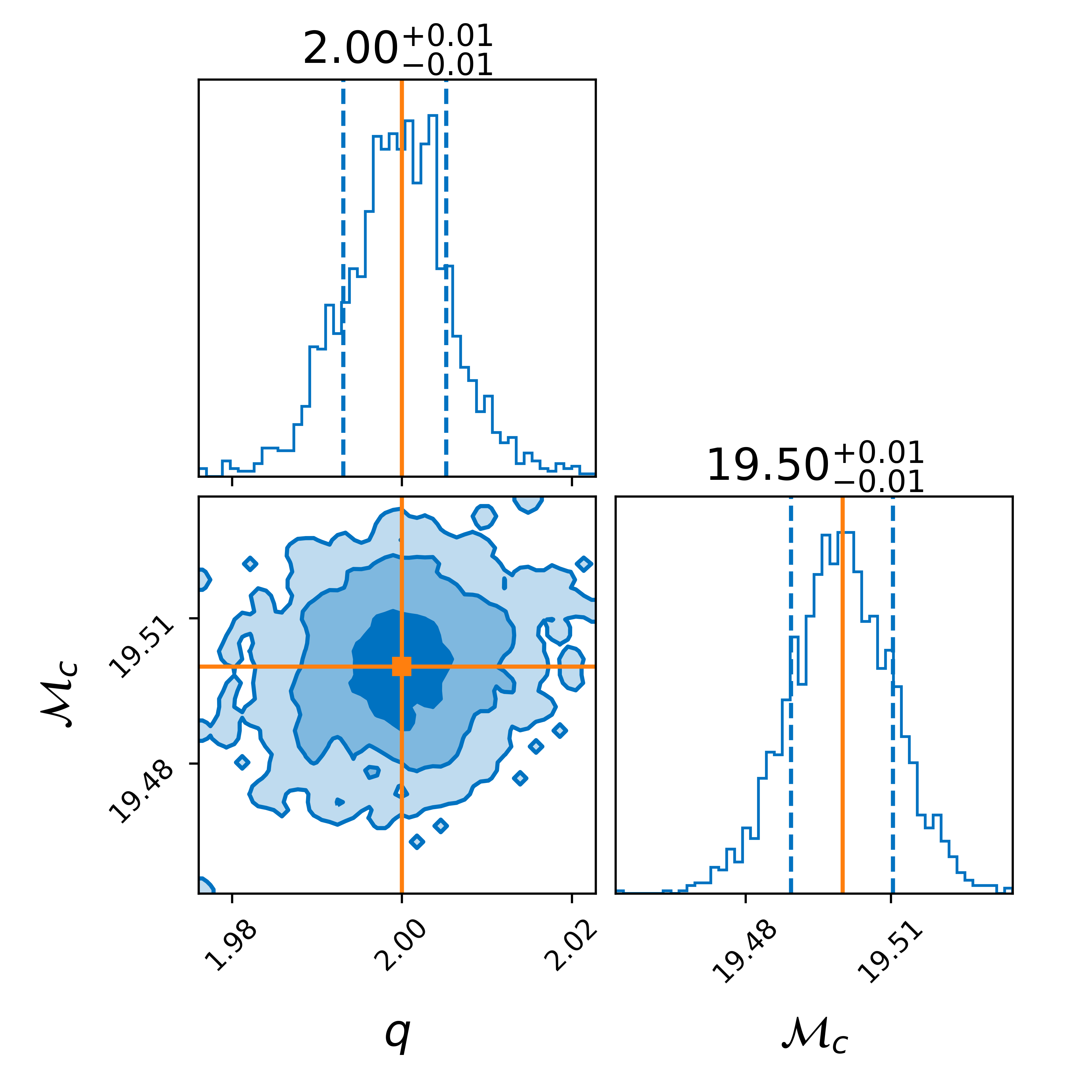}
        \caption{Posterior distributions obtained with \textit{qBIRD} for the chirp mass $\mathcal{M}_c$ and mass ratio $q$ of a simulated BBH gravitational wave signal injected into Gaussian-noise using \textit{PyCBC}. The injected values are $\mathcal{M}_c = 19.50\ {\text{M}_{\odot}}$ and $q = 2.00$, shown in orange.}
        \label{fig:injections_2param}
    \end{figure}

\begin{figure}[t]
        \centering
        \includegraphics[width=0.6\textwidth]{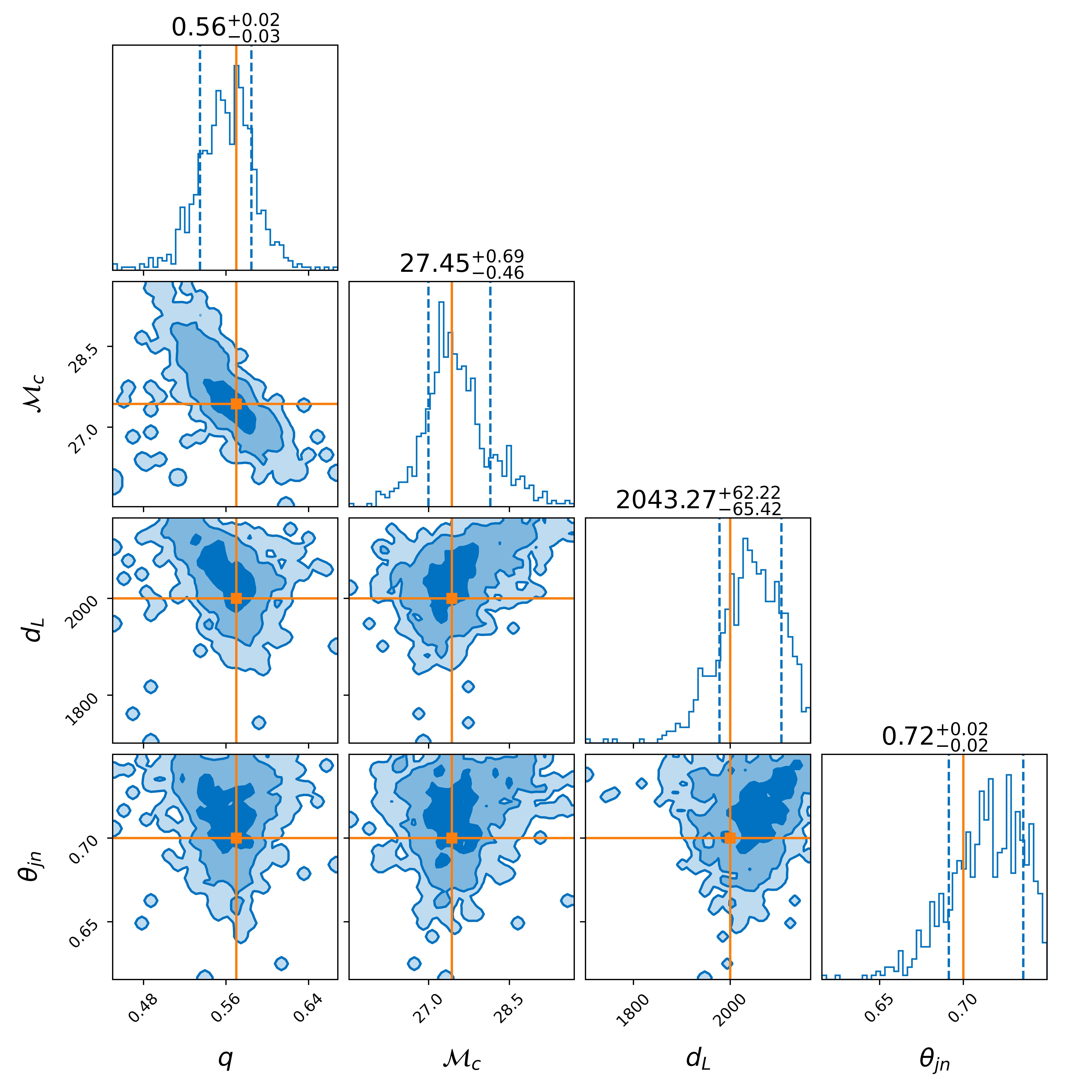}
        \caption{Posterior distributions obtained with \textit{qBIRD} for a synthetic BBH signal characterized with 4 unknown parameters. The simulated gravitational wave was injected into zero-noise using \textit{Bilby}, and the injected values are $\mathcal{M}_c = 27.43\ {\text{M}_{\odot}}$, $q = 0.57$, $d_L = 2000 \text{ Mpc}$, and $\theta_{jn} = 0.70$, shown in orange.}
    \label{fig:injections_4param}
    \end{figure}

\begin{figure}[H]
            \centering
            \includegraphics[width=0.9\textwidth]{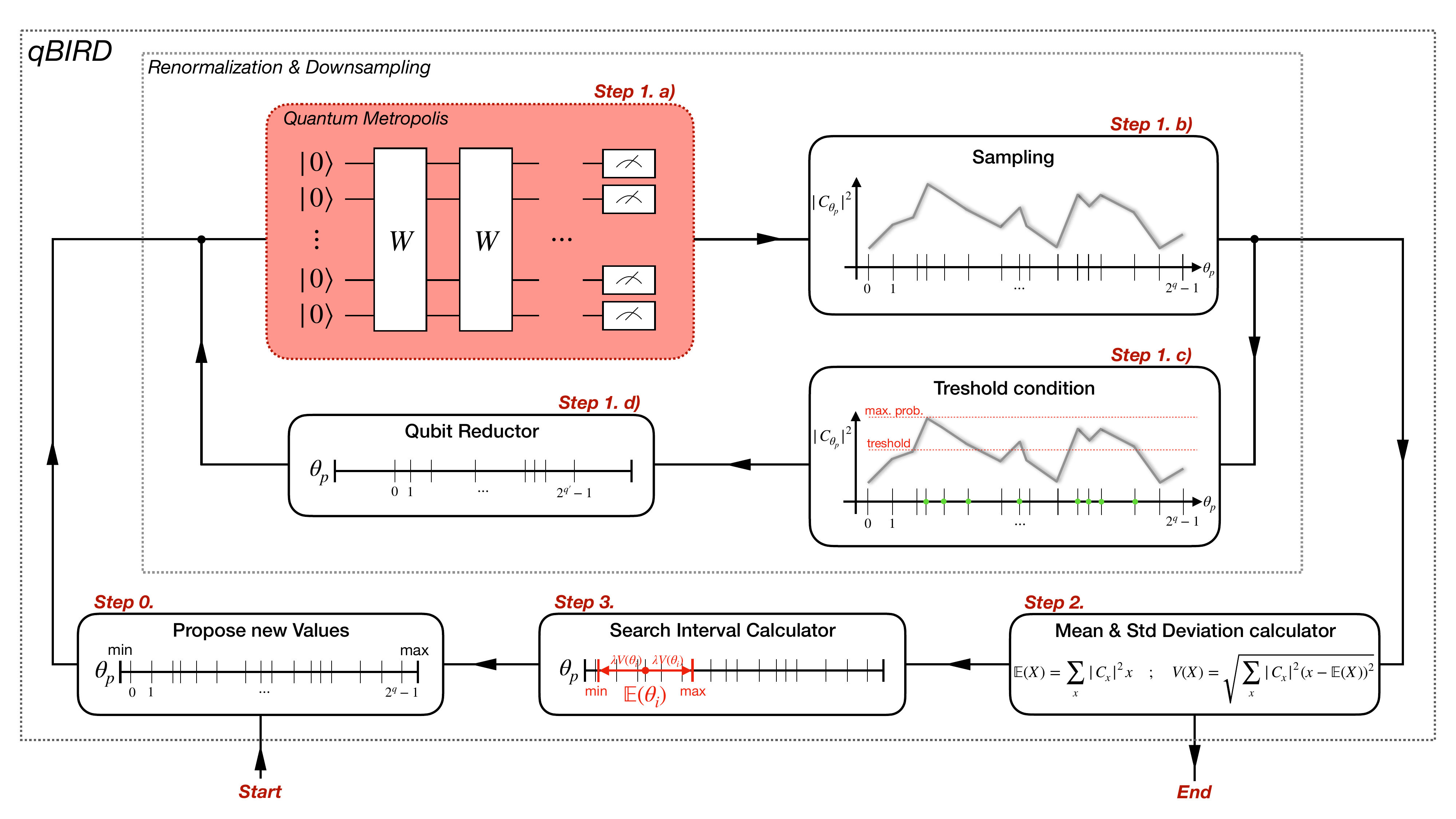}
            \caption{\label{fig:qbird_scheme} Flowchart of \textit{qBIRD} algorithm. This algorithm is divided in three steps. Step 1 is focus on renormalization process. This renormalization executes the quantum metropolis circuit and reduce the search. Then, steps 2 and 3 calculates the new intervals for the variables and the process starts again.}
        \end{figure}

    In the figures \ref{fig:injections_2param} and \ref{fig:injections_4param} it is shown the results of building corner plots using \textit{qBIRD}. Both cases are generated using injected parameters. The injection option is very used in the GWs community to test the precision of the algorithm. In both figures, the injection value (an orange cross) is the middle of the inference, falling always in the dark blue region, the region marked by \textit{qBIRD} as the more likelihood or the more posterior values region. Both corner plots are tests of the accuracy of the algorithm.

    Finally, if the algorithm has demonstrated an advantage in execution speed and good accuracy, the final test to ensure that the quantum algorithm can outperform a classical one, once the quantum hardware limitations are solved, is to show its scalability. It is possible to prove that the scalability of \textit{qBIRD} is polynomial in the number of parameters, $\mathcal{O}(P)$. The point is to analyze the qubits required by the quantum walk.

    The quantum walk employs a total of $PQ + \lceil \log_2 P \rceil + a + 2 $ qubits: $PQ$ represents the product of $P$, the number of inferred parameters and $Q$, the number of discretization qubits, with $2^Q$ states represented for each parameter; $\lceil \log_2 P \rceil$ qubits to represent the register of each parameter in binary encoding; $a$ qubits to represent the auxiliary register for the acceptance probability. Finally, 2 qubits are needed, one for the movement register, and another for the coin register to encode the accept/reject probability of all states.

    All terms in the scalabilty formula are dominated by the number of parameters, and it results in a good scalability, which has a lower consumption in the number of qubits than the qubit growth trend of quantum hardware.

    The execution in quantum hardware following these numbers and using two parameters with three qubits of discretization and three qubits of ancilla will result in a 12 qubit execution. If qBIRD uses four $W$ operators at each repetition, the number of gates will be close to 6.000 gates including H, X, CX, CCX, MCX, CU and RY gates.

        \subsection{Conclusions}\label{apt:res_gws}

        \begin{itemize}[label=\textcolor{green}{\checkmark}]
        
            \item QMS has been correctly adapted to a new use case in \cite{escrig2023} and \cite{escrig2024}, the inference of gravitational wave parameters, thus demonstrating the high versatility of QMS as a generic tool to adapt to other optimization problems.
            
            \item Quantum advantage results have been obtained using data from real gravitational wave events in \cite{escrig2023} and \cite{escrig2024}.
            
            \item Results have been obtained with an accuracy similar to that of state-of-the-art methods using real data from gravitational wave events in \cite{escrig2023} and \cite{escrig2024}.
            
            \item Generation of corner plots with accuracy similar to those generated by state-of-the-art techniques and using data from real gravitational wave events.
            
            \item \textit{qBIRD} has good scalability in the number of qubits and is better than the state of the art, which is essential for estimating 17 parameters.
            
            \item Modular software tool to make any module classical or quantum or to integrate new functionalities. In this case, the classical module that can be quantized is the calculation of likelihoods.
            
            \item Viable solution and new research line based on quantum software for an open problem in LIGO collaborations.
            
        \end{itemize}

    \newpage
    \newpage\null\thispagestyle{empty}\newpage

    \section{S\&S applied to Protein Folding}\label{ch:SSchem}
    \thispagestyle{empty}

        \lettrine[lines=1, findent=2pt]{\resizebox{!}{1.2\baselineskip}{T}}{}he research field of quantum chemistry includes those processes of chemistry in which quantum phenomena occur. The cases treated by quantum chemistry must be modeled in order to study them and predict the processes that will occur. For example, to predict the behavior of a molecule under certain conditions, it is necessary to create a model that explains the interactions of the molecule with the environment.

        The main limitations of the classical algorithms are to represent the superposition and entanglement effects between particles. The only solution for these cases is to create very simplified models that, although they do not predict accurately, it allows to know some characteristics of the systems. Although this is a very limited approach, it was the only possibility that existed until the birth of quantum computing.

        The advantage of quantum computing applied to quantum chemistry models is that it can inherently represent many of the processes that occur without having to use extra storage and processing resources. Examples include calculating the minimum energy of a molecule, calculating the intermediate states in a chemical reaction, predicting the best structure of a system to have certain properties, or simulating the behavior of complex molecules in a certain environment, for example, the protein folding problem.

        Quantum chemistry is fundamental in the development of both industrial processes and research. A large part of the research lines for the development of new drugs, new materials, carbon footprint reduction, etc. is directly related to quantum chemistry processes.

        The fact that classical computation is very limited in this field of research and that the basis of quantum computation is perfectly adapted to the processes to be modeled, makes quantum chemistry the most promising candidate for quantum advantage in the short term. This means that the first practical application of quantum computing is expected to be in a problem that, at some point, involves a quantum chemistry process.

        This chapter has been approached as a use case of QMS and, therefore, of the application of the QS\&S philosophy to different problems. The interesting quantum chemistry problems in this chapter are those that can be understood as optimization problems. Within quantum chemistry, there are problems that can be seen as a set of variables that can take different values and it is necessary to adapt these variables to find a solution. Specifically, the central problem that has been chosen is the protein folding problem.

        \subsection{Introduction to Protein Folding problem}\label{apt:qfold}
        The problem of protein folding is to find the three-dimensional structure of an amino acid chain that is stable and functional in the human body. For this to happen, it is necessary for the amino acid chain to adopt the three-dimensional structure of minimum energy. Naturally, the human body generates proteins folded in this minimum energy form, but introducing artificially a protein into the body, it is necessary to calculate its structure in such a way that it fulfills its task perfectly.

        In the human body, there are many functions that are performed by a protein that attaches to a particular tissue and captures other substances. This is the case with insulin and glucose. Insulin is a protein capable of capturing free glucose in the bloodstream and transporting it to a storage point for metabolization. If the body cannot generate insulin due to a pathology known as diabetes, hyperglycemia occurs, which can be very harmful to the human body. To solve this problem, insulin proteins have been artificially designed, which when injected into the blood perform the same function as if they had been generated by the human body.

        Research in the field has revelled the effects of diabetes by folding the amino acids that make up this protein so that it could perform its function in the human body. This structure is relatively simple because insulin is a small protein with few amino acids, but in other cases, proteins are much more complex, and the calculation of its structure can only be done by an algorithm.

        Classical protein folding algorithms must calculate, for each pair of amino acids, all possible ways of binding. This extended to all the amino acids of a protein, which can be more than 100, scales exponentially and is uncomputable for a classical computer, at least with exhaustive methods.

        Protein folding is the basis for the development of new drugs. In the human body there are only 20 amino acids that combine to create proteins. The functions of each of these amino acids are well known. Therefore, and simplifying the problem, if a protein to perform a specific function is required, it is only necessary to create it using certain amino acids. The next complicated step is to go from a linear chain of amino acids to a three-dimensional structure. Once the protein structure is calculated, the response of the human body to certain stimuli can be tested, measure the degradation of cells after exposure to certain products, and so on.

        In 2021, a historic milestone in the development of algorithms applied to protein folding occurred. The company Deep Mind released AlphaFold, a tool that uses deep learning technology to predict the structure of proteins given a chain of amino acids. AlphaFold proved to work very well for proteins of large complexity and size, which earned it the recognition of the entire community. While it has some limitations, such as training time or solution quality for some proteins with several possible structures, it is a breakthrough in the field.

        AlphaFold makes a prediction based on convolutional neural networks equivalent to those used for image recognition in computer vision. This type of network does not take into account the energy of the predicted protein, therefore, the solution offered by AlphaFold or any other DL-based algorithm can be refined to find a structure of lower energy and, therefore, more stable. That is the motivation of using protein folding as a use case for QMS, to create a search algorithm that refines the solution given by a method that makes an initial guess. In this case, QMS has been adapted to protein folding in a quantum classical hybrid algorithm called QFold.

        QFold has a hybrid architecture, apart from the internal architecture of QMS itself, in which a classical DL algorithm is run first, which gives a first structure of the protein that is not very stable, but serves as a starting point for the classical algorithm to search for the three-dimensional structure of the protein with minimum energy and maximum stability.

        \subsection{Methodology}\label{apt:metho_qfold}

\begin{figure}[t]
            \centering
            \includegraphics[width=0.75\textwidth]{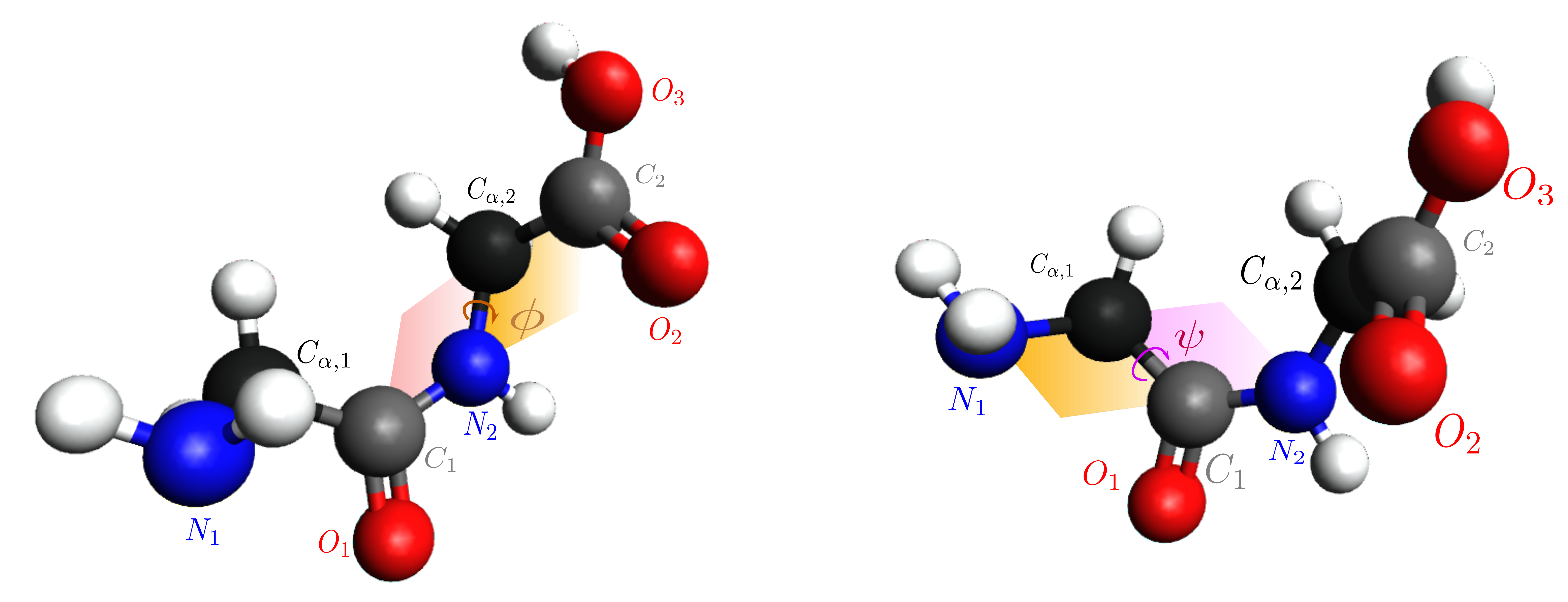}
            \caption{\label{fig:protein} It shows the dihedral angles $\phi$ (yellow) and $\psi$ (pink) of protein. Both angles are key to represent the bond between aminoacids. These angles are discretized with a number of qubits to calculate the 3D structure of the protein.}
        \end{figure}

        To define the overall structure of the protein, the pairwise bonds of the amino acids can be defined. Each amino acid bond can be defined as two torsion angles, $\phi$ and $\psi$. The torsion angles, also called dihedral, are angles between the atoms in the backbone structure of the protein, that determine its folding. An example with the smallest of the dipeptides, the glycylglycine, can be found in Figure \ref{fig:protein}. These angles are usually three per amino acid, $\phi$, $\psi$ and $\omega$, but the latter is almost always fixed at value $\pi$ and for that reason, not commonly taken into account.

        By discretizing the torsion angles between each of the amino acid pairs, the overall structure of the protein can be described. Then, all possible angle values of the protein are evaluated, calculating for each possible configuration its energy. In this way, the configuration with the lowest value in the evaluation function can be obtained following a search guided by energy minimization.

        The protein folding problem has been extensively studied both from the perspective of classical and quantum algorithms. The latter, due to the limitations of the algorithms and their executability, have had to resort to very simplified models of the proteins. In fact, all models prior to QFold, use lattice representations for proteins. This means that the angles between amino acids can only be represented at right angles, 4 possible values. This representation is totally unrealistic and greatly limits the results obtained.
        
        One of the main innovations of QFold, along with the results obtained, is the totally realistic representation of proteins. QFold represents both phi and psi angles and does so at any value between 0 and 360, namely between $-\pi$ and $\pi$, with a user-defined discretization. This discretization is limited by the runtime capability of the devices, but can be as large or small as chosen, as the performance of QFold is maintained. Obviously, using a discretization with a larger number of states, better is the performance of the quantum algorithm versus the classical one because it increases the complexity of the problem.

        The main objective of QFold in the short term is to be able to refine the search done by another module. In the medium to long term, QFold can be considered as a stand-alone tool, but currently, classical algorithms have greater capabilities. Following this objective, the QFold architecture has two main modules, initializer and search algorithm.

        The initializer is a classic deep learning module. Right now it is a module that uses a simplified AlphaFold algorithm called MiniFold. However, this initializer could use any other module that is proven to perform a good first approximation of the protein structure. This first initializer module can be understood as a fast and low quality solution, which allows one to guide the quantum search to make it simpler. This type of mechanisms are very common classically when the search space is too large and a fast first solution can be obtained.

        The search module is an adaptation of the QMS algorithm to the protein folding problem. As in the previous cases, this module runs a QM-H circuit in which the energy values of the proteins are loaded and a search is performed until the minimum energy structure is found.

        To calculate the energy of each possible protein structure, using the evaluation function, the PSI4 software tool \cite{turney2012} was used. This tool receives a set of amino acids with a given structure and returns its energy. As in previous use cases, this energy calculation could be done by a quantum algorithm and integrated into the algorithm itself, but the capabilities of quantum devices do not allow it. 

        However, and unlike other use cases, this quantum algorithm of protein energy calculation is really complex and expensive to implement, being the development of these algorithms a line of research in itself. For this reason, the following part \ref{pt:qchem} is devoted to study the complexity of quantum chemistry algorithms with classical algorithms and their application in a use case.

        In order to validate QFold as a valid solution, it was executed in real quantum hardware. Due to the limitations, only a reduced size version of the problem was tested but it is useful as a proof-of-concept. This simulation was done using the quantum device IBMQ Casablanca with 7 qubits and quantum volume of 32. Around 20.000 circuits were run just to ensure that the results were valid, avoiding the noise. Besides, it was necessary to use a library, \textit{MITIQ} to mitigate the errores in the quantum hardware.

        \subsection{Results}\label{apt:res_qfold}

        \begin{itemize}[label=\textcolor{green}{\checkmark}]
        
            \item Application of QMS to a high impact use case where classical computing is very limited.
            
            \item Polynomial quantum advantage over the classical M-H algorithm in the protein folding problem.
            
            \item Quantum algorithm with potential to assist or refine the solutions of the classical algorithm.
            
            \item The first quantum algorithm that uses a realistic model in the protein folding problem, as opposed to quantum algorithms that represent the protein with lattice models.
            
            \item QFold architecture is fully modular for both the classical DL initializer and the quantum search module, allowing the use of different initialization and search algorithms.
            
            \item Execution of M-H quantum algorithm with different beta schemes that allow to study its evolution and behavior on an optimization problem.
            
            \item Test of the QMS algorithm adapted to a use case, called QFold, on a NISQ quantum chip to confirm the advantage results observed in the simulator.

            \item Inclusion of QFold into Amazon-AWS repository as a case of study. QFold was consired by AWS experts, as a valuable use case of quantum computing applied to the protein folding problem. It allows any user to use it for their own research.
            
        \end{itemize}
        
        \addcontentsline{toc}{subsection}{\numberline{}Publication P2. \textit{QFold: Quantum Walks and Deep Learning to Solve Protein Folding}}
        \includepdf[pages=-]{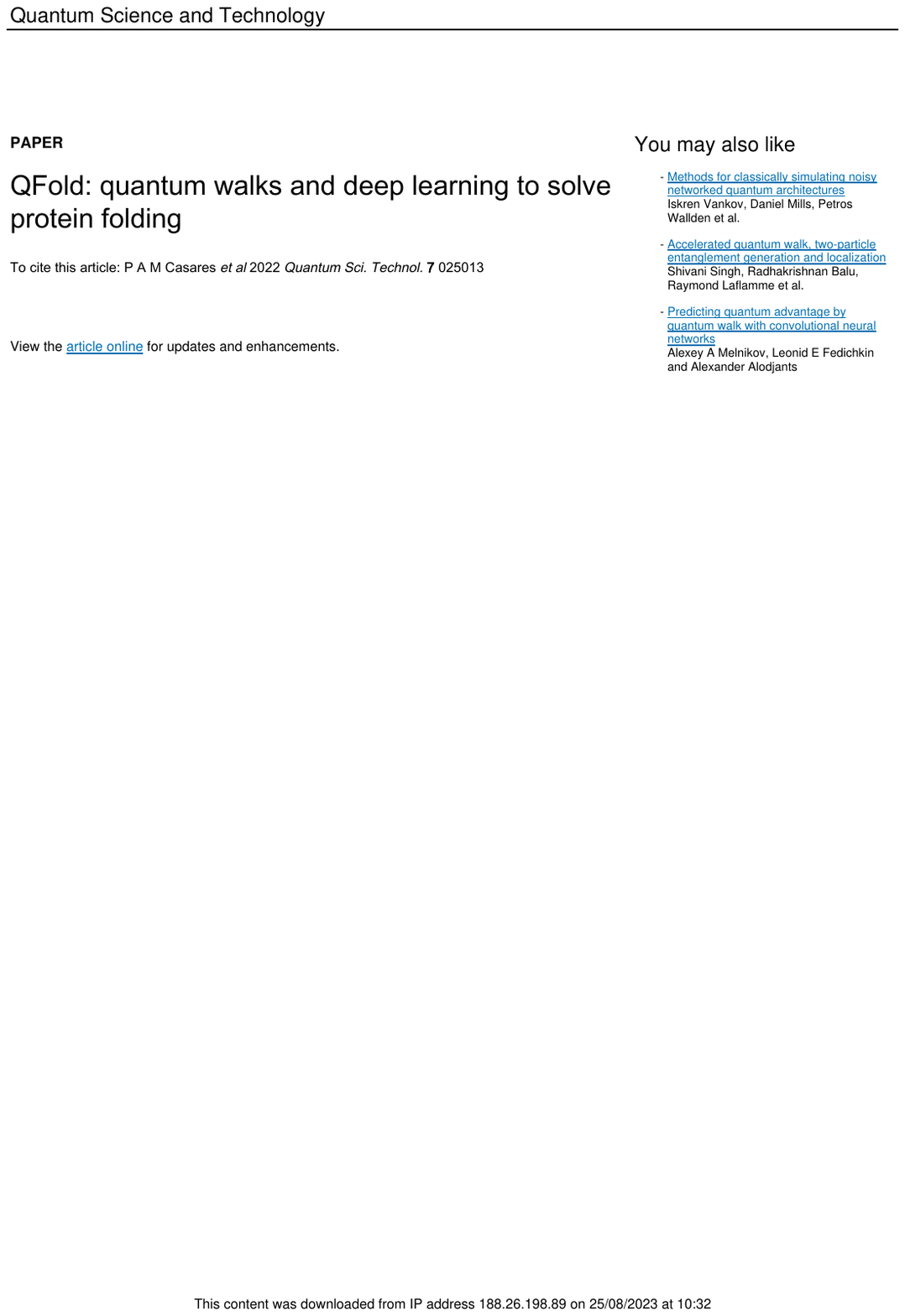}\label{pub:p2}

\part[Quantum Chemistry]{Part II\\Quantum Chemistry}\label{pt:qchem}
\thispagestyle{empty}
\vspace*{\fill}
\newpage\null\thispagestyle{empty}\newpage

\newpage

    \section{TFermion}\label{ch:tfermion}
    \thispagestyle{empty}

        \lettrine[lines=1, findent=2pt]{\resizebox{!}{1.2\baselineskip}{O}}{}ne of the processes in quantum chemistry that can have the greatest impact is to estimate the ground state energy of a quantum system (GSEE). The ground state is the lowest possible energy state in a physical system, according to quantum mechanics. This fundamental energy is of particular importance because it affects the overall behavior of the system and can be crucial to understanding its properties.

        To solve the GSEE problem, there are some methods to efficiently estimate the energy of a state. However, the problem arises because of the difficulty in preparing the ground state (GSP) of a quantum system. The GSP problem has been classically studied and some algorithms exist for it. However, these algorithms do not efficiently represent the information contained in a quantum system and, therefore, performing the evolution operations is so costly that simplifications far removed from reality are made. These simplifications are not able to produce reliable results, making both GSP and GSEE classically very limited.
        
        Quantum computing has two fundamental advantages in quantum chemical algorithms. First, it allows a direct and realistic representation of quantum states without approximations. Second, a QC algorithm inherently simulates evolution. These two advantages are automatic because of the nature of quantum algorithms but, in addition, there may be other advantages related to state space exploration or data set processing. These advantages convert quantum chemistry in a solid candidate for quantum advantage. These advantages position quantum chemistry as a strong candidate for achieving quantum advantage. Nonetheless, it is crucial to investigate the quantum gate requirements of quantum chemistry algorithms to assess the current capability of quantum devices to execute them. For that purpose, TFermion library is presented. 

        \subsection{Introduction to GSP and GSEE}\label{apt:intro_gsp}
        One of the fundamental purposes of GSEE is to understand and predict the behavior of quantum systems, such as molecules in quantum chemistry or materials in condensed matter physics. The ability to estimate the ground state energy is essential for practical problems, such as the simulation of molecules for the design of new drugs or materials, where understanding the ground state is crucial to understanding their properties and behaviors.
        
        As seen in other sections, GSP and GSEE algorithms can combine classical and quantum modules to create solutions that are executable in the near term on the quantum devices at hand. However, while this can be done, it is not a mainstream development in the scientific community. This is because the advantage of quantum algorithms for GSEE is very large. The effort is more focused on getting algorithms that require fewer and fewer resources than on combining them with other classical ones. In fact, if the concept of hybrid algorithm exists, it is simply because the objective is to find a quantum advantage of this type of algorithms as soon as possible. An example of hybrid quantum chemistry algorithms that will be discussed later is the Variational Quantum Eigensolver (VQE) that served as the basis for the variational algorithms explained in chapter \ref{ch:qaiSS}.

        In this thesis, a different approach is applied to GSEE quantum algorithms. This approach is not intended to result in a new algorithm, neither quantum nor hybrid. The goal of this new approach is to accelerate the development process of quantum algorithms and to better understand their performance. 
        
        In this work, a classical computing software library has been developed that gives information about the cost of several GSEE quantum algorithms to compare them with each other and to understand which techniques are better. Basides, to give information about the needs of quantum hardware to be able to execute these algorithms. All this has resulted in the creation of a software tool for measuring the performance of GSEE quantum algorithms called TFermion.

        To explain how TFermion works and to show a practical example of its usefulness, as a use case, it is necessary to understand some intermediate concepts. First, the most important classical algorithms for GSP and GSEE will be briefly explained. Then, it is necessary to explain some Hamiltonian simulation techniques, which is the key process to represent the molecule and various forms of representation needed. Finally, some quantum algorithms for GSEE based on the representations explained above will be reviewed.


        As discussed above, classical algorithms for GSP and GSEE make sense as long as there are no quantum algorithms that can be run on real devices. Still, some of these classical algorithms and their approximations have served as a fundamental basis for other quantum algorithms. The main limitations of classical quantum chemistry algorithms are found in the representation of states, especially when they need to describe in great detail the behavior of some particles with quantum effects. Of this type, three theoretical descriptions stand out: Hartree-Fock (HF), density functional theory (DFT), and coupled cluster (CC).

        The Hartree-Fock (HF) method is an approximate form of the quantum mechanical equations for fermions. HF method uses equations based on single-particle orbitals that are more computationally accessible than methods based on many-particle wave functions. It takes into account the interaction between electrons and the influence of their electromagnetic fields.

        The HF method receives as input the electronic Hamiltonian and returns the molecular orbitals and their energies. It is an iterative method until convergence of the orbital values. It is important to note that the Hartree-Fock method is an approximation, it does not take into account the complete electronic correlation. Although it is an approximation, it has been useful as a basis for developing more sophisticated methods to describe molecular systems with greater accuracy.

        Density Functional Theory (DFT) \cite{parr1983, orio2009} is a theoretical framework in quantum chemistry used to describe the electronic structure of atomic and molecular systems. Unlike the HF method seen above, it does not deal directly with electron wave functions but focuses on electron density as the fundamental quantity.

        In DFT, the electron density represents the probability of finding an electron at a particular point in space. Upon that, DFT tries to minimize the total energy of the system by varying the electron density. DFT is based on the Kohn-Sham equation, which decomposes the system into non-interacting dummy electrons, called Kohn-Sham electrons, moving in an effective potential. The main limitation of DFT is that due to the computational complexity of calculating the quantum effects between the electrons, approximations are required that cause the system to lose accuracy.
        
        DFT has proven to be a powerful and efficient tool for describing a variety of molecular and solid systems. It is widely used in theoretical chemistry and materials simulation, and has contributed significantly to the understanding of electronic structures and chemical properties.


        The basic idea of the Coupled Cluster procedure is to expand the wave function as excitations of the HF state, which is an approximation to the fundamental state of a quantum system. The method employs cluster excitation operators that act on the reference state to create electronic excitations. These cluster operators are decomposed in terms of single-electron excitations (singles), double excitations (doubles) and so on. The cluster wave function is expanded exponentially using the cluster operator. This means that higher order terms take into account the interactions between multiple electrons.

        CC is known for its high accuracy, especially in systems where electronic correlation is crucial. However, it is also computationally intensive and its practical use is limited to systems of moderate size due to the complexity of the calculations. This method is also the basis for the Unitary Coupled-Cluster ansatz used for the Variational Quantum Eigensolver.
        
        \subsection{Hamiltonian simulation, representation, mapping and encoding for GSP and GSEE}\label{apt:repre}

        In order to design quantum algorithms to simulate the evolution of the quantum chemistry system to be evaluated, a definition of that evolution is necessary. One such way of defining evolution is Hamiltonian simulation. This type of simulation gives the algorithm the ability to simulate the time evolution of a system described by a given Hamiltonian. This simulation makes it possible to study how the quantum system evolves in time.

        By describing a quantum system with its Halmiltonian and being able to use it to describe a behavior, a more detailed study of its properties is performed. This results in an advantage in accuracy. In addition, quantum computers allow Hamiltonian simulation in a natural way, so this is another aspect in which a quantum advantage is expected to be obtained in a practical problem.

        The Trotter-Suzuki decomposition was the first method proposed to implement a Hamiltonian simulation. It is based on the decomposition of the time evolution operator into a series of more manageable operators. The main idea is to divide the time into small steps and express the time evolution as a product of simpler operators. 
        
        In this way, it is possible to apply the simple operators separately and then combine the result. This algorithm tries to make a trade-off between accuracy and performance. In such a way that higher precision implies lower performance, higher complexity of execution. The Trotter-Suzuki algorithm can be generalized to include higher order terms, improving the accuracy of the approximation. However, as the order increases, the computational complexity also increases.

        Trotter's method is the most widespread and widely used method for Hamiltonian simulation. However, it has some points that can be improved such as the scalability of its error. For this purpose, methods known as post-Trotter have emerged. One of them is the Taylor series decomposition for Hamiltonian simulation.

        The terms of the Taylor series can be realized by introducing an additional superposition and conducting controlled operations. The temporal evolution is divided into segments, each sufficiently short to allow for an accurate approximation of the evolution using a certain number of terms in the Taylor series. Each segment is then performed using oblivious amplitude amplification \cite{brassard2002}.

        Another post-Trotter method is qubitization. Qubitization refers to the expression of the time evolution operator in terms of qubits. The key idea behind this algorithm is to find an efficient way to express the time evolution operator in local terms of quantum operations that can be efficiently implemented in a quantum circuit. 

        An example of representation, in the context of molecule simulation, is that the Hamiltonian terms can represent interactions between atoms or between electrons and nuclei, and the qubits represent the occupation of an orbital in a molecular simulation. Once the quantum system has been mapped to qubit space, other algorithms such as the Trotter-Suzuki algorithm can be applied. The main advantage of qubitization is that it is an errorless Hamiltonian simulation, but it is costly for a quantum computer.

        In order to reduce the complexity of the implementation of the qubitization method, the coupled cluster method was developed. In this method, the time evolution is divided into two parts: the free part and the interaction part. 

        The free part describes the time evolution of the system without any interaction, using an unperturbed Hamiltonian. The interaction part describes how the system interacts, and its time evolution operator is constructed by combining the free part and the interaction part. This approach facilitates the treatment of complex systems by dividing the problem into more manageable parts.
        
        In order to run the GSP and GSEE quantum algorithms, it is necessary to use the Hamiltonian describing the evolution, as seen above. In addition, it is necessary to represent the problem, with first or second quantization, to make a mapping between operators and qubits, by means of Jordan-Wigner or Bravyi-Kitaev, and to make a choice of a set of basis functions to generate the wave function, which can be gaussian functions or plane waves.
        
        The representation of the states in a quantum circuit is one of the key decisions when implementing an algorithm. Classically, it is already an important decision because it largely determines whether the algorithm is valid or not; quantumly, it becomes even more important because of the need to optimize the number of qubits and operations. In this case, it should be discussed how to use the logical qubits, noise-free and fully connected to each other, to represent the possible quantum states of the system.

        The first option is to use the first quantization. This representation indicates the state of each electron, i.e., for each electron it indicates in which orbital it is located. It is a compact representation because it only requires a number of qubits proportional to the number of electrons and the logarithm in base 2 of the number of orbitals. However, operating in this quantization introduces a certain computational overhead, which slows down the calculation. 

        The second option is to use the second quantization. This represents the occupation of each orbital, i.e., in each orbital it is indicated how many electrons there are. This representation requires a number proportional to the number of orbitals in the system. If the number of orbitals is much larger than the number of electrons, the second quantization is much less efficient than the first. However, this representation does not require any additional calculations, so it is more efficient in performing operations.

        To simulate the quantum system, once the representation has been chosen, it is necessary to map the system to the operators. In other words, designate a set of qubit operators (matrices) which satisfy the canonical anticommutation relations. In order to map fermions to qubits, and depending on the chosen quantization, there are two main methods to perform the mapping, Jordan-Wigner or Bravyi-Kitaev.

        The Jordan-Wigner mapping is a theoretical method that uses second quantization to transform spin operators into fermionic creation and annihilation operators. This operation transforms ``up'' spins into fermions or occupied states, and ``down'' spins into unoccupied states. The cost of Jordan-Wigner mapping and parity mapping is $\mathcal{O}(N)$, where $N$ is the number of qubits \cite{tranter2018}.

        Alternatively, the Bravyi-Kitaev method ensures that the mapping can be done at a complexity equivalent to the logarithm in base 2 of the number of qubits. Balances the locality of the occupancy and parity information to improve the efficiency of the simulation. The qubit stores the parity of the set of occupancy numbers corresponding to that set of orbitals \cite{seeley2012}.

        Once the representation and mapping methods were explained, it is necessary to define the set of basis functions to generate the wave function. The basis set can either be composed of atomic orbitals (yielding the linear combination of atomic orbitals approach), which is the usual choice within the quantum chemistry community; plane waves, which are typically used within the solid-state community; or real-space approaches. Several types of atomic orbitals can be used: Gaussian-type orbitals are by far the most often used, as they allow efficient implementations of post-Hartree–Fock methods.

        In molecular calculations, a common practice is to utilize a basis comprising atomic orbitals centered at each nucleus within the molecule (linear combination of atomic orbitals ansatz). The most physically motivated basis sets are Slater-type orbitals (STOs), which are solutions to the Schrödinger equation for hydrogen-like atoms. However, computing integrals with STOs poses computational challenges, leading to the approximation of STOs as linear combinations of Gaussian-type orbitals (GTOs). Numerous Gaussian-type orbital basis sets have been documented in the literature \cite{basch1969}, often organized in hierarchies of increasing size to provide a controlled approach for obtaining more accurate solutions, albeit at a higher computational cost.

        The alternative to Gaussian basis functions is the use of plane waves, which are better suited for periodic materials because of their periodicity. Generally, the selection of the plane wave basis set is based on a cutoff energy. Plane waves within the simulation cell that satisfy the energy criterion are incorporated into the calculation. Moreover, certain integrals and operations are more straightforward to program and execute with plane-wave basis functions compared to their localized counterparts. The number of plane waves is typically around 100 times larger than the Gaussian functions, as recommended in Appendix E of \cite{babbush2018}.

        Now that all the elements necessary to run a quantum algorithm based on the Quantum Phase Estimation (QPE) technique for GSP and GSEE have been explained, some of these algorithms will be briefly mentioned. Alternatively, the Variational Quantum Eigensolver (VQE) algorithm will also be reviewed because it is an alternative, although less efficient, method to the QPE-based algorithms.

        The concept of variational algorithms for QML problems arises from the creation of a variational algorithm applied to quantum chemistry, the VQE. The Variational Quantum Eigensolver (VQE) is built upon the concept that it is possible ton prepare the ground state by variationally determining a state that minimizes the ground state energy.

        In the variational algorithms scheme, a quantum ansatz and a classical optimizer are needed. In VQE, the classical optimizer is still a gradient descent, and the original quantum ansatz was Unitary Coupled-Cluster. Later, other ansatz derived from the original such as Qubit Coupled-Cluster was chosen. In VQE, as in all variational algorithms, the selection of the ansatz is key for the algorithm to find a solution close to the optimum. Otherwise, the barren plateaus phenomenon appears and taking a step requires an exponential number of intermediate operations.

        The Quantum Phase Estimation (QPE) technique allows to know the ground state of a quantum system and its energy with greater precision than other classical and quantum techniques. However, it is very expensive to implement. Some methods that implement it and that have been analyzed in the TFermion software are: Sparsity low-rank \cite{berry2019}, Linear T \cite{babbush2018b} or Taylor naive \cite{babbush2016}.

        \subsection{TFermion library}\label{apt:tfermion}

        As seen above, quantum algorithms applied to chemistry offer great advantages over classical algorithms. In fact, the only reason why they are not currently used is because there are no quantum devices large enough in number of gates to be executed. Therefore, it is interesting to know what is the real cost in gates of the execution of certain algorithms and to understand if this cost can be reduced by introducing certain optimizations.

        In order to obtain detailed data from the analysis of the main state-of-the-art quantum algorithms of GSP and GSEE, a classical software tool has been created following the open source scheme called TFermion. This tool contains the implementation of the T-gate cost of the analyzed algorithms. In this way, the user enters a molecule to be analyzed and the method among those available, and the software returns the cost in T gates.  

        This gate cost is calculated according to the formulas detailed in the subsection \ref{apt:intro_gsp} and an optimized margin of error according to parameters that can be set by the user in the configuration file. In addition, there are other parameters that can be adjusted, such as the basis functions, the conversion between plane and Gaussian waves, the chemical accuracy, etc.

        The effort in quantum computing is divided between an attempt to create algorithms with practical quantum advantage and the ability to run those algorithms on quantum hardware. That leads to lines of research that are more focused on running the algorithms even though there is noise on NISQ-type devices and trying to improve those algorithms. On the other hand, there are lines more focused on developing error correction and fault-tolerant computers. This dichotomy provokes the discussion of how far it is to run certain quantum algorithms on fault-tolerant computers or, seen in another way, what resources a quantum device must have to be suitable for running quantum algorithms.

        In that vein, TFermion focuses on making a reliable estimate of how many quantum bridges are needed to run the quantum algorithms that are best positioned to achieve practical quantum advantage, the quantum chemistry algorithms. Software tools such as TFermion are needed to size resource needs, compare lines of research and make technology estimates. Right now, in quantum computing, the quality of the solution, in this case the accuracy with which it simulates a quantum chemistry system, is almost as important as the resources used for it. A high-precision solution may be so expensive to run that it makes no sense to continue its development, or even if it is inexpensive with simple molecules, its scalability is abysmal. All of these metrics can be applied using TFermion.

        TFermion can be useful when developing new algorithms and applying them to specific problems. For example, a scientist who wants to implement a quantum algorithm to obtain the ground state energy of a molecule will be able to decide which encoding or which mapping to use based on the results of other algorithms applied to that molecule. Once implemented, he will be able to see how his method compares with the state-of-the-art and know where he has advantages and where he needs to improve. Before TFermion, this possibility did not exist.

        In order for TFermion to have continuity and remain valid as research progresses, it has been developed and published following the open-source philosophy. Thus, it is publicly available on GitHub. It is a modular software that is easy to update. Thus, when a scientist develops a new algorithm and wants it to be in TFermion, he can create an issue and ask for it to be included or, directly, include it in the code. Then, when the user goes to run TFermion, he will see that the list of available methods has been increased.

        One of the TFermion use cases to be analyzed in the next chapter \ref{ch:cars} is the application of the library to the development of electric batteries. In this case, TFermion allows for a detailed analysis of the cost of the algorithms for a specific problem, of great complexity and with numerous industrial applications. As before, a specific algorithm for the development of a system with quantum properties is analyzed.
        
        In order to follow a fully modular philosophy and to facilitate external collaborations on the TFermion code, each implemented method has been split into different python files. In this way, if a scientist wants to include his own algorithm, he only has to create a concrete class that, respecting a basic structure, calculates the count of T gates of his algorithm. To do this, he must analyze the cost of each of the steps required for its implementation. The extraction of the system configuration to be analyzed, the error optimization, and other technical aspects are automatically performed by TFermion. Once the code is implemented, an administrator is asked to verify it, and, if approved, the new quantum algorithm is incorporated into TFermion.

        The gate count for TFermion must be expressed in T gates. T-gates are single-qubit gates that, in combination with other gates, can do what is called universal quantum computation. Therefore, a given result in T gates is as general as possible and serves as an upper bound and can be optimized by using other gates. To understand this, it is necessary to explain some concepts beforehand.

        DiVincenzo's criteria \cite{divincenzo2009} assert that for a quantum computer to be scalable, it must demonstrate the ability to execute a universal set of quantum gates. This comprehensive set should encompass all the gates essential for conducting any quantum computation, ensuring that any computation can be expressed as a finite sequence of these universal gates. Fundamentally, a quantum computer, as per these criteria, should not only manipulate individual qubits through single-qubit gates, but also introduce entanglement in the system, a functionality achieved through multi-qubit gates.

        Following this definition, two sets of gates can be distinguished, Clifford and non-Clifford. The Clifford gate set consists of Hadamard, phase S, and CNOT gates. This gate set is considered minimal because excluding any one gate renders the inability to implement certain Clifford operations. Specifically, omitting the Hadamard gate eliminates the capability to represent powers of $\frac{1}{\sqrt{2}}$ in the unitary matrix, excluding the phase gate disallows the presence of $i$ in the unitary matrix, and removing the CNOT gate restricts the controlled operations. Given that all Pauli matrices can be constructed from the phase and Hadamard gates, each Pauli gate is inherently an element of the Clifford group.

        In addition, a quantum circuit consisting only of gates of the Clifford ensemble can be simulated efficiently. Therefore, to achieve quantum advantage it is necessary to use gates of the non-Clifford set, such as the T-gate. 

        Therefore, measuring the complexity of running an algorithm on T-gates better represents the bottleneck that exists when running it on quantum hardware. Clifford type gates are more dependent on the topology of the chip and therefore a more variable result. Another important point is that the real limitation of today's quantum computers is not in the number of qubits they have, but in the number of gates, circuit depth, that they can run in quantum coherence and without noise effects.

        The gate count performed by TFermion is exclusively adjusted to the operations performed by the quantum algorithm to calculate the ground state energy, leaving aside auxiliary operations such as ground state preparation or the execution by some algorithms of quantum Fourier transform techniques.

        Each of the algorithms analyzed in TFermion has a set of parameters that can be varied to obtain the final gate cost. The algorithms allow a chemical accuracy value to be set at which the energy is calculated. In this way, certain parameters of the algorithms can be optimized so that, although the result is always within the desired accuracy, the cost of running the algorithm in gates is minimized. For this purpose, TFermion uses different optimizers that allow to calculate the minimum gate cost T of the algorithm when it is executed with a given accuracy.
        
        In order to test the results of TFermion with a molecule that was used as a standard by other methods and to compare their results, FeMoCo was chosen. It is a molecule with well-known and widely studied properties. Since other authors have also analyzed the cost of applying some of the TFermion algorithms on FeMoCo, this molecule can also be used to validate the TFermion results. FeMoCo was the molecule used to compare the cost of the quantum algorithms used in TFermion.

        \subsection{Results}\label{apt:res_tfermion}

        \begin{itemize}[label=\textcolor{green}{\checkmark}]
            \item The use of QROM techniques in the plane wave naïve Taylorization method makes it particularly efficient, and there are indications that employing plane waves could be more effective than Gaussian for the same Taylorization techniques in isolated molecules.

            \item TFermion has been created as a software tool to help make decisions about quantum algorithms for quantum chemistry simulations.

            \item The main state-of-the-art quantum algorithms for GSP and GSEE have been analyzed, making an exhaustive comparison between them.

            \item A quantitative comparison, in terms of implementation cost, between different encoding, mapping and representation methods has been established.

            \item A line of hybrid algorithms has been developed following the model of using classical computation as benchmark and assistance of quantum algorithms.

            \item It has been shown that the most promising Hamiltonian simulation methods are, on the one hand, to use qubitization, gaussian basis and rank factorization and, on the other hand, Qubitization or Interaction Picture with a plane wave basis and an encoding in first quantization.

            \item Algorithm implementation costs have been provided to understand the gate count requirements of quantum algorithms, which is useful in guiding the development of fault-tolerant quantum devices.

            \item As shown in figure 3 of \ref{pub:p3}, using TFermion different molecules have been analyzed to each of the studied methods. Initially, simple molecules such as $H_2$ and $HF$ were analyzed just to test the algorithm. Then, more complex molecules such as $NH_4$, $CO_2$ or $NaCl$ were analyzed to show that TFermion is useful.
            
        \end{itemize}
        
        \addcontentsline{toc}{subsection}{\numberline{}Publication P3. \textit{TFermion: A non-Clifford gate cost assessment library of quantum phase estimation algorithms for quantum chemistry}}
        \includepdf[pages=-]{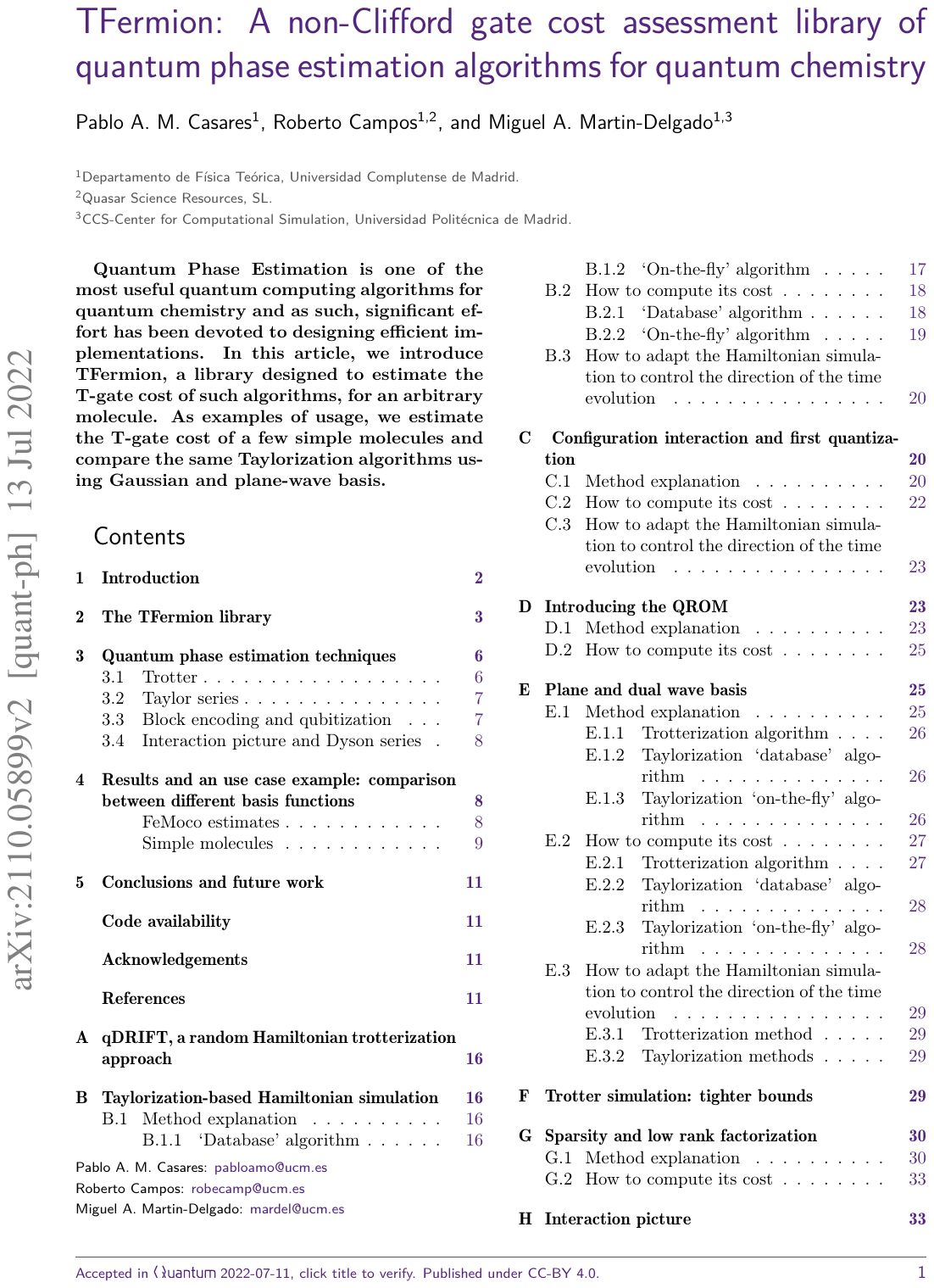}\label{pub:p3}

\newpage\null\thispagestyle{empty}\newpage

    \newpage
    \section{Applications to Electric Batteries Design}\label{ch:cars}
    \thispagestyle{empty}

        \lettrine[lines=1, findent=2pt]{\resizebox{!}{1.2\baselineskip}{T}}{}he electric battery industry includes any process related to the manufacture, marketing, use or repair of a battery. This chapter is especially focused on manufacturing. A key process in the manufacture of an electric battery is the selection and design of its materials. For this, simulation of the behavior of materials is a key process and can be achieved using quantum chemistry algorithms. 

        Problems such as the design of stronger, lighter, or cheaper materials, the chemical reaction to achieve propulsion, or the architecture of new electric batteries with better properties are concrete examples of the challenges faced by manufacturing companies. These are problems of enormous complexity in which the most advanced computational techniques must be used to obtain results. For this reason, these companies always look to the research field with great interest in the new techniques it can offer them. These techniques include high-performance computing, deep learning, and quantum computing. 

        One of the main questions companies in the electric battery sector, as in other sectors, are asking themselves is when the technology needed to run quantum computing algorithms will become available. The interest is twofold, first because they want to make breakthroughs of their own, second because no company wants to lag behind its competitors' advances. At this point, TFermion can become a fundamental tool because it allows estimating the quantum hardware resources needed to run high-impact algorithms in battery design problems.

        In many cases, for companies, it is not so important what is created, but rather when it is created and what is needed to create it. In order to get ahead of the future era of quantum computing battery design, companies are interested in sizing the amount of resources that running quantum algorithms can cost them. This can be used to make strategic decisions in the medium term, based on whether or not quantum algorithms will exist with practical advantage, deciding whether to purchase a quantum computer, depending on the amount of resources needed, it may be more efficient to outsource a quantum computer to the cloud, etc.

        \subsection{Introduction to electric batteries design}\label{apt:intro_cars}
        
        The current situation in electric battery manufacturing is especially changing with the arrival of many electric devices, as laptops, smartphones, autonomous vehicles, etc. For example, an electric car is a major paradigm shift, not only for the car itself, but also for its needs. For example, if the vehicle fleet is mostly electric, gas stations will see their business dwindle and will have to convert to charging points. Therefore, any tool that provides information on the timescale of this change is very useful.

        The optimization of lithium-ion battery performance is pivotal for the advancement of next-generation energy storage systems. Progress in this field relies not only on the discovery of novel materials but also on the refinement of simulation methods to accurately characterize key properties of lithium-ion batteries. The spectrum of properties influencing their performance is diverse and encompasses mechanical and electrochemical attributes, thermal stability of the cathode, the electrochemical windows of the electrolyte, formation of the solid-electrolyte interphase, and ionic mobility, among others. This chapter is based on publication \cite{delgado2022}.

        Today, the limiting factor in the deployment of electric vehicles is their range, that is, the number of kilometers they are able to travel without recharging the battery. Although this measure depends on numerous factors, such as aerodynamics, vehicle weight, driving style, tires, etc., the factor with the greatest impact is the capacity of the electric batteries. These batteries have been increasing their capacity to have autonomies that range from 400 to 800 km. However, it is expected that this range will be much greater with the development of technology.

        To increase the capacity of an electric battery, it must be possible to simulate the chemical processes that occur inside the battery. Understanding the optimum structure of the anode and cathode, how to achieve the best storage and discharge rate, or how to avoid battery degradation are fundamental processes for extending the autonomy and life of these batteries.

        All these processes have in common the need to use a GSP and GSEE algorithm. As seen above, the classical algorithms that do these tasks have low accuracy due to their limited scalability. Therefore, quantum algorithms have become the most promising solution in the medium term. There are numerous approaches to accelerate the development of new algorithms, TFermion proposes to study their cost and compare the techniques that implement each of them to obtain certain conclusions.

        Following this motivation, this paper reports the collaboration with the start-up Xanadu and the company Volkswagen for the application of TFermion to the study of quantum algorithms that allow to know and design with detail and precision electric batteries.

        \subsection{Methodology}\label{apt:metho_cars}

        The task of simulating all the processes that occur in an electric battery is too complex, not to mention that there are numerous technologies for creating electric batteries. In this case, the work is focused on the simulation of a lithium-ion battery, and the study focused on the cathode.

        The unit cell material, Li2FeSiO4, is studied because it is a common candidate material for the cathode of batteries. Other properties such as the ionic mobility inside the cathode, and the thermal stability of the material, can also be predicted from the ground state energy of the different phases of the material. In summary, this work is focused on the simulation of cathode materials, which is crucial for predicting important properties of a battery cell.

        The battery cell comprises a positive electrode (cathode) and a negative electrode (anode), separated by a porous membrane (separator) and immersed in an ion-conducting material (electrolyte). Chemical reactions occurring at the electrode-electrolyte interface drive the conversion of chemical energy into electrical energy. During discharge, an oxidation reaction at the anode generates electrons and lithium ions. The electrons flow through an external circuit, while the lithium ions migrate through the electrolyte and intercalate into the cathode material. Conversely, during charging, an external voltage is applied to reverse this process. The lithium ions are extracted from the cathode, transported back through the electrolyte, and intercalated into the anode material.

        In this chapter, three fundamental aspects of electric batteries are investigated: the equilibrium cell voltage, the ionic mobility, and the thermal stability of the cathode material. These aspects are critical for enhancing the capacity of lithium-ion batteries and are instrumental in studying the electronic structure of cathode materials.

        The equilibrium voltage plays a crucial role in defining the energy storage capacity of a battery relative to its volume (energy density) and weight (specific energy). It represents the average voltage (V) of an electrochemical cell, which converts chemical energy into electrical energy. The Nernst equation governs the equilibrium voltage, taking into account factors such as the number of charges transferred, the Faraday constant, and the change in free energy associated with the cell reaction \cite{feiner1994}.

        The chemical diffusivity (D) serves as a key parameter characterizing the mobility of lithium ions within a material. In scenarios where diffusion mechanisms remain independent of temperature, a microscopic model can effectively illustrate the process of lithium ion hopping from its initial position to an adjacent vacant site within the crystal structure of the host material.

        Numerous processes contribute to the degradation of battery performance over time, encompassing the formation of the solid electrolyte interphase, degradation of cathode active materials, lithium plating on the anode, and the growth of lithium dendrites, among others \cite{leng2017}. Simulating these phenomena poses a significant challenge, as it requires a bottom-up approach spanning from the atomic level to the macroscopic scale \cite{hausbrand2015}.
        
        Anticipating the thermal stability of cathode materials is crucial for enhancing the safety of lithium-ion batteries, particularly as they can become unstable in their charged state. The removal of more lithium ions from oxide-based cathode materials can lead to their degradation into other phases of the material \cite{wang2007}.
    
        At present, first-principles calculations of the electronic structure of cathode materials are largely performed using density functional theory (DFT) methods \cite{urban2016}, explained in chapter \ref{ch:tfermion}. This chapter is dedicated to exploring methods suitable for calculating the ground-state energies of cathode materials, a crucial parameter for simulating battery properties, specifically quantum phase estimation algorithm. A quantum phase estimation (QPE) algorithm has been used with simulation of the Hamiltonian by qubitization, plane waves to represent the quantum state, and first quantization to encode the system. 

        The quantum algorithm begins by taking the Hamiltonian, which describes the interactions among electrons within the material's unit cell, as input. It then generates an approximation of the smallest eigenvalue of this Hamiltonian, which corresponds to the ground-state energy. To facilitate this process, a method for representing and constructing Hamiltonians is necessary, tailored specifically to the quantum algorithm.

        Studying the electronic Hamiltonian is crucial to understand why employing a first-quantization approach in a plane-wave basis is well-suited for simulating battery materials \cite{babbush2018}. This approach allows for an effective representation of the electronic structure and interactions within the material, making it particularly suitable for quantum simulations.

        However, preparing an approximate ground state to serve as input for the quantum phase estimation algorithm presents challenges, especially in the case of periodic materials and within the context of first quantization. Careful consideration is required to identify suitable methods for achieving this task effectively.
        
        Once the ground state is prepared, the qubitization formalism comes into play, enabling the encoding of the Hamiltonian into a suitable unitary transformation \cite{low2019}. This step is essential for implementing the quantum phase estimation algorithm.
        
        To analyze the overall complexity of the algorithm and ensure compatibility with fault-tolerant architectures, TFermion is employed. TFermion facilitates the compilation of all necessary operations into a universal set of quantum gates, ensuring the algorithm's feasibility and scalability on quantum hardware.

        Plane waves offer a well-suited framework for studying periodic systems, providing concise representations of Hamiltonians. However, achieving high accuracy often demands a significant number of plane waves, resulting in an impractically large number of qubits in second quantization. To address this challenge, first-quantization techniques are favored for materials simulation. This approach leverages qubitization-based quantum phase estimation algorithms, specifically designed for first-quantized Hamiltonians represented in a plane-wave basis.

        The atomic configuration of a cathode material is characterized by its unit cell, comprising a collection of atoms that can be spatially translated to cover the entire crystal. To determine the material's electronic structure, the Schrödinger equation is solved within the unit cell, subject to periodic boundary conditions. While any comprehensive set of basis functions can represent the first-quantized Hamiltonian $H$, for periodic systems, employing plane waves with the lattice's periodicity emerges as a fitting selection.

        In the quantum phase estimation algorithm, achieving an input state with substantial overlap with the actual ground state is crucial. In many quantum chemistry algorithms, including quantum phase estimation, this is accomplished by generating a state of non-interacting electrons characterized by single-particle wave functions, or orbitals, which are optimized using the Hartree-Fock method.

        The situation is more complicated when studying periodic materials in first quantization using a plane-wave basis. Here it is necessary to apply the Hartree-Fock method to extended materials and provide an algorithm to prepare the resulting Hartree-Fock state in a plane-wave basis, which must be explicitly antisymmetrized.

        Studying periodic materials in first quantization using a plane-wave basis presents a more intricate scenario. In this context, applying the Hartree-Fock method to extended materials becomes necessary. Moreover, an algorithm is required to prepare the resulting Hartree-Fock state in a plane-wave basis, with explicit antisymmetrization.

         For the Quantum Phase Estimation (QPE) algorithm is necessary to select a method to encode the Hamiltonian into an appropriate unitary operator. The qubitization method stands out due to its ability to achieve exact implementation without any approximation. Quantum phase estimation, executed on the qubitization operator, relies on operators of preparation, $PREP_H$, and selection, $SEL_H$. The compilation strategy for implementing these operators is derived from \cite{su2021}.

        Unlike the methods analyzed in TFermion, for this particular use case, the cost of preparing the ground state approach has been taken into account. This cost has been divided into cost in T-gates, Toffoli gates and an estimate of the algorithm execution time in quantum hardware.

        To carry out this study, it has been necessary to make slight modifications to the TFermion library. First, the combination of techniques to be used for the simulation was included in a concrete way. In addition, new performance metrics, Toffoli gates, and execution time have been included. Finally, the code has been optimized to include a calculation of the cost of the algorithm for a molecule of a higher complexity than those analyzed in the TFermion work.

\begin{figure}[H]
            \centering
            \includegraphics[width=1\textwidth]{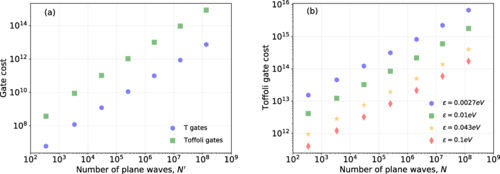}
            \caption{\label{fig:gate_batteries} Non-Clifford gate cost for initial state preparation and quantum phase estimation. (a) The non-Clifford gate cost due to Givens rotations used in the circuit for initial state preparation. (b) Toffoli gate cost of the quantum phase estimation algorithm. All calculations are done for the unit cell of Li2FeSiO4 with 156 electrons. The total number of qubits is 2,375 for np = 4 and 6,652 for np = 9. In the right figure it only depicts Toffoli gate count, as the number of T gates is much smaller (< 3 × 105). The total error includes contributions from different approximations throughout the algorithm, but it does not take into account the error derived from a finite basis set. The slope of the Toffoli gate cost for fixed target precision is a consequence of the leading cost term. These calculations were performed with the T-Fermion library.}
            
        \end{figure}

        Figure \ref{fig:gate_batteries} illustrates the relationship between the overall cost of the algorithm, quantified by the number of Toffoli gates, and varying parameters such as the number of plane waves (N) and the error in estimating the ground-state energy. Given the discrete nature of qubit requirements for representing the quantum state, which directly influences the Toffoli gate cost (set at 100), it is advisable to select a number of plane waves (N) corresponding to integer values of np.

\begin{figure}[H]
            \centering
            \includegraphics[width=0.6\textwidth]{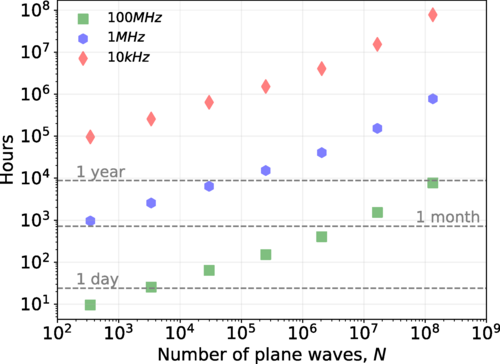}
            \caption{\label{fig:time_batteries}Estimation of the time required to run the algorithm. This figure illustrates total runtime for synthesizing all the Toffoli gates indicated in figure \ref{fig:gate_batteries}. All calculations are done for the unit cell of Li2FeSiO4 with 156 electrons, and it is assumed that the number of plane waves used in the state preparation and quantum phase estimation are the same. The total number of qubits is 2,375 for np = 4 and 6,652 for np = 9. The distillation time is computed as the product of the number of Toffoli gates, the surface code distance d, and the clock frequency, all divided by a small np factor that parallelize the CSWAPs and arithmetic computations. It is emphasized that these are rough estimates whose main purpose is to provide a method to interpret the gate cost.}
        \end{figure}

        The insights from figure \ref{fig:time_batteries} shed light on the computational demands of executing a quantum algorithm, particularly in relation to precision requirements and clock frequency. At a modest clock rate of 10 kHz, even with a low precision of np = 4, the estimated runtime extends over several years. Conversely, with an optimistic but feasible clock rate of 100 MHz, the runtime for np = 4 is reduced to less than a day. However, for np = 9, the estimated runtime extends to approximately a year. It may be prudent to limit the number of plane waves to N = $10^6$, corresponding to np = 7, as this adequately captures the system's accuracy even in large basis sets like cc-pV5Z. With this choice and a 100 MHz clock rate, the runtime is projected to be approximately a few weeks.

        \newpage

        \newpage

        \subsection{Results}\label{apt:res_cars}

        \begin{itemize}[label=\textcolor{green}{\checkmark}]

            \item Application of TFermion to a highly complex industrial problem, with results that help to learn more about the needs of quantum devices.

            \item Extension of the TFermion library to obtain two new metrics, Toffoli gates and execution time. In addition, the analysis of ground state preparation cost has been included.

            \item Analysis of how quantum chemistry algorithms should be improved to approach the capabilities of quantum hardware.

            \item Demonstration that the resources required to run Quantum Phase Estimation algorithms on quantum hardware and gain practical quantum advantage, while far from the capabilities of current devices, are in line with medium-term device growth estimates.
            
        \end{itemize}

\pagenumbering{roman}

\newpage
\newpage\null\thispagestyle{empty}\newpage

\section*{Conclusions}\label{pt:conclusions}
\addcontentsline{toc}{section}{Conclusions}
\thispagestyle{empty}
\markboth{Conclusions}{Conclusions}

\lettrine[lines=1, findent=2pt]{\resizebox{!}{1.2\baselineskip}{T}}{}his thesis has focused on the study of algorithms that follow the hybrid classical-quantum paradigm. As has been explained, this paradigm represents a middle way that tries to take advantage of the widely studied classical algorithms and the new possibilities offered by quantum algorithms. Although it may not be a definitive solution, it seems more logical to move towards a scenario in which the algorithms are either classical or quantum, it is an intermediate step that facilitates the execution of quantum algorithms and possible scenarios in which existing algorithms are accelerated in highly complex problems.

Hybrid algorithms encompass a heterogeneous group of techniques ranging from classical algorithms with a quantum module in charge of performing a complex but bounded task, algorithms that only include a classical preprocessing module to reduce the volume of data to be encoded in the classical circuit, to architectures in which the quantum circuit is optimized by a classical algorithm. There are even other formulations in which a classical algorithm analyzes a quantum algorithm to optimize its cost and analyze its properties.

The first problems that quantum computing turned its attention to were simulation problems. Later, it was realized that the best scenario for quantum simulation with practical applications was quantum chemistry. In the next generation of quantum computing development and with more algorithms proposed, optimization problems also began to be thought of as good candidates to be accelerated by quantum algorithms. Therefore, in this work, these two families of problems, optimization and quantum chemistry, have been chosen to apply the developed algorithms. The motivation behind this choice was to create algorithms that can have a high impact on the industrial sector.

In relation to optimization problems, the study has focused on a specific subcategory, namely search \& sample algorithms. This subcategory identifies algorithms capable of performing a search in a state space to find the distribution of the data, or a minimum or a maximum. For this purpose, the main classical algorithms belonging to this category have been analyzed, and their limitations have been identified.

Different quantum proposals and their limitations were analyzed. Then, one has chosen the one that has until now had aroused less interest in the world of quantum optimization, the quantum walks. Using quantum walks as the central algorithm, other authors have built on them a quantum version of the Metropolis-Hastings algorithm. In this work, these versions of the M-H quantum algorithm were taken as a reference and modified to implement it in a quantum circuit. This implementation has resulted in a software tool called quantum Metropolis Solver, QMS, which has been applied to different use cases.

QMS has been applied to the artificial intelligence use case because the fundamental process of many AI algorithms is search, especially machine learning algorithms. ML algorithms perform an abstraction and inference process, in which they require a search process to find the best hypothesis to explain the input examples. Since performing tests on these algorithms is too complex due to the limited capabilities of quantum hardware, a reduced problem that is classically used as a benchmark for these problems, the N-Queen problem, has been used. The results show that there is a polynomial advantage of the hybrid algorithm over the purely quantum algorithm.

 Another field in which there are numerous problems that can be understood in the context of optimization and where classical algorithms have strong scalability limitations is the analysis of space exploration data. In this use case, this work focused on the analysis of gravitational waves. For this purpose, QMS was adapted to the gravitational wave parameter estimation problem, observing quantum advantage over a classical algorithm used by the LIGO collaboration to obtain its results.

The last use case in which QMS was applied is the protein folding problem. In this case, there are many processes in nature that can only be simulated by trying many combinations until the one that best fits the description of the problem is found. In this case, QMS was applied to protein folding as an assistant to a classical deep learning algorithm. Results were obtained that validated the initial intuition that a quantum search algorithm was able to improve the initial results offered by a deep learning algorithm.

In the quantum chemistry part \ref{pt:qchem}, a different methodology for hybrid algorithms has been presented. In this case, the hybrid algorithm allows a detailed analysis of the cost of quantum algorithms when executed on a quantum device.

For this purpose, a software tool, TFermion, has been designed to calculate the T-gate cost of a quantum algorithm applying Quantum Phase Estimation. The results have made it possible to understand and compare the cost of state-of-the-art quantum chemistry algorithms and the evolution required by quantum devices to run these algorithms.

Finally, the application of TFermion on a real problem with industrial interest, the design of electric batteries, has allowed to understand the usefulness of this tool. The results of TFermion in this problem have shown that the algorithms based on simulation of the Hamiltonian by qubitization, plane waves to represent the quantum state and first quantization to encode the system, obtain the best results for the design of the cathode.

This thesis aims to demonstrate that hybrid algorithms offer, at present, many advantages for the main problems that are candidates for the early fault-tolerant quantum advantage. In addition, this hybrid paradigm enables the implementation of the algorithms in current and future devices in the near term by reducing the size and depth of quantum circuits. Hybrid algorithms do not seem to be the ultimate solution but they are a small step for algorithm design but a big step for the quantum computing giant.

\newpage

\thispagestyle{empty}
\bibliographystyle{ieeetr}
\bibliography{bibliography}
\addcontentsline{toc}{section}{References}

\newpage\null\thispagestyle{empty}\newpage
\newpage

\end{document}